\documentclass[10pt]{article}
\pdfoutput=1

\usepackage[utf8]{inputenc}
\usepackage[T2A]{fontenc}

\input{epsf}
\usepackage{epsfig}
\usepackage{amssymb}
\usepackage{amsfonts}
\usepackage{amsbsy}
\usepackage[cmtip,all]{xy}
\usepackage{amsmath}
\usepackage{esint}
\usepackage{collectbox}
\usepackage{amscd}
\usepackage{mathrsfs}
\usepackage{amsmath,amsthm}
\usepackage{enumitem}
\usepackage{dsfont}
\usepackage{soul}
\usepackage{comment}
\usepackage{cite} 
\usepackage[most]{tcolorbox}
\usepackage{stmaryrd}
\usepackage{accents}
\usepackage{xfrac}
\usepackage{bm}

\usepackage{xcolor}
\definecolor{color1}{RGB}{113,86,96}   
\definecolor{color2}{RGB}{132,100,112} 
\definecolor{color3}{RGB}{86,96,113}   
\definecolor{color4}{RGB}{114,127,150} 
\definecolor{color5}{RGB}{195,159,114} 
\definecolor{color6}{RGB}{213,191,149} 
\definecolor{color7}{RGB}{102,119,98}  
\definecolor{color8}{RGB}{135,158,130} 
\usepackage[linktocpage=true,colorlinks=true,linkcolor=color4,citecolor=black!10!color5,urlcolor=color7]{hyperref}

\usepackage{tikz}
\usepackage{tikz-cd}
\usetikzlibrary{arrows}
\usetikzlibrary{arrows.meta}
\usetikzlibrary{positioning}
\usetikzlibrary{shapes}
\usetikzlibrary{fit}
\usetikzlibrary{decorations.pathmorphing,decorations.pathreplacing,decorations.markings}
\usetikzlibrary{calc}

\usepackage[font=footnotesize,labelfont=bf]{caption}

\theoremstyle{definition}

\DeclareMathAlphabet{\mathpzc}{OT1}{pzc}{m}{it}

\def\I{{\rm i}}

\def\Tr{{\rm Tr}}

\def\p{\partial}

\def\CA{{\cal A}}
\def\CB{{\cal B}}

\def\CF {{\cal F}}

\def\CN {{\cal N}}
\def\CO {{\cal O}}

\def\CV {{\cal V}}

\def\CO {{\cal O}}

\def\CB {{\cal B}}

\def\CS {{\cal S}}

\def\IC{\mathbb{C}}

\def\IP{\mathbb{P}}

\def\IZ{{\mathbb{Z}}}

\def\fa{\mathfrak{a}}

\def\ff{\mathfrak{f}}

\def\fg{\mathfrak{g}}

\def\fl{\mathfrak{l}}

\def\fs{\mathfrak{s}}

\def\fs{\mathfrak{s}}

\def\fA{\mathfrak{A}}

\def\fL{\mathfrak{L}}

\def\fQ{\mathfrak{Q}}

\def\myA{{\mathsf{A}}}

\def\myY{{\mathsf{Y}}}
\def\myT{{\mathsf{T}}}

\def\sD{{\mathscr{D}}}
\def\sH{{\mathscr{H}}}
\def\sU{{\mathscr{U}}}
\def\sV{{\mathscr{V}}}

\def\lm{\limits}

\hoffset= - 1.0in

\numberwithin{equation}{section}

\tcbset{boxrule=0.6pt,colback=white!97!orange}

\DeclareSymbolFont{bbsymbol}{U}{bbold}{m}{n}
\DeclareMathSymbol{\bbzero}{\mathbin}{bbsymbol}{"30}
\DeclareMathSymbol{\bbone}{\mathbin}{bbsymbol}{"31}
\DeclareMathSymbol{\bbtwo}{\mathbin}{bbsymbol}{"32}
\DeclareMathSymbol{\bbthree}{\mathbin}{bbsymbol}{"33}
\DeclareMathSymbol{\bbfour}{\mathbin}{bbsymbol}{"34}
\DeclareMathSymbol{\bbfive}{\mathbin}{bbsymbol}{"35}
\DeclareMathSymbol{\bbsix}{\mathbin}{bbsymbol}{"36}
\DeclareMathSymbol{\bbseven}{\mathbin}{bbsymbol}{"37}
\DeclareMathSymbol{\bbeight}{\mathbin}{bbsymbol}{"38}
\DeclareMathSymbol{\bbnine}{\mathbin}{bbsymbol}{"39}

\newcommand\sqbox[1]{{
		\setbox0=\hbox{\mbox{$\Box$}}
		\setbox1=\hbox{\mbox{\raisebox{0.35ex}{\small #1}}}
		\mbox{\raisebox{-0.2ex}{\rlap{\hbox to \wd0{\hss{\box1}\hss}}\box0}}
}}
\newcommand\ssqbox[1]{{
		\setbox0=\hbox{\mbox{$\scriptstyle\Box$}}
		\setbox1=\hbox{\mbox{\raisebox{0.35ex}{\tiny #1}}}
		\mbox{\raisebox{-0.2ex}{\rlap{\hbox to \wd0{\hss{\box1}\hss}}\box0}}
}}

\begin{document}

\hfill MIPT/TH-33/26

\hfill ITEP/TH-43/26

\hfill IITP/TH-36/26

\vskip 1.5in
\begin{center}
	
    {\bf\Large On Tunnel Fusion and Yangian $R$-matrices}
	
	\vskip 0.2in
	\renewcommand{\thefootnote}{\fnsymbol{footnote}}
	{Dmitry Galakhov
		\footnote[2]{e-mail: d.galakhov.pion@gmail.com, galakhov@itep.ru}}\\
	
	\vskip 0.2in
	\renewcommand{\thefootnote}{\roman{footnote}}
	{\small{
			\textit{
				MIPT, 141701, Dolgoprudny, Russia
			}
			\vskip 0 cm
			\textit{
				NRC “Kurchatov Institute”, 123182, Moscow, Russia
			}
			\vskip 0 cm
			\textit{
				IITP RAS, 127051, Moscow, Russia}
			\vskip 0 cm
			\textit{
				ITEP, Moscow, Russia}
	}}
\end{center}

\vskip 0.2in
\baselineskip 16pt

\centerline{ABSTRACT}

\bigskip

{\footnotesize
	XXX spin-chain Hamiltonians can be derived within the $R$-matrix formalism.
	For a simple Lie algebra $\mathfrak{g}$, the rational $R$-matrix $R$ intertwines the coproduct $\Delta$ in the Yangian $\mathsf{Y}(\mathfrak{g})$.
	$R$ admits a particular Gauss decomposition $R=LDU$, where the \textit{upper-triangular} matrix $U$ can be obtained from the lower-triangular matrix $L$.
	We treat $U$ in three ways.
	Algebraically, $U$ is known as \textit{a factorization twist}.
	It plays the role of a ``universalizer'': it conjugates a \textit{non-universal} pseudo-coproduct $\mathscr{D}$ in the molten crystal chain basis to the true Yangian coproduct $\Delta=U^{-1}\mathscr{D} U$.
	Physically, $U$ can be obtained as a Berry phase in SUSY quiver QFT for a quiver associated with the Dynkin diagram of $\mathfrak{g}$.
	In this case, $U$ is calculated as a sum over \textit{tunneling} amplitudes.
	Geometrically, $U$ can be interpreted as a renormalized Maulik-Okounkov stable envelope localized at a fixed point.
	In the physical setting, the Yang-Baxter equation (YBE), which ensures integrability, follows from the flatness of the Berry connection, which in turn follows from supersymmetry.
	However, the YBE imposes subtle constraints on the tunneling amplitudes, which we examine here.
	We construct recursion relations that allow higher collective amplitudes to be obtained from more elementary ones.
	We observe that different tunneling channels are in fact intertwined, leading to finer cancellations among amplitudes and to wall-crossing phenomena: depending on the parameters, the same amplitude is represented as a sum over different channels.
	We focus on $\mathsf{Y}(\mathfrak{sl}_2)$ with higher-spin and non-evaluation representations, as well as on the fundamental representation of $\mathsf{Y}(\mathfrak{sl}_{n+1})$.

}

\bigskip

\bigskip

\tableofcontents

\bigskip


\section{Introduction}

In modern language, the description of any physical system begins with a notion of \emph{symmetry}.
In most cases, symmetry is the only available tool for approaching non-perturbative phenomena.
The notion of symmetry has, of course, been transformed over recent decades, and now incorporates hidden symmetries of string theory \cite{Morozov:1992ea,Mironov:1993wi} as well as categorical and higher-form symmetries \cite{Bah:2022wot,Cordova:2022qga,Brennan:2023rmb,Luo:2024nsj,Freed:2022qja}.
The canonical mathematical realization of a physical symmetry is a group, or a Lie algebra when the symmetry is continuous.
This identification has likewise been transformed toward broader categorical concepts.

Here we are interested in a relatively intermediate situation, namely that of quantum algebras.
Non-perturbative effects render the original symmetries anomalous, but they do not break them completely; rather, the underlying Lie algebra is deformed, or ``quantized.''
Usually this signals the presence of a larger and deeper symmetry (possibly supersymmetry) in the system, one that allows some perturbative aspects to survive the non-perturbative RG flow.
As a consequence, such systems are quite often related to the notion of \emph{integrability} (see e.g. \cite{Morozov:1994hh,Mironov:1994sf,Gorsky:1995zq,Mironov:2022fsr,Yamazaki:2025gti,Bombardelli:2016prim,Komatsu:2026tdi}).

For supersymmetric QFTs, the notion of integrability can be viewed within the broader framework of the gauge/Bethe correspondence \cite{Nekrasov:2009uh,Nekrasov:2009ui}.
The quantum algebra in this setting is the Yangian \cite{Drinfeld1,Drinfeld2} or one of its higher-loop deformations, such as a quantum loop algebra, a quantum toroidal algebra \cite{Moody:1990toroidal}, or an elliptic algebra \cite{Feigin:1995elliptic}.
The physical side of the correspondence provides avatars for geometric and algebraic quantities in terms of QFT observables.
The very notion of the gauge/Bethe correspondence begins with an association of Bethe roots with classical isolated vacua in  2d $\CN=(2,2)$ quiver theories.

Here we concentrate on the properties of Yangian $R$-matrices $R$.
Like any matrix, $R$ has a Gauss decomposition $R=LDU$, where we normalize the lower-triangular matrix $L$ and the upper-triangular matrix $U$ so that both are unipotent.
For Yangian $R$-matrices, this decomposition has certain peculiarities.
These $R$-matrices depend on the spectral parameters $u_{1,2}$ of the two channels only through their difference $u_{12}:=u_1-u_2$.
In the decomposition $R(u)=L(u)D(u)U(u)$, the matrix-valued functions $L(u)$ and $U(u)$ are not independent; rather, $U(u)=\left(PL(-u)P\right)^{-1}$, where $P$ is the permutation operator.

This decomposition has a threefold meaning: algebraic, physical, and geometric.
Algebraically, $U$ is known as a factorization (Drinfeld) twist \cite{drinfeld1990quasi,Pfeiffer2000Yangian,Terras2000,Pfeiffer2001QuantumAffine,Albert2000XYZ,Yang2004,Torrielli2023,Maillet2000}.
It transforms the complicated universal Yangian coproduct into a simpler, though non-universal, one that depends explicitly on the representation.
Physically, the full $R$-matrix is represented as an interface \cite{Brunner:2007Btype,Brunner:2008Defects,Gaiotto:2015aoa,Galakhov:2020upa,Dedushenko:2021mds,Brunner:2021ulc,Dedushenko:2023qjq,Haouzi:2023doo,Khan:2024yiy,Cordova:2024iti,Brunner:2024quk} in a supersymmetric \emph{quiver} QFT, in which certain real parameters $x_1$ and $x_2$ are interchanged as one moves from $x_1<x_2$ to $x_2<x_1$.
Thus $U$ and $L$ correspond to the ``halves'' of such an interface, at which the spectral parameters ``fuse'' with one another.
Certain properties of such a quiver theory concerning its IR behavior can be described in terms of the geometry of \emph{quiver} varieties.
In these terms, $U$ is the decomposition matrix relating the (renormalized) Maulik-Okounkov stable basis \cite{MO} to the fixed-point basis in equivariant cohomology.

From a practical point of view, the algebraic construction of the matrices $U$, $L$, and ultimately $R$ is the simplest.
As we will discuss, recursion relations allow one to fix $U$ explicitly for any Yangian $\myY(\fg)$.
The stable-envelope side, however, comes with several complications.
Its definition is involved and difficult to implement from first principles.
It was designed primarily for Nakajima quiver varieties, which correspond to minuscule highest weights; a corrected version that extends to non-Nakajima varieties exists \cite{Ishtiaque:2023acr,Cao:2025yng,Cao:2026pir,Cao:2026vwf}, but it is recent and remains an area of active research.
Moreover, the stable envelope is hard to translate into the language of physics, since it has no natural physical counterpart.
This is in contrast with the fixed-point basis, which corresponds simply to the classical vacua, and with the cohomology classes built on it, which correspond to localized ground-state wave functions \cite{Witten:1982im}.
An alternative available in the literature is the rather involved notion of supersymmetric interfaces \cite{Dedushenko:2021mds,Bullimore:2017lwu}.\footnote{It is also expected \cite{Crew:2020psc} that a specific choice of boundary conditions for the hemisphere partition function of 3d $\mathcal{N}=4$ theories produces elliptic stable envelopes.}
We hope that the present discussion will help to make this transition from geometry to physics simpler and more transparent.

The physics side has accompanying complications of its own.
Berry transport corresponding to an interface is constructed as an evolution operator and expanded over saddle-point field configurations, namely tunneling instantons; effectively, each saddle point yields a tunneling amplitude.
Classical vacua in this model are represented by crystal chains: effective quasiparticles (cells) occupy the minima of an external potential.
Inside each minimum, the cells are strongly correlated and generate distinct formations -- crystals, or partitions \cite{Okounkov:2003sp,Ooguri:2009ijd,Li:2020rij,Bao:2024ygr} -- described combinatorially and pictorially as Young diagrams.
In this picture, an instanton describes the migration of cells from one partition to another.
The main difficulty is that multi-cell tunneling amplitudes are hard to describe and compute, because the cells become strongly correlated in the flow.
The recursion relations for $U$ can be derived from the YBE, which the $R$-matrix satisfies.
On the physics side, the YBE follows from the local flatness of the Berry connection, which in turn follows from supersymmetry.
Here we translate these relations into recursion relations for tunneling amplitudes, allowing higher multi-particle amplitudes to be obtained from a few elementary ones.
We then examine whether the corresponding relations are produced on the tunneling trajectories themselves, and whether multi-cell trajectories can be constructed from elementary pieces.
Such a gluing construction does indeed take place.
However, we observe that the different tunnels entering the sum over trajectories for a given amplitude are intertwined, leading to finer cancellations among amplitudes, as required for the recursion relations to hold.
We also observe wall-crossing phenomena, in which the same amplitude, in different parameter windows, is represented by sums over different tunnels.

Here we encounter the following tunneling phenomena:
\begin{itemize}
	\item $\myY(\fs\fl_{n+1})$, fundamental representation (Sec.~\ref{sec:sl(n+1)_fu}).
	In this case, the tunnels are represented by multi-cell jumps.
	They are, however, relatively simple to describe, and they are the elementary tunnels from which the others are glued.
	Multiple cells in a tunneling flow are correlated into a single jumping ``blob'', reflecting the presence of roots of $\fs\fl_{n+1}$ other than the simple ones.
	In other words, the simplicity of these amplitudes reflects an emergent Weyl symmetry.
	\item $\myY(\fs\fl_{2})$, non-evaluation representation (Sec.~\ref{sec:sl(2)_nonev}).
	In this case, there are multiple tunneling paths between two vacua.
	We observe that these paths can be glued from more elementary ones: for example, in a single process, $k+m$ cells migrate by first moving $k$ cells and then the remaining $m$.
	We also observe finer cancellations among amplitudes and wall-crossing phenomena.
	\item $\myY(\fs\fl_{2})$, higher-spin representation (Sec.~\ref{sec:sl(2)_hs}).
	In this case, the tunneling trajectories that constitute the attracting sets supporting the corresponding stable envelopes are more involved, yet describable in simple terms.
	Here the two ways of obtaining the amplitudes can be compared explicitly.
	A single $(k+1)$-cell tunnel is obtained by recursively gluing a one-cell tunnel and a $k$-cell tunnel.
\end{itemize}

The paper is organized as follows.
In Sec.~\ref{sec:Yangian}, we give a brief review of Yangians and spin chains and introduce the definition of the $R$-matrix.
In Sec.~\ref{sec:fact_twist}, we review the construction of crystal representations for quiver Yangians, including crystal chains.
We define a pseudo-coproduct as a natural representation on crystal chains and introduce the factorization twist $U$ as an intertwining operator between the pseudo-coproduct and the true coproduct.
In Sec.~\ref{sec:quiver}, we review the QFT and geometric approaches to algebraic quantities.
We explain the construction of the $R$-matrix and of the factorization twist from tunneling amplitudes and stable envelopes.
In Sec.~\ref{sec:fusion}, we construct the recursion relations for tunneling amplitudes and examine several examples in detail.


\section{Yangian side}\label{sec:Yangian}

\subsection{Spin chains}\label{sec:spin_chains}

 We start with a brief review of the canonical $R$-matrix approach to constructing eigenstates of Heisenberg XXX spin-chain Hamiltonians.
 This is standard textbook material \cite{Sklyanin:1979pfu,TakhtajanFaddeev1979,slavnov2019algebraicbetheansatz}, yet we use this review section to introduce our notation, which we will use throughout the paper.
 
 We consider a tensor product of representations of the Yangian $\myY(\fg)$\footnote{
 	Correspondingly, XXZ and XYZ spin chains require considering the quantum loop algebra $U_q(\hat{\fg})$ and the elliptic quantum group $E_{\tau,q}(\fg)$, constructed from $\fg$ by extra layers of loops and quantizations \cite{Konno2020,Jimbo1999}.
 	We do not consider them here.
 } of a simple Lie algebra $\fg$.
 Canonically, $\fg$ is contained in $\myY(\fg)$.
 Moreover, there are evaluation representations of $\myY(\fg)$ that are pullbacks of the corresponding $\fg$-modules along the evaluation homomorphism.
 In plain language, when viewed as a representation of the subalgebra $\fg\subset\myY(\fg)$, it is a canonical representation of $\fg$.
 Evaluation representations of $\myY(\fg)$ can be equipped with a parameter $u$ referred to as a spectral parameter.
 Other non-evaluation modules of $\myY(\fg)$ can be parameterized by a tuple of spectral parameters $\vec u$.
 Although non-evaluation modules are not usually considered in applications to spin chains, we will discuss them in Sec.~\ref{sec:sl(2)_nonev}.
 The $R$-matrix machinery can also be extended to these modules.
 In this section, when we write a spectral parameter $u$, we allow the whole tuple $\vec u$ to be used in principle. 
 Graphically, we depict such a tensor product as a collection of parallel strands labeled by spectral parameters, so that operators acting on this space are depicted as crossings.
 We assume that operators are applied in order from bottom to top:
 \begin{equation}
 	\rho^{(1)}_{u_1}\otimes \rho^{(2)}_{u_2}\otimes \rho^{(3)}_{u_3}\otimes \rho^{(4)}_{u_4}\otimes \rho^{(5)}_{u_5} \quad\rightsquigarrow\quad \begin{array}{c}
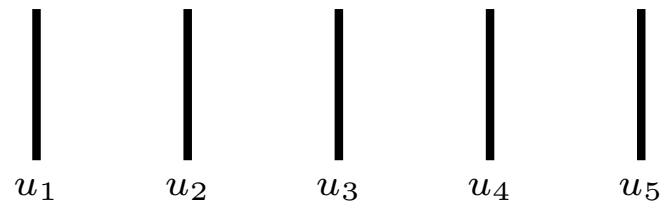
 		\begin{tikzpicture}
 			\foreach \i/\t in {0/1, 1/2, 2/3, 3/4, 4/5}
 			{
 				\draw[ultra thick] (\i,0) -- (\i,1);
 				\node[below] at (\i,0) {$\scriptstyle u_{\t}$};
 			}
 		\end{tikzpicture}
 	\end{array}\,.
 \end{equation}
 
 The $R$-matrix $\check R_{12}(u_1,u_2):\; \rho_{u_1}^{(1)}\otimes \rho_{u_2}^{(2)}\to \rho_{u_2}^{(2)}\otimes \rho_{u_1}^{(1)}$ acts on a pair of neighboring representations in a product and permutes them.
 One considers also a version of the $R$-matrix combined with the permutations operator $P:\; a\otimes b\mapsto b\otimes a$, so that $R_{12}(u_1,u_2)=P_{12}\check R_{12}(u_1,u_2):\; \rho_{u_1}^{(1)}\otimes \rho_{u_2}^{(2)}\to \rho_{u_1}^{(1)}\otimes \rho_{u_2}^{(2)}$.
 The latter version of the $R$-matrix is extended naturally $R_{ij}(u_i,u_j)$ to any pair of representations with indices $i<j$.
 We depict them accordingly:
 \begin{equation}
 	\check R_{12}(u_1,u_2)\;\rightsquigarrow\;\begin{array}{c}
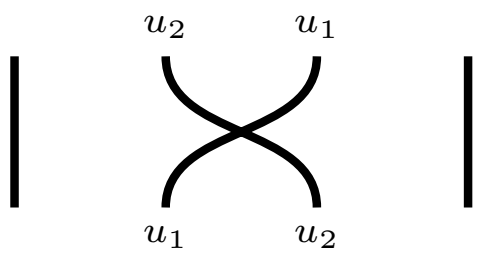
 		\begin{tikzpicture}
 			\draw[ultra thick] (0,0) -- (0,1) (1,0) to[out=90,in=270] (2,1) (2,0) to[out=90,in=270] (1,1) (3,0) -- (3,1);
 			\node[below] at (1,0) {$\scriptstyle u_1$};
 			\node[below] at (2,0) {$\scriptstyle u_2$};
 			\node[above] at (1,1) {$\scriptstyle u_2$};
 			\node[above] at (2,1) {$\scriptstyle u_1$};
 		\end{tikzpicture}
 	\end{array},\quad R_{ij}(u_i,u_j)\;\rightsquigarrow\;\begin{array}{c}
 		
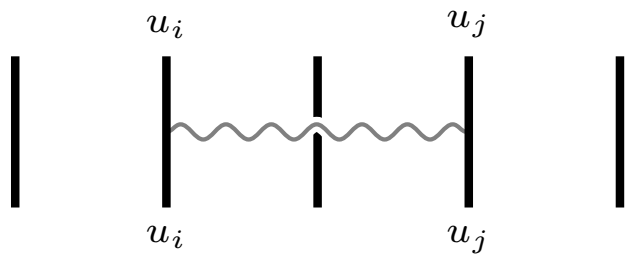
\begin{tikzpicture}
 			\draw[ultra thick] (0,0) -- (0,1) (2,0) -- (2,1) (4,0) -- (4,1);
 			\node[below] at (1,0) {$\scriptstyle u_i$};
 			\node[below] at (3,0) {$\scriptstyle u_j$};
 			\node[above] at (1,1) {$\scriptstyle u_i$};
 			\node[above] at (3,1) {$\scriptstyle u_j$};
 			\draw[line width=1mm, white, decorate,
 			decoration={snake, amplitude=0.5mm, segment length=3mm}] (1,0.5) -- (3,0.5);
 			\draw[thick, gray, decorate,
 			decoration={snake, amplitude=0.5mm, segment length=3mm}] (1,0.5) -- (3,0.5);
 			\draw[ultra thick] (1,0) -- (1,1) (3,0) -- (3,1);
 		\end{tikzpicture}
 	\end{array}\,.
 \end{equation}
 
The $R$-matrix satisfies the Yang-Baxter equation (YBE):
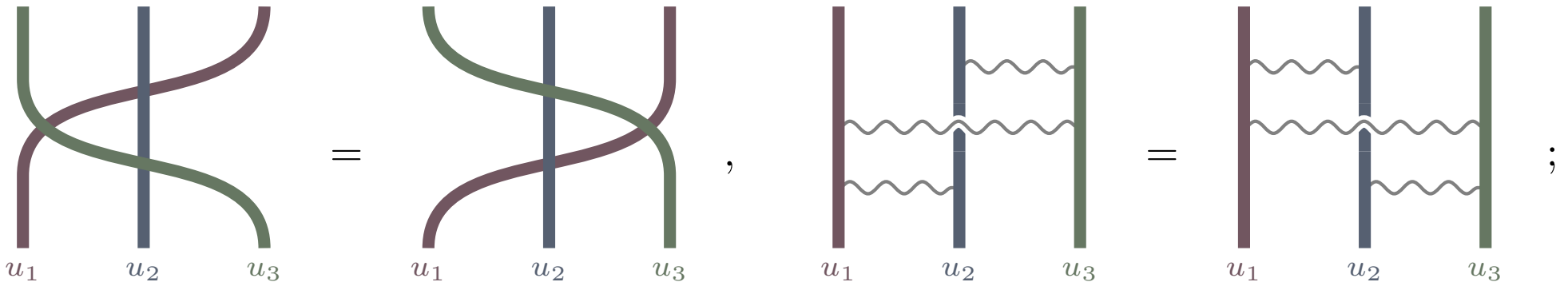
\begin{equation}\label{YBE}
	\begin{aligned}
	&\begin{array}{c}
		\begin{tikzpicture}
			\draw[line width = 1mm, color1] (0,0) -- (0,0.6) to[out=90,in=270] (2,2);
			\draw[line width = 1mm, color3] (1,0) -- (1,2);
			\draw[line width = 1mm, color7] (2,0) to[out=90,in=270] (0,1.4) -- (0,2);
			\node[below, color1] at (0,0) {$\scriptstyle u_1$};
			\node[below, color3] at (1,0) {$\scriptstyle u_2$};
			\node[below, color7] at (2,0) {$\scriptstyle u_3$};
		\end{tikzpicture}
	\end{array}=\begin{array}{c}
	\begin{tikzpicture}
		\draw[line width = 1mm, color1] (0,0) to[out=90,in=270] (2,1.4) -- (2,2);
		\draw[line width = 1mm, color3] (1,0) -- (1,2);
		\draw[line width = 1mm, color7] (2,0) -- (2,0.6) to[out=90,in=270] (0,2);
		\node[below, color1] at (0,0) {$\scriptstyle u_1$};
		\node[below, color3] at (1,0) {$\scriptstyle u_2$};
		\node[below, color7] at (2,0) {$\scriptstyle u_3$};
	\end{tikzpicture}
	\end{array},\quad \begin{array}{c}
	\begin{tikzpicture}
		\draw[line width = 1mm, color3] (1,0.8) -- (1,1.2);
		\draw[line width=1mm, white, decorate,
		decoration={snake, amplitude=0.5mm, segment length=3mm}] (0,1) -- (2,1);
		\draw[thick, gray, decorate,
		decoration={snake, amplitude=0.5mm, segment length=3mm}] (0,1) -- (2,1);
		\draw[thick, gray, decorate,
		decoration={snake, amplitude=0.5mm, segment length=3mm}] (0,0.5) -- (1,0.5);
		\draw[thick, gray, decorate,
		decoration={snake, amplitude=0.5mm, segment length=3mm}] (1,1.5) -- (2,1.5);
		\draw[line width = 1mm, color1] (0,0) -- (0,2);
		\draw[line width = 1mm, color3] (1,0) -- (1,0.8) (1,1.2) -- (1,2);
		\draw[line width = 1mm, color7] (2,0) -- (2,2);
		\node[below, color1] at (0,0) {$\scriptstyle u_1$};
		\node[below, color3] at (1,0) {$\scriptstyle u_2$};
		\node[below, color7] at (2,0) {$\scriptstyle u_3$};
	\end{tikzpicture}
	\end{array}=\begin{array}{c}
	\begin{tikzpicture}
		\draw[line width = 1mm, color3] (1,0.8) -- (1,1.2);
		\draw[line width=1mm, white, decorate,
		decoration={snake, amplitude=0.5mm, segment length=3mm}] (0,1) -- (2,1);
		\draw[thick, gray, decorate,
		decoration={snake, amplitude=0.5mm, segment length=3mm}] (0,1) -- (2,1);
		\draw[thick, gray, decorate,
		decoration={snake, amplitude=0.5mm, segment length=3mm}] (0,1.5) -- (1,1.5);
		\draw[thick, gray, decorate,
		decoration={snake, amplitude=0.5mm, segment length=3mm}] (1,0.5) -- (2,0.5);
		\draw[line width = 1mm, color1] (0,0) -- (0,2);
		\draw[line width = 1mm, color3] (1,0) -- (1,0.8) (1,1.2) -- (1,2);
		\draw[line width = 1mm, color7] (2,0) -- (2,2);
		\node[below, color1] at (0,0) {$\scriptstyle u_1$};
		\node[below, color3] at (1,0) {$\scriptstyle u_2$};
		\node[below, color7] at (2,0) {$\scriptstyle u_3$};
	\end{tikzpicture}
	\end{array}\,;\\
	&\check R_{23}(u_1,u_2)\check R_{12}(u_1,u_3)\check R_{23}(u_2,u_3)=\check R_{12}(u_2,u_3)\check R_{23}(u_1,u_3)\check R_{12}(u_1,u_2)\,,\\
	& R_{23}(u_2,u_3)R_{13}(u_1,u_3)R_{12}(u_1,u_2)=R_{12}(u_1,u_2)R_{13}(u_1,u_3)R_{23}(u_2,u_3)\,.
\end{aligned}
\end{equation}

The transfer matrix $T(v)$ acting on a ``chain'' (tensor product) of $n$ representations of $\fg$ can be compactly depicted in this graphical language as a collection of $n$ strands crossed by a single ``control'' strand with the spectral parameter $v$:
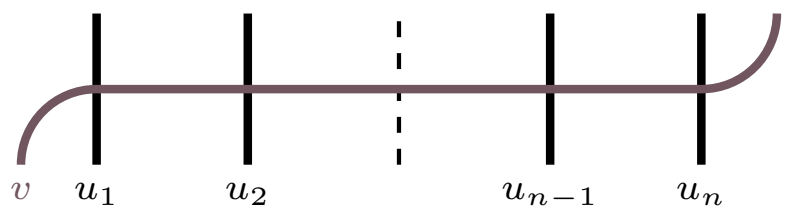
\begin{equation}\label{transfer}
	T({\color{color1} v}|u_1,\ldots, u_n)=\begin{array}{c}
		\begin{tikzpicture}
			\draw[ultra thick] (0,0) -- (0,1) (1,0) -- (1,1) (3,0) -- (3,1) (4,0) -- (4,1);
			\draw[thick, dashed] (2,0) -- (2,1);
			\draw[ultra thick, color1] (-0.5,0) to[out=90,in=180] (0,0.5) -- (4,0.5) to[out=0,in=270] (4.5,1);
			\node[below, color1] at (-0.5,0) {$\scriptstyle v$};
			\node[below] at (0,0) {$\scriptstyle u_1$};
			\node[below] at (1,0) {$\scriptstyle u_2$};
			\node[below] at (3,0) {$\scriptstyle u_{n-1}$};
			\node[below] at (4,0) {$\scriptstyle u_n$};
		\end{tikzpicture}
	\end{array}\,.
\end{equation}

Partial traces $\tau(v)$ of the transfer matrix over the space corresponding to the strand labeled by the spectral parameter $v$ commute for different values of $v$:
\begin{equation}
	\left[\tau(v),\tau(v')\right]=0\,.
\end{equation}
This relation follows directly from the YBE.
Therefore, if the transfer matrix is expanded as a series $\tau(v)=1+\sum\lm_{k=0}^{\infty} H_k v^{-k-1}$, the corresponding operators $H_k$ form sufficiently many commuting Hamiltonians on the tensor product $\myY(\fg)^{\otimes n}$.
If the $R$-matrix solution to the YBE takes the form $R(z)=1+r/z+O(z^2)$, where $r$ is the classical $R$-matrix, then $H_0$ is the canonical Hamiltonian of an XXX spin chain written in terms of $\fg\subset \myY(\fg)$.

Individual elements of the transfer matrices $T(v)$ are used to generate Bethe vectors and to obtain Bethe ansatz equations for the Hamiltonian eigenvalues.
These relations also have a nice pictorial realization \cite{Litvinov:2020zeq,Yang:2024hsp}.
However, this subject is beyond our current scope, and we do not consider it here.


\subsection{Yangian \texorpdfstring{$\myY(\fs\fl_{n+1})$}{Y(sl(n+1))}}

We use the second Drinfeld (current) presentation of $\myY(\fs\fl_{n+1})$ \cite{Drinfeld2} (sometimes denoted as $\myY_{\hbar}(\fs\fl_{n+1})$ to make the deformation parametrization $\hbar$ explicit).
The algebra is generated by the modes of the Chevalley triplets of the raising, lowering and Cartan  generators $({}^{+}e_k^a,{}^{-}e_k^a,h_k^a)$, where $k\in \IZ_{\geq 0}$.
The commutation relations can be expressed in terms of the Cartan matrix $A_{ab}=2\delta_{a,b}-\delta_{a,b+1}-\delta_{a+1,b}$ of $\fs\fl_{n+1}$:
\begin{equation}\label{Yangian}
	\begin{aligned}
		& \left[h_k^a,h_m^b\right]=0,\quad \left[h_0^a,{}^{\pm}e_k^b\right]=\pm A_{ab}{}^{\pm}e_k^b,\quad \left[{}^+e_k^a,{}^-e_m^b\right]=\delta_{ab}h_{k+m}^a\,,\\
		& \left[{}^{\pm}e_{k+1}^a,{}^{\pm}e_{m}^b\right]-\left[{}^{\pm}e_{k}^a,{}^{\pm}e_{m+1}^b\right]=\pm\hbar A_{ab}\left\{{}^{\pm}e_{k}^a,{}^{\pm}e_{m}^b\right\}\,,\\
		& \left[h_{k+1}^a,{}^{\pm}e_{m}^b\right]-\left[h_{k}^a,{}^{\pm}e_{m+1}^b\right]=\pm\hbar A_{ab}\left\{h_k^a,{}^{\pm}e_{m}^b\right\}\,,\\
		& \mathop{\rm Sym}\lm_{k,l,m}\left[{}^{\pm}e_k^{a},\left[{}^{\pm}e_l^{a},{}^{\pm}e_m^{a\pm 1}\right]\right]=0,\quad \left[{}^{\pm}e_{k}^a,{}^{\pm}e_{m}^b\right]=0\mbox{ for }A_{ab}=0\,.
	\end{aligned}
\end{equation}
The subalgebra $\fs\fl_{n+1}\subset \myY(\fs\fl_{n+1})$ is generated by the Chevalley generators ${}^{\pm}e_0^a$ and $h_0^a$.

In representations it is sufficient to construct explicitly only generators ${}^{\pm}e^a_0$, $h_0^a$, and $h_1^k$.
The remaining generators can be reconstructed from the algebraic relations, in particular:
\begin{equation}\label{mode-shift}
	{}^{\pm}e_{k+1}^a=\pm\frac{1}{2}\left[h_1^a,{}^{\pm}e_k^a\right]-\hbar\left\{h_0^a,e_k^a\right\},\quad h_k^a=\left[{}^+e_k^a,{}^-e_0^a\right]\,.
\end{equation}

$\myY(\fs\fl_{n+1})$ is a Hopf algebra. 
Here again it suffices to define the coproduct for the lower-mode generators ${}^{\pm}e^a_0$, $h_0^a$, and $h_1^k$.
For all other generators, the coproduct can be obtained from the mode-shift relations \eqref{mode-shift}.
For the lower-mode generators, we obtain:\footnote{
These formulas can be derived explicitly by passing to the first Drinfeld presentation (J-presentation) of the Yangian \cite{Drinfeld1}, which has a clear and simple coproduct.
However, the transition formulas given in \cite{Drinfeld2} are somewhat involved.
Instead, in practice, we checked that $\Delta$ in \eqref{Y_co} yields the true coproduct (a Yangian algebra homomorphism) in numerous examples.
Some alternative expressions for the coproduct in the current presentation can be found in \cite{Crampe2004,GuayNakajimaWendlandt2018}.
}
\begin{equation}\label{Y_co}
	\begin{aligned}
		& \Delta\left( {}^{\pm}e^{a}_0\right)={}^{\pm}e^{a}_0\otimes 1+1\otimes {}^{\pm}e^{a}_0\,,\\
		& \Delta\left( h_0^a\right)=h_0^a\otimes 1+1\otimes h_0^a\,,\\
		& \Delta\left( h_1^a\right)=h_1^a\otimes 1+2\hbar\, h_0^a\otimes h_0^a+1\otimes h_1^a+2\hbar\,\Omega_+^a\,,
	\end{aligned}
\end{equation}
where $\Omega_+^a$ is a sum over positive roots:
\begin{equation}\label{Omega}
	\Omega_+^a=\sum\lm_{b}\sum\lm_{c>b}\left(\delta_{b,a+1}+\delta_{c,a}-\delta_{b,a}-\delta_{c,a+1}\right)E_{cb}\otimes E_{bc}\,,
\end{equation}
where $E_{ab}$ is an operator corresponding to the matrix  unit $e_{ab}$ expressed via the Lie bracket of $\fs\fl_{n+1}$.
For example, for $\myY(\fs\fl_5)$:
\begin{equation}
	E_{ab}=\begin{pmatrix}
			0 & {}^+e_0^1 & [{}^+e_0^1,{}^+e_0^2] & [{}^+e_0^1,[{}^+e_0^2,{}^+e_0^3]] & [{}^+e_0^1,[{}^+e_0^2,[{}^+e_0^3,{}^+e_0^4]]] \\
			{}^-e_0^1 & 0 & {}^+e_0^2 & [{}^+e_0^2,{}^+e_0^3] & [{}^+e_0^2,[{}^+e_0^3,{}^+e_0^4]] \\
			[{}^-e_0^2,{}^-e_0^1] & {}^-e_0^2 & 0 & {}^+e_0^3 & [{}^+e_0^3,{}^+e_0^4] \\
			[[{}^-e_0^3,{}^-e_0^2],{}^-e_0^1] & [{}^-e_0^3,{}^-e_0^2] & {}^-e_0^3 & 0 & {}^+e_0^4 \\
			[[[{}^-e_0^4,{}^-e_0^3],{}^-e_0^2],{}^-e_0^1] & [[{}^-e_0^4,{}^-e_0^3],{}^-e_0^2] & [{}^-e_0^4,{}^-e_0^3] & {}^-e_0^4 & 0 \\
		\end{pmatrix}\,.
\end{equation}

An explicit solution to the YBE \eqref{YBE}, the $R$-matrix $R$, can be constructed as an intertwining operator for the Yangian coproduct:
\begin{tcolorbox}
\begin{equation}\label{intertw}
	R\circ \Delta(\fa)=\Delta^{\rm op}(\fa)\circ R,\quad \forall\fa\in \myY(\fg)\,,
\end{equation}
\end{tcolorbox}
\noindent where $\Delta^{\rm op}:=P\Delta P$.
It is not obvious from \eqref{intertw}. 
However, if the representations for which the $R$-matrix is constructed depend only on a single spectral parameter $u$, then the $R$-matrix is a function of $u_{12}:=u_1-u_2$, where $u_i$ is the spectral parameter of the $i^{\rm th}$ tensor component.
This follows from a Yangian automorphism, namely an overall boost for the spectral parameters $u_i\mapsto u_i+w$:
\begin{equation}
	{}^{\pm}e_k^a\mapsto\sum\lm_{j=0}^k\frac{k!}{j!(k-j)!} w^{k-j} \;{}^{\pm}e_j^a,\quad h_k^a\mapsto\sum\lm_{j=0}^k\frac{k!}{j!(k-j)!} w^{k-j}\; h_j^a\,.
\end{equation}


\section{Factorization twist}\label{sec:fact_twist}

\subsection{Quiver Yangians and crystal representations}

\emph{Quiver} Yangians,  together with their shifted and double generalizations \cite{Li:2020rij,Galakhov:2021xum,Noshita:2021dgj,Li:2023zub,Bao:2025hfu}, are associative (super)algebras with relations defined by quiver data.
For a quiver resembling the corresponding \emph{Dynkin diagram} of a simple Lie algebra $\fg$, the \emph{quiver} Yangian is expected to coincide with the Yangian $\myY(\fg)$.
Quiver Yangians admit canonical representations on molten crystals\footnote{
	Originally, as a combinatorial object, molten crystals were introduced to count BPS states in D-brane systems in type IIA compactifications on toric Calabi-Yau manifolds \cite{Okounkov:2003sp,Ooguri:2009ijd}.
	Nowadays this topic is rather broad,  and the literature is too vast to cover here.
	The interested reader can consult the survey in \cite{Bao:2024ygr}.
}.
The parameters of these representations are fixed by a choice of quiver framing.
Here we restrict ourselves to $\fg=\fs\fl_{n+1}$, i.e., to the Dynkin series $\myA_n$, although an explicit general construction is available for both simply-laced and non-simply-laced cases \cite{Cecotti:2012gh,Bao:2025hfu,Chen:2025xoe,Gavshin:2025akn,Galakhov:2026cwv}.
Let us briefly recall this correspondence.

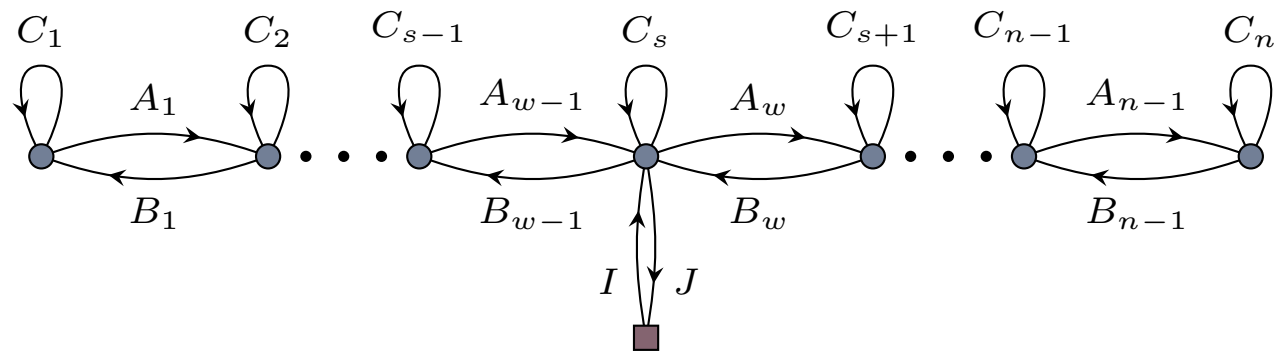
\begin{figure}[ht!]
	\centering
	\begin{tikzpicture}
		\draw[postaction=decorate, decoration={markings, mark= at position 0.7 with {\arrow{stealth}}}] (0,0) to[out=20,in=160] node[pos=0.5,above] {$\scriptstyle A_1$} (1.5,0);
		\draw[postaction=decorate, decoration={markings, mark= at position 0.7 with {\arrow{stealth}}}] (1.5,0) to[out=200,in=340] node[pos=0.5,below] {$\scriptstyle B_1$} (0,0);
		\draw[postaction=decorate, decoration={markings, mark= at position 0.8 with {\arrow{stealth}}}] (0,0) to[out=60,in=0] (0,0.6) to[out=180,in=120] (0,0);
		\node[above] at (0,0.6) {$\scriptstyle C_1$};
		\begin{scope}[shift={(1.5,0)}]
			\draw[postaction=decorate, decoration={markings, mark= at position 0.8 with {\arrow{stealth}}}] (0,0) to[out=60,in=0] (0,0.6) to[out=180,in=120] (0,0);
			\node[above] at (0,0.6) {$\scriptstyle C_2$};
		\end{scope}
		\draw[fill=color4] (0,0) circle (0.08) (1.5,0) circle (0.08);
		\begin{scope}[shift = {(4,0)}]
			\draw[postaction=decorate, decoration={markings, mark= at position 0.7 with {\arrow{stealth}}}] (0,-1.2) to[out=100,in=260] node[pos=0.3, left] {$\scriptstyle I$} (0,0);
			\draw[postaction=decorate, decoration={markings, mark= at position 0.7 with {\arrow{stealth}}}] (0,0) to[out=280,in=80] node[pos=0.7, right] {$\scriptstyle J$} (0,-1.2);
			\draw[postaction=decorate, decoration={markings, mark= at position 0.7 with {\arrow{stealth}}}] (0,0) to[out=20,in=160] node[pos=0.5,above] {$\scriptstyle A_w$} (1.5,0);
			\draw[postaction=decorate, decoration={markings, mark= at position 0.7 with {\arrow{stealth}}}] (1.5,0) to[out=200,in=340] node[pos=0.5,below] {$\scriptstyle B_w$} (0,0);
			\begin{scope}[shift={(-1.5,0)}]
				\draw[postaction=decorate, decoration={markings, mark= at position 0.7 with {\arrow{stealth}}}] (0,0) to[out=20,in=160] node[pos=0.5,above] {$\scriptstyle A_{w-1}$} (1.5,0);
				\draw[postaction=decorate, decoration={markings, mark= at position 0.7 with {\arrow{stealth}}}] (1.5,0) to[out=200,in=340] node[pos=0.5,below] {$\scriptstyle B_{w-1}$} (0,0);
			\end{scope}
			\draw[postaction=decorate, decoration={markings, mark= at position 0.8 with {\arrow{stealth}}}] (0,0) to[out=60,in=0] (0,0.6) to[out=180,in=120] (0,0);
			\node[above] at (0,0.6) {$\scriptstyle C_s$};
			\begin{scope}[shift={(1.5,0)}]
				\draw[postaction=decorate, decoration={markings, mark= at position 0.8 with {\arrow{stealth}}}] (0,0) to[out=60,in=0] (0,0.6) to[out=180,in=120] (0,0);
				\node[above] at (0,0.6) {$\scriptstyle C_{s+1}$};
			\end{scope}
			\begin{scope}[shift={(-1.5,0)}]
				\draw[postaction=decorate, decoration={markings, mark= at position 0.8 with {\arrow{stealth}}}] (0,0) to[out=60,in=0] (0,0.6) to[out=180,in=120] (0,0);
				\node[above] at (0,0.6) {$\scriptstyle C_{s-1}$};
			\end{scope}
			\draw[fill=color4] (-1.5,0) circle (0.08) (0,0) circle (0.08) (1.5,0) circle (0.08);
			\begin{scope}[shift={(0,-1.2)}]
				\draw[fill=color2] (-0.08,-0.08) -- (-0.08,0.08) -- (0.08,0.08) -- (0.08,-0.08) -- cycle;
			\end{scope}
		\end{scope}
		\begin{scope}[shift = {(8,0)}]
			\begin{scope}[shift={(-1.5,0)}]
				\draw[postaction=decorate, decoration={markings, mark= at position 0.7 with {\arrow{stealth}}}] (0,0) to[out=20,in=160] node[pos=0.5,above] {$\scriptstyle A_{n-1}$} (1.5,0);
				\draw[postaction=decorate, decoration={markings, mark= at position 0.7 with {\arrow{stealth}}}] (1.5,0) to[out=200,in=340] node[pos=0.5,below] {$\scriptstyle B_{n-1}$} (0,0);
			\end{scope}
			\draw[postaction=decorate, decoration={markings, mark= at position 0.8 with {\arrow{stealth}}}] (0,0) to[out=60,in=0] (0,0.6) to[out=180,in=120] (0,0);
			\node[above] at (0,0.6) {$\scriptstyle C_{n}$};
			\begin{scope}[shift={(-1.5,0)}]
				\draw[postaction=decorate, decoration={markings, mark= at position 0.8 with {\arrow{stealth}}}] (0,0) to[out=60,in=0] (0,0.6) to[out=180,in=120] (0,0);
				\node[above] at (0,0.6) {$\scriptstyle C_{n-1}$};
			\end{scope}
			\draw[fill=color4] (-1.5,0) circle (0.08) (0,0) circle (0.08);
		\end{scope}
		\draw[fill=black] (1.75,0) circle (0.03) (2,0) circle (0.03) (2.25,0) circle (0.03) (5.75,0) circle (0.03) (6,0) circle (0.03) (6.25,0) circle (0.03);
		\node at (4,-2) {$	W=\Tr\left[A_1C_1B_1+\sum\lm_{a=2}^{n-1}\left(A_aC_aB_a-B_{a-1}C_aA_{a-1}\right)-B_{n-1}C_{n}A_{n-1}+{\color{color2}J C_s^{\ell}I}\right]$};
	\end{tikzpicture}
	\caption{A framed quiver example corresponding to the Dynkin diagram $\myA_n$.}\label{fig:quiver}
\end{figure}

In the case of the Dynkin diagram $\myA_n$, the quiver is depicted in Fig.~\ref{fig:quiver}.
We assign the following equivariant weights (flavor fugacities) to the quiver arrows, so that the deformation parametrization coincides with that in \eqref{Yangian}:
\begin{equation}\label{fcharges}
	\begin{array}{c|c|c|c|c|c}
		\mbox{Arrow} & A_a & B_a & C_a & I & J\\
		\hline
		\mbox{Weight/flavor charge }{\bf f} & -\hbar & -\hbar & 2\hbar & 0 & -2\ell \hbar
	\end{array}\,.
\end{equation}

\begin{figure}[ht!]
	\centering
	\begin{tikzpicture}
		\tikzset{col1/.style={fill=white!40!color1}}
		\tikzset{col2/.style={fill=white!40!color8}}
		\tikzset{col3/.style={fill=white!40!color3}}
		\tikzset{col4/.style={fill=color6}}
		\tikzset{col5/.style={fill=white!40!color2}}
		\tikzset{col6/.style={fill=white!40!color4}}
		\node at (6,2.5) {$\begin{array}{c}
				\begin{tikzpicture}
					\draw[thick] (0,0) -- (4,0);
					\foreach \x/\c/\t in {0/col1/1,1/col2/2,2/col3/3,3/col4/4,4/col5/5,5/col6/6}
					{
						\draw[\c] (0.8*\x,0) circle (0.1);
						\node at (0.8*\x,0.4) {\scriptsize \t};
					}
				\end{tikzpicture}
			\end{array}$};
		\node at (0,0) {$\begin{array}{c}
				\begin{tikzpicture}[scale=0.8]
					\draw[-stealth] (0,0) -- (-1.81596,-2.09488) node[left,pos=1] {$\scriptstyle A$};
					\draw[-stealth] (0,0) -- (2.67302,-0.800547) node[above,pos=1] {$\scriptstyle B$};
					\draw[-stealth] (0,0) -- (0.,2.42705) node[left,pos=1] {$\scriptstyle C$};
					\foreach \a/\b/\c/\d in {0./0./-0.45399/-0.52372,0./0.809017/-0.45399/0.285296,-0.45399/-0.52372/-0.907981/-1.04744,-0.45399/0.285296/-0.907981/-0.238424,-0.907981/-1.04744/-1.36197/-1.57116,0.891007/-0.266849/0.437016/-0.790569,0.891007/0.542168/0.437016/0.0184476,0.437016/-0.790569/-0.0169745/-1.31429,0./0./0.891007/-0.266849,0./0.809017/0.891007/0.542168,0./1.61803/0.891007/1.35119,-0.45399/-0.52372/0.437016/-0.790569,-0.45399/0.285296/0.437016/0.0184476,-0.907981/-1.04744/-0.0169745/-1.31429,0.891007/-0.266849/1.78201/-0.533698,0.891007/0.542168/1.78201/0.275319,0./0./0./0.809017,0./0.809017/0./1.61803,-0.45399/-0.52372/-0.45399/0.285296,-0.907981/-1.04744/-0.907981/-0.238424,0.891007/-0.266849/0.891007/0.542168,0.891007/0.542168/0.891007/1.35119,0.437016/-0.790569/0.437016/0.0184476,1.78201/-0.533698/1.78201/0.275319}
					{
						\draw[postaction=decorate, decoration={markings, mark= at position 0.7 with {\arrow{stealth}}}] (\a,\b) -- (\c,\d);
					}
					\foreach \x/\y/\c in {0./0./col4,0./0.809017/col4,0./1.61803/col4,-0.45399/-0.52372/col3,-0.45399/0.285296/col3,-0.907981/-1.04744/col2,-0.907981/-0.238424/col2,-1.36197/-1.57116/col1,0.891007/-0.266849/col5,0.891007/0.542168/col5,0.891007/1.35119/col5,0.437016/-0.790569/col4,0.437016/0.0184476/col4,-0.0169745/-1.31429/col3,1.78201/-0.533698/col6,1.78201/0.275319/col6}
					{
						\draw[\c] (\x,\y) circle (0.1);
					}
				\end{tikzpicture}
			\end{array}$};
		\node at (4,0) {$\begin{array}{c}
				\begin{tikzpicture}[scale=0.5]
					\foreach \a/\b/\c/\d/\e/\f/\g/\h/\k in {-0.672499/1.08509/-0.672499/1.89411/0.218508/1.62726/0.218508/0.818241/col4,0.218508/0.818241/0.218508/1.62726/1.10951/1.36041/1.10951/0.551392/col5,1.10951/-1.06664/1.10951/-0.257625/2.00052/-0.524474/2.00052/-1.33349/col6,1.10951/-0.257625/1.10951/0.551392/2.00052/0.284543/2.00052/-0.524474/col6,-0.235482/-0.514497/-0.235482/0.29452/0.655524/0.0276714/0.655524/-0.781346/col4,-1.58048/-0.771368/-1.58048/0.0376487/-0.689473/-0.2292/-0.689473/-1.03822/col2,-0.689473/-1.84723/-0.689473/-1.03822/0.201534/-1.30507/0.201534/-2.11408/col3,-2.03447/-2.10411/-2.03447/-1.29509/-1.14346/-1.56194/-1.14346/-2.37095/col1,1.56351/1.07511/1.56351/1.88413/1.10951/1.36041/1.10951/0.551392/col5,2.45451/-0.809771/2.45451/-0.000753557/2.00052/-0.524474/2.00052/-1.33349/col6,2.45451/-0.000753557/2.45451/0.808263/2.00052/0.284543/2.00052/-0.524474/col6,1.10951/-1.06664/1.10951/-0.257625/0.655524/-0.781346/0.655524/-1.59036/col4,1.10951/-0.257625/1.10951/0.551392/0.655524/0.0276714/0.655524/-0.781346/col4,-0.235482/-0.514497/-0.235482/0.29452/-0.689473/-0.2292/-0.689473/-1.03822/col2,0.655524/-1.59036/0.655524/-0.781346/0.201534/-1.30507/0.201534/-2.11408/col3,-0.689473/-1.84723/-0.689473/-1.03822/-1.14346/-1.56194/-1.14346/-2.37095/col1,-0.218508/2.41783/0.672499/2.15098/0.218508/1.62726/-0.672499/1.89411/col4,0.672499/2.15098/1.56351/1.88413/1.10951/1.36041/0.218508/1.62726/col5,1.56351/1.07511/2.45451/0.808263/2.00052/0.284543/1.10951/0.551392/col6,-0.672499/1.08509/0.218508/0.818241/-0.235482/0.29452/-1.12649/0.561369/col3,0.218508/0.818241/1.10951/0.551392/0.655524/0.0276714/-0.235482/0.29452/col4,-1.12649/0.561369/-0.235482/0.29452/-0.689473/-0.2292/-1.58048/0.0376487/col2,-0.235482/-0.514497/0.655524/-0.781346/0.201534/-1.30507/-0.689473/-1.03822/col3,-1.58048/-0.771368/-0.689473/-1.03822/-1.14346/-1.56194/-2.03447/-1.29509/col1}
					{
						\draw[\k, rounded corners=1] (\a,\b) -- (\c,\d) -- (\e,\f) -- (\g,\h) -- cycle;
					}
				\end{tikzpicture}
			\end{array}$};
		\node at (8,0) {$\begin{array}{c}
				\begin{tikzpicture}[scale=0.4]
					\foreach \x/\y/\c in {0/0/col4,1/0/col5,2/0/col6,0/-1/col3,1/-1/col4,0/-2/col2,1/-2/col3,0/-3/col1}
					{
						\draw[\c] (\x,\y) -- (\x+1,\y) -- (\x+1,\y-1) -- (\x,\y-1) -- cycle;
					}
					\foreach \a/\b/\c in {0.5/-0.5/3,1.5/-0.5/3,2.5/-0.5/2,0.5/-1.5/2,1.5/-1.5/2,0.5/-2.5/2,1.5/-2.5/1,0.5/-3.5/1}
					{\node at (\a,\b) {\small\c};}
				\end{tikzpicture}
			\end{array}$};
		\node at (12,0) {$\begin{array}{c}
				\begin{tikzpicture}[scale=0.3]
					\foreach \i/\c in {0/col1, 1/col2, 2/col3, 3/col4, 4/col5, 5/col6}
					{
						\draw[\c, rounded corners = 1] (-\i-0.1,\i-0.4) -- (-\i-0.9,\i+0.4) -- (\i+0.9,\i+0.4) -- (\i+0.1,\i-0.4) -- cycle;
					}
					\foreach \a/\b/\c in {0/0/1,-1/1/2,-2/2/2,-3/3/3,-4/4/$r$,-5/5/$r$,-6/6/$r$,1/1/0,0/2/1,-1/3/2,-2/4/3,-3/5/$r$,-4/6/$r$,2/2/0,1/3/0,0/4/0,-1/5/2,-2/6/$r$,3/3/0,2/4/0,1/5/0,0/6/0,4/4/0,3/5/0,2/6/0,5/5/0,4/6/0,6/6/0}
					{\node at (\a,\b) {\footnotesize \c};}
					\begin{scope}[shift = {(-4,6.5)}]
						\draw[ultra thick, gray, rounded corners=3] (-3.5,0) -- (3.5,0) -- (0,-3.5) -- cycle;
					\end{scope}
					\begin{scope}[shift = {(3,6.5)}]
						\draw[ultra thick, gray, rounded corners=3] (-4.5,0) -- (4.5,0) -- (0,-4.5) -- cycle;
					\end{scope}
				\end{tikzpicture}
			\end{array}$};
		\node at (3,2.5) {(a)};
		\node at (0,-1.5) {(b)};
		\node at (4,-1.5) {(c)};
		\node at (8,-1.5) {(d)};
		\node at (12,-1.5) {(e)};
		\draw[-stealth] (5.5,0) -- (7,0);
		\draw[-stealth] (9,0) to[out=315,in=225] node[below,pos=0.7] {\scriptsize $\begin{array}{c}
				\mbox{rotate}\\
				\mbox{counterclockwise}\\
				\mbox{by }45^{\circ}
			\end{array}$} (11,-0.5);
	\end{tikzpicture}
	\caption{Equivalences of basis presentations: (a) coloring code for Dynkin nodes; (b) molten crystal; (c) plane partition (box pile); (d) plane partition (view from the top); (e) GT diagram.}\label{fig:parti}
\end{figure}
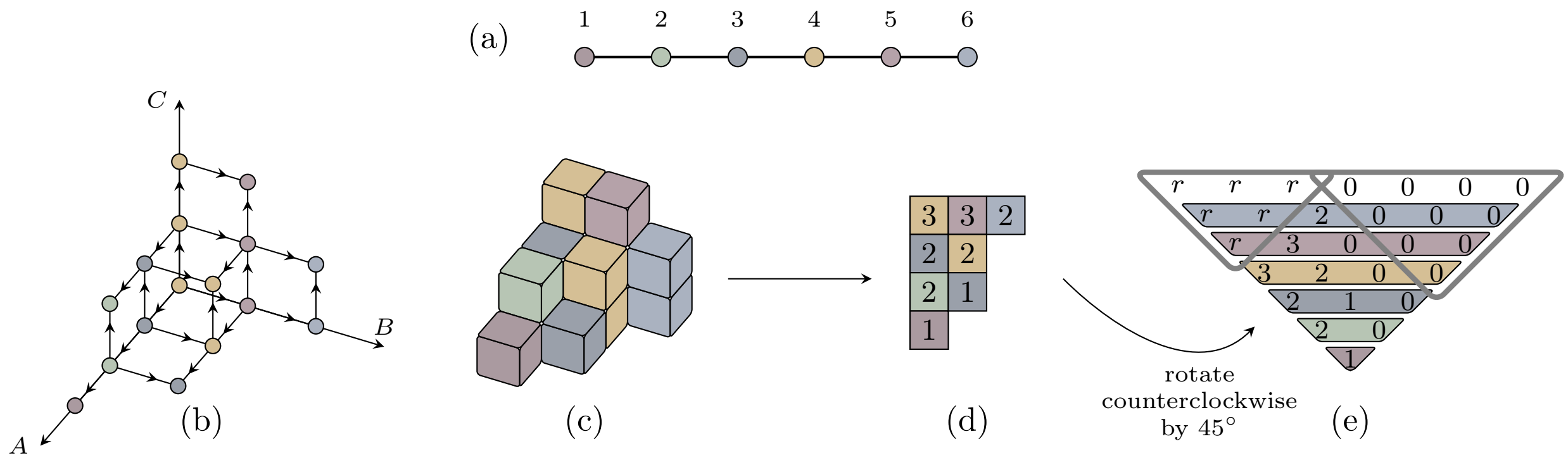

The choice of framing for the quiver depicted in Fig.~\ref{fig:quiver} defines a representation of $\myY(\fs\fl_{n+1})$ that is an evaluation representation, i.e., a pullback of a representation of the subalgebra $\fs\fl_{n+1}\subset\myY(\fs\fl_{n+1})$ with lowest weight $w_l=-\ell\omega_s$ (the highest weight is $w_h=\ell\omega_{n-s}$), where $\omega_s$ is the fundamental weight corresponding to the $s^{\rm th}$ root.
The vectors of this representation are labeled by molten crystals, which are in one-to-one correspondence with 3d Young diagrams (plane partitions) contained in a box of dimensions $s\times (n-s)\times \ell$ (see Fig.~\ref{fig:parti}) \cite{Galakhov:2024bzs,Gavshin:2025akn,Galakhov:2026cwv}.
In this case, the molten crystal basis is equivalent to the Gelfand-Tsetlin (GT) basis \cite{Gavshin:2025akn}.

Vectors in the GT basis are labeled by triangular tables of integes $\lambda_{i,j}$, with $1\leq j \leq i \leq n+1$, arranged in rows one below the other starting with the row $i=n+1$.
See \cite{molev2006gelfand} for details.
The entries in the rows must satisfy the inequalities $\lambda_{i,j}\geq \lambda_{i-1,j}\geq \lambda_{i,j+1}$.
The top row $\lambda_{n+1,j}$ represents a partition labeling the irreducible representation with highest weight $w_h$ in question, i.e., the corresponding Young diagram is a rectangle of dimensions $(n-s)\times \ell$ (see Fig.~\ref{fig:parti}(e)).
The correspondence between plane partitions and GT tables is as follows.
One projects the 3d Young diagram onto the horizontal plane, obtaining a Young diagram with the vertical column heights written in the cells (see Fig.~\ref{fig:parti}(c,d)).
These height numbers are exactly the entries appearing in the corresponding GT table (see Fig.~\ref{fig:parti}(e)).
The remaining numbers in the GT table (those inside the gray triangles in Fig.~\ref{fig:parti}(e)) are uniquely fixed by the top row and the inequalities $\lambda_{i,j}\geq \lambda_{i-1,j}\geq \lambda_{i,j+1}$.

All cells in the diagrams are colored.
The colors $1,\ldots,n$ correspond to the nodes of  the Dynkin diagram $\myA_n$, i.e., to the simple roots (see Fig.~\ref{fig:parti}(a)).
We assume that the position of each cell $\Box$ in the 3d partition is defined by integral Cartesian coordinates $(A_{\Box},B_{\Box},C_{\Box})$ (see Fig.~\ref{fig:parti}(b)).
The corner cell of the 3d partition is assigned coordinates $(0,0,0)$.
Then we define the cell color as:
\begin{equation}
	c_{\Box}=s-A_{\Box}+B_{\Box}\,.
\end{equation}
To emphasize that a particular cell has color $c$, we denote it by $\sqbox{$c$}$.
In the corresponding GT table $\lambda_{i,j}$, the color is determined by the row number $i$.
In addition to the color function, we assign to each cell $\Box$ a complex-valued function called the cell content:
\begin{equation}\label{content}
	\phi_{\Box}=u-\hbar\, A_{\Box}-\hbar \, B_{\Box}+2\hbar\, C_{\Box}\,,
\end{equation}
where $u$ is the spectral parameter.
The spectral parameter $u$ is part of the quiver data associated with the framing node in Fig.~\ref{fig:quiver}.
We will discuss this assignment in more detail in Sec.~\ref{sec:QSQM}.

The corresponding representation of the Yangian $\myY(\fs\fl_{n+1})$ in the molten crystal basis $|\lambda\rangle$ reads:
\begin{equation}\label{Y_rep}
	\begin{aligned}
		&{}^{\pm} e_k^{a}|\lambda\rangle_u=\sum\lm_{\ssqbox{$a$}\in\lambda^{\pm}}\CA_{\lambda\to\lambda\pm\ssqbox{$a$}}\times\phi_{\ssqbox{$a$}}^k|\lambda\pm\sqbox{$a$}\rangle_u\,,\\
		& \left(1+2\hbar\sum\lm_{k=0}^{\infty}\frac{h_k^a}{z^{k+1}}\right)|\lambda\rangle_u=\eta_{\lambda}^a(z)|\lambda\rangle_u\,.
	\end{aligned}
\end{equation}
Here by $\lambda^+$ ($\lambda^-$) we denote the set of cells in the positive 3d octant, that can be added to (subtracted from) partition $\lambda$, so that the new set of cells denoted $\lambda+\Box$ ($\lambda-\Box$, respectively) is again a partition inside a box of dimension $s\times (n-s)\times \ell$.
The rational functions $\eta_{\lambda}^a(z)$ are so-called charge functions \cite{Li:2020rij}; their Taylor expansions at $z=\infty$ determine the eigenvalues of $h_k^a$ in the molten crystal basis:
\begin{equation}
	\eta_{\lambda}^a(z)=\left(\frac{z-u-2\hbar \ell}{z-u}\right)^{\delta_{a,s}}\prod\lm_{b=1}^n\prod\lm_{\ssqbox{$b$}\in\lambda}\frac{z-\phi_{\ssqbox{$b$}}+\hbar A_{ab}}{z-\phi_{\ssqbox{$b$}}-\hbar A_{ab}}\,.
\end{equation}
Finally, the matrix coefficients $\CA_{\lambda\to\lambda'}$ are \emph{``amplitudes''} describing transitions $\lambda\to\lambda'$.
Their explicit values depend on a choice of vector norms $\CN_{\lambda}$, so that $\langle\lambda|\lambda'\rangle=\delta_{\lambda,\lambda'}\CN_{\lambda}$ and $\left({}^{\pm}e_k^a\right)^{\dagger}={}^{\mp}e_k^a$.
For \eqref{Y_rep} to define a Yangian representation, it suffices for $\CA_{\lambda\to\lambda'}$ to satisfy the following \emph{hysteresis} relations \cite{Galakhov:2024bzs}:\footnote{In \cite{Li:2020rij} the ratios on the r.h.s. are called \emph{bonding factors}; they depend only on the quiver data.}
\begin{equation}\label{hysteresis}
	\begin{aligned}
		&\frac{\CA_{\lambda\to\lambda+\ssqbox{$b$}_2}\CA_{\lambda+\ssqbox{$b$}_2\to\lambda+\ssqbox{$a$}_1+\ssqbox{$b$}_2}}{\CA_{\lambda\to\lambda+\ssqbox{$a$}_1}\CA_{\lambda+\ssqbox{$a$}_1\to\lambda+\ssqbox{$a$}_1+\ssqbox{$b$}_2}}=\frac{\CA_{\lambda+\ssqbox{$a$}_1+\ssqbox{$b$}_2\to\lambda+\ssqbox{$b$}_2}\CA_{\lambda+\ssqbox{$b$}_2\to\lambda}}{\CA_{\lambda+\ssqbox{$a$}_1+\ssqbox{$b$}_2\to\lambda+\ssqbox{$a$}_1}\CA_{\lambda+\ssqbox{$a$}_1\to\lambda}}=\frac{\phi_{\ssqbox{$a$}_1}-\phi_{\ssqbox{$b$}_2}+\hbar A_{ab}}{\phi_{\ssqbox{$a$}_1}-\phi_{\ssqbox{$b$}_2}-\hbar A_{ab}}\,,\\
		&\CA_{\lambda+\ssqbox{$a$}_1\to\lambda}\CA_{\lambda\to\lambda+\ssqbox{$b$}_2}=\CA_{\lambda+\ssqbox{$a$}_1\to\lambda+\ssqbox{$a$}_1+\ssqbox{$b$}_2}\CA_{\lambda+\ssqbox{$a$}_1+\ssqbox{$b$}_2\to\lambda+\ssqbox{$b$}_2}\,,\\
		&\CA_{\lambda\to\lambda+\ssqbox{$a$}}\CA_{\lambda+\ssqbox{$a$}\to\lambda}=\mathop{\rm res}\lm_{z=\phi_{\ssqbox{$a$}}}\eta_{\lambda}^a(z)\,.
	\end{aligned}
\end{equation}
There are two choices of norms $\CN_\lambda$ canonically used in the literature: a square-root representation \cite{Prochazka:2015deb,Li:2020rij}, where $\CN_{\lambda}=1$, and a cohomological representation \cite{Nakajima1998,Rapcak:2018nsl,Rapcak:2020ueh,Galakhov:2020vyb}, where $\CA_{\lambda\to\lambda'}$ describes a Hecke correspondence between moduli spaces of quiver representations associated with fixed points labeled by $\lambda$ and $\lambda'$.
In the latter case $\CN_{\lambda}$ is given by the Euler class of the tangent space at the fixed point $\lambda$.
However, for $\myY(\fs\fl_{n+1})$, since the representation is a pullback of an evaluation representation, we can choose as amplitudes $\CA_{\lambda\to\lambda'}$ simple matrix coefficients $\CA_{\lambda\to\lambda\pm\ssqbox{$a$}}=\langle \lambda\pm\sqbox{$a$}|{}^{\pm}e_0^a|\lambda\rangle$ of the subalgebra $\fs\fl_{n+1}$.
For example, the canonical matrix coefficients in the GT basis (see \cite[Th.~2.3]{molev2006gelfand}) suffice.


\subsection{Crystal chain representations}\label{sec:cry_chain}

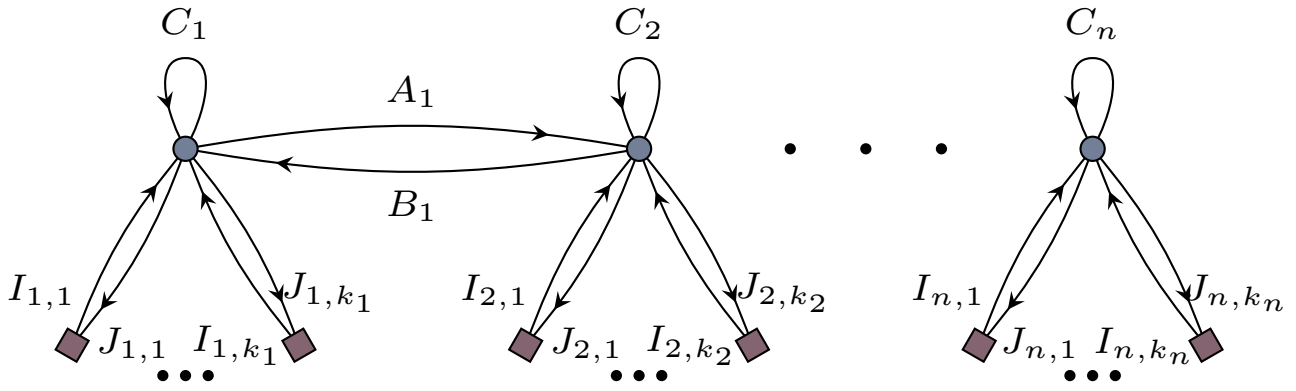
\begin{figure}[ht!]
	\centering
	\begin{tikzpicture}
		\begin{scope}
			\draw[postaction=decorate, decoration={markings, mark= at position 0.8 with {\arrow{stealth}}}] (0,0) to[out=60,in=0] (0,0.6) to[out=180,in=120] (0,0);
			\node[above] at (0,0.6) {$\scriptstyle C_1$};
			\begin{scope}[rotate=-30]
				\draw[postaction=decorate, decoration={markings, mark= at position 0.8 with {\arrow{stealth}}}] (0,-1.5) to[out=100,in=260] (0,0);
				\draw[postaction=decorate, decoration={markings, mark= at position 0.8 with {\arrow{stealth}}}] (0,0) to[out=280,in=80] (0,-1.5);
				\begin{scope}[shift={(0,-1.5)}]
					\draw[fill=color2] (-0.08,-0.08) -- (-0.08,0.08) -- (0.08,0.08) -- (0.08,-0.08) -- cycle;
				\end{scope}
				\node at (-0.35,-1.3) {$\scriptstyle I_{1,1}$}; 
				\node at (0.35,-1.3) {$\scriptstyle J_{1,1}$}; 
			\end{scope}
			\begin{scope}[rotate=30]
				\draw[postaction=decorate, decoration={markings, mark= at position 0.8 with {\arrow{stealth}}}] (0,-1.5) to[out=100,in=260] (0,0);
				\draw[postaction=decorate, decoration={markings, mark= at position 0.8 with {\arrow{stealth}}}] (0,0) to[out=280,in=80] (0,-1.5);
				\begin{scope}[shift={(0,-1.5)}]
					\draw[fill=color2] (-0.08,-0.08) -- (-0.08,0.08) -- (0.08,0.08) -- (0.08,-0.08) -- cycle;
				\end{scope}
				\node at (-0.35,-1.3) {$\scriptstyle I_{1,k_1}$}; 
				\node at (0.35,-1.3) {$\scriptstyle J_{1,k_1}$};
			\end{scope}
			\draw[fill=black] (-0.15,-1.5) circle (0.03) (0,-1.5) circle (0.03) (0.15,-1.5) circle (0.03);
		\end{scope}
		\begin{scope}[shift = {(3,0)}]
			\draw[postaction=decorate, decoration={markings, mark= at position 0.8 with {\arrow{stealth}}}] (0,0) to[out=60,in=0] (0,0.6) to[out=180,in=120] (0,0);
			\node[above] at (0,0.6) {$\scriptstyle C_2$};
			\begin{scope}[rotate=-30]
				\draw[postaction=decorate, decoration={markings, mark= at position 0.8 with {\arrow{stealth}}}] (0,-1.5) to[out=100,in=260] (0,0);
				\draw[postaction=decorate, decoration={markings, mark= at position 0.8 with {\arrow{stealth}}}] (0,0) to[out=280,in=80] (0,-1.5);
				\begin{scope}[shift={(0,-1.5)}]
					\draw[fill=color2] (-0.08,-0.08) -- (-0.08,0.08) -- (0.08,0.08) -- (0.08,-0.08) -- cycle;
				\end{scope}
				\node at (-0.35,-1.3) {$\scriptstyle I_{2,1}$}; 
				\node at (0.35,-1.3) {$\scriptstyle J_{2,1}$}; 
			\end{scope}
			\begin{scope}[rotate=30]
				\draw[postaction=decorate, decoration={markings, mark= at position 0.8 with {\arrow{stealth}}}] (0,-1.5) to[out=100,in=260] (0,0);
				\draw[postaction=decorate, decoration={markings, mark= at position 0.8 with {\arrow{stealth}}}] (0,0) to[out=280,in=80] (0,-1.5);
				\begin{scope}[shift={(0,-1.5)}]
					\draw[fill=color2] (-0.08,-0.08) -- (-0.08,0.08) -- (0.08,0.08) -- (0.08,-0.08) -- cycle;
				\end{scope}
				\node at (-0.35,-1.3) {$\scriptstyle I_{2,k_2}$}; 
				\node at (0.35,-1.3) {$\scriptstyle J_{2,k_2}$};
			\end{scope}
			\draw[fill=black] (-0.15,-1.5) circle (0.03) (0,-1.5) circle (0.03) (0.15,-1.5) circle (0.03);
		\end{scope}
		\begin{scope}[shift = {(6,0)}]
			\draw[postaction=decorate, decoration={markings, mark= at position 0.8 with {\arrow{stealth}}}] (0,0) to[out=60,in=0] (0,0.6) to[out=180,in=120] (0,0);
			\node[above] at (0,0.6) {$\scriptstyle C_n$};
			\begin{scope}[rotate=-30]
				\draw[postaction=decorate, decoration={markings, mark= at position 0.8 with {\arrow{stealth}}}] (0,-1.5) to[out=100,in=260] (0,0);
				\draw[postaction=decorate, decoration={markings, mark= at position 0.8 with {\arrow{stealth}}}] (0,0) to[out=280,in=80] (0,-1.5);
				\begin{scope}[shift={(0,-1.5)}]
					\draw[fill=color2] (-0.08,-0.08) -- (-0.08,0.08) -- (0.08,0.08) -- (0.08,-0.08) -- cycle;
				\end{scope}
				\node at (-0.35,-1.3) {$\scriptstyle I_{n,1}$}; 
				\node at (0.35,-1.3) {$\scriptstyle J_{n,1}$}; 
			\end{scope}
			\begin{scope}[rotate=30]
				\draw[postaction=decorate, decoration={markings, mark= at position 0.8 with {\arrow{stealth}}}] (0,-1.5) to[out=100,in=260] (0,0);
				\draw[postaction=decorate, decoration={markings, mark= at position 0.8 with {\arrow{stealth}}}] (0,0) to[out=280,in=80] (0,-1.5);
				\begin{scope}[shift={(0,-1.5)}]
					\draw[fill=color2] (-0.08,-0.08) -- (-0.08,0.08) -- (0.08,0.08) -- (0.08,-0.08) -- cycle;
				\end{scope}
				\node at (-0.35,-1.3) {$\scriptstyle I_{n,k_n}$}; 
				\node at (0.35,-1.3) {$\scriptstyle J_{n,k_n}$};
			\end{scope}
			\draw[fill=black] (-0.15,-1.5) circle (0.03) (0,-1.5) circle (0.03) (0.15,-1.5) circle (0.03);
		\end{scope}
		\draw[postaction=decorate, decoration={markings, mark= at position 0.8 with {\arrow{stealth}}}] (0,0) to[out=10,in=170] node[pos=0.5,above] {$\scriptstyle A_1$} (3,0);
		\draw[postaction=decorate, decoration={markings, mark= at position 0.8 with {\arrow{stealth}}}] (3,0) to[out=190,in=350]  node[pos=0.5,below] {$\scriptstyle B_1$} (0,0);
		\draw[fill=black] (4,0) circle (0.03) (4.5,0) circle (0.03) (5,0) circle (0.03);
		\draw[fill=color4] (0,0) circle (0.08) (3,0) circle (0.08) (6,0) circle (0.08);
		\node[right] at (7.5,-0.5) {$\displaystyle ,\quad W=W_0+\sum\lm_{j=1}^n\sum\lm_{i=1}^{k_j}\Tr\, J_{j,i}C_j^{\ell_{j,i}}I_{j,i}\,.$};
	\end{tikzpicture}
	\caption{Multiple framing nodes}\label{fig:multi_fr}
\end{figure}

Now consider the situation where the framing is a ``tuple'' of elementary framings presented in Fig.~\ref{fig:quiver}.
In other words, we have a collection of framing nodes connected to various gauge nodes by pairs of arrows $R$ and $S$; there are also specific deformations of the superpotential with parameters $r$ (see Fig.~\ref{fig:multi_fr}).	
In this case, the crystals $\lambda_{\ff}$ grow from a ``nucleation center'' associated with each framing node $\ff$ \cite{Galakhov:2021xum,Galakhov:2022uyu}.
We call these cell arrangements \emph{crystal chains}.
The cell content $\phi_{\Box}$ for $\Box\in\lambda_{\ff}$ is calculated in the same way as in \eqref{content}, where the spectral parameter $u$ is substituted by the spectral parameter  $u_{\ff}$ associated with $\ff$.
If the spectral parameters $u_{\ff}$ are in general position, then the grown crystals never intersect in the complex $\phi$-plane and remain independent.
A basis labeled by such crystal chains is in one-to-one correspondence with a basis of the tensor product of the representation spaces associated with the framings $\ff$:
\begin{equation}
	|\Lambda({\vec u})\rangle=|\lambda_1\rangle_{u_1}\otimes |\lambda_2\rangle_{u_2}\otimes \ldots \otimes |\lambda_m\rangle_{u_m}\,.
\end{equation}

Following \cite{Galakhov:2022uyu}, we naturally extend the crystal representation \eqref{Y_rep} to crystal chains:
\begin{equation}\label{pseudo}
	\begin{aligned}
		&\sD\left({}^{+} e_k^{a}\right)|\lambda_1\rangle_{u_1}\otimes\ldots\otimes|\lambda_m\rangle_{u_m}=\sum\lm_{\ff=1}^m\sum\lm_{\ssqbox{$a$}\in\lambda^{+}_{\ff}}\CA_{\lambda_{\ff}\to\lambda_{\ff}+\ssqbox{$a$}}\times\phi_{\ssqbox{$a$}}^k\times\left(\prod\lm_{\ff'<\ff}\eta_{\lambda_{\ff'}}^a(\phi_{\ssqbox{$a$}})\right)\times\\
		&\hspace{6cm}\times\ldots\otimes|\lambda_{\ff-1}\rangle_{u_{\ff-1}}\otimes|\lambda_{\ff}+\sqbox{$a$}\rangle_{u_{\ff}}\otimes |\lambda_{\ff-1}\rangle_{u_{\ff+1}}\otimes\ldots\,,\\
		&\sD\left({}^{-} e_k^{a}\right)|\lambda_1\rangle_{u_1}\otimes\ldots\otimes|\lambda_m\rangle_{u_m}=\sum\lm_{\ff=1}^m\sum\lm_{\ssqbox{$a$}\in\lambda^{-}_{\ff}}\CA_{\lambda_{\ff}\to\lambda_{\ff}-\ssqbox{$a$}}\times\phi_{\ssqbox{$a$}}^k\times\left(\prod\lm_{\ff'>\ff}\eta_{\lambda_{\ff'}}^a(\phi_{\ssqbox{$a$}})\right)\times\\
		&\hspace{6cm}\times\ldots\otimes|\lambda_{\ff-1}\rangle_{u_{\ff-1}}\otimes|\lambda_{\ff}-\sqbox{$a$}\rangle_{u_{\ff}}\otimes |\lambda_{\ff-1}\rangle_{u_{\ff+1}}\otimes\ldots\,,\\
		& \left(1+2\hbar\sum\lm_{k=0}^{\infty}\frac{\sD\left(h_k^a\right)}{z^{k+1}}\right)|\lambda_1\rangle_{u_1}\otimes\ldots\otimes|\lambda_m\rangle_{u_m}=\left(\prod\lm_{\ff=1}^m\eta_{\lambda_\ff}^a(z)\right)\times|\lambda_1\rangle_{u_1}\otimes\ldots\otimes|\lambda_m\rangle_{u_m}\,.
	\end{aligned}	
\end{equation}
In many respects, the structure $\sD$ shares properties with the coproduct $\Delta$; therefore, we refer to it as a \emph{chain pseudo-coproduct}.
The prefix ``pseudo'' is appropriate here because the structure $\sD$ cannot be universalized to the true coproduct on the Yangians.
In other words, it cannot be promoted to an algebra homomorphism $\myY(\fg)\to \myY(\fg)^{\otimes m}$ (see \cite[Sec.~3.1]{Galakhov:2022uyu} for a detailed discussion).
However, we should stress that for Yangian doubles, quantum loop deformations, and elliptic deformations this obstruction disappears, and \eqref{pseudo} becomes one of the coproducts realized explicitly in a representation.

The present representation is \emph{non-evaluation}; it cannot be obtained as a pullback of a representation of $\fs\fl_{n+1}$.
Yet it is a finite-dimensional representation satisfying Drinfeld's criterion \cite[Th.~2]{Drinfeld2}.
In this case, the vacuum (empty crystal) charge functions read (setting $\hbar=\sfrac{1}{2}$ to comply with the notation in \cite{Drinfeld2}):
\begin{equation}
	\eta_{\varnothing}^a(z)=\prod\lm_{\ff\in{\rm fr}(a)}\frac{z-u_{\ff}-r_{\ff}}{z-u_{\ff}}=\frac{P_a(z-1)}{P_a(z)},\quad\mbox{where }P_a(z)=\prod\lm_{\ff\in{\rm fr}(a)}\prod\lm_{k=0}^{r_{\ff}-1}(z-u_{\ff}-k)\,,
\end{equation}
where ${\rm fr}(a)$ is the list of framing nodes for gauge node $a$.


\subsection{Factorization twist}

For some generators the pseudo-coproduct can be represented in a universal form, for example:
\begin{equation}
	\sD(h_0^a)=h_0^a\otimes 1+1\otimes h_0^a,\quad \sD(h_1^a)=h_1^a\otimes 1+2\hbar\, h_0^a\otimes h_0^a+1\otimes h_1^a\,.
\end{equation}
Comparing these relations with \eqref{Y_co}, we find:
\begin{equation}\label{small_sD}
	\Delta\left(h_0^a\right)=\sD\left(h_0^a\right),\quad \Delta\left(h_1^a\right)=\sD\left(h_1^a\right)+2\hbar\,\Omega_+^a\,.
\end{equation}
Although the structures $\Delta$ and $\sD$ are quite different, they both describe valid extensions of the Yangian algebra on tensor products of representations.
Thus, it is natural to expect them to be related \cite{Galakhov:2022uyu}.
In particular, for tensor squares these representations are related by an isomorphism $U$:
\begin{tcolorbox}
\begin{equation}\label{univ}
	\sD(\fa)\circ U=U\circ \Delta(\fa),\quad \forall \fa\in \myY(\fg)\,.
\end{equation}
\end{tcolorbox}
\noindent Isomorphism $U$ is a ``universalizer''  for the pseudo-coproduct $\sD$; it turns it into the true universal  coproduct $\Delta=U^{-1}\sD U$.

Matrix $U$ has a useful property: it is an \emph{unipotent upper-triangular} matrix with respect to a specific order of vectors in the tensor square.
Let us note that \eqref{small_sD} together with \eqref{univ} imply that $\left[\sD(h_0^a),U\right]=\left[\Delta(h_0^a),U\right]=0$.
Thus, $U$ also has a block-diagonal structure and does not mix vectors from distinct total weight spaces.
For a total weight space of weight $w=w(\lambda_1)+w(\lambda_2)=w(\lambda_1')+w(\lambda_2')$, we order the vectors in it according to the following rule:
\begin{equation}\label{ord}
	|\lambda_1\rangle\otimes |\lambda_2\rangle < |\lambda_1'\rangle\otimes |\lambda_2'\rangle\;\Longleftrightarrow\; w(\lambda_1)>w(\lambda_1')\,.
\end{equation}
With respect to this order, matrices $\Omega_+^a$ are \emph{nilpotent}, strictly upper-triangular.
Substituting $h_1^a$ into \eqref{univ} allows us to derive the following relations, which determine $U$ completely:
\begin{tcolorbox}
	\begin{equation}\label{main_alg}
		\left[\sD(h_1^a),U\right]=U\,\Omega_+^a\,.
	\end{equation}
\end{tcolorbox}
Indeed, by decomposing $U$ as follows:
\begin{equation}\label{ansatz}
	U=\bbone+\sum\lm_{k=1}^{\infty}\Upsilon_k\,,
\end{equation}
where $\Upsilon_k$ is a matrix having nonzero entries above the $k^{\rm th}$ diagonal, we arrive at the following system of recursion equations (here we use the fact that $\sD(h_1^a)$ acts diagonally):
\begin{equation}\label{recursion}
		\left[\sD(h_1^a),\Upsilon_1\right]=\Omega_+^a,\quad 	\left[\sD(h_1^a),\Upsilon_{k+1}\right]=\Upsilon_k\,\Omega_+^a\,.
\end{equation}
In fact, if an operator $X$ commutes with $\sD\left(h_0^a\right)$, one can easily compute the matrix elements of its commutators with $\sD\left(h_1^a\right)$ in terms of the matrix elements of $X$ itself:
\begin{equation}\label{factor}
	\langle\Lambda'|\left[\sD\left(h_1^a\right),X\right]|\Lambda\rangle=z_{\Lambda\to \Lambda'}\times\langle\Lambda'|X|\Lambda\rangle,\quad z_{\Lambda\to\Lambda'}=\sum\lm_b\sum\lm_{\ssqbox{$b$}'\in\Lambda'}A_{ab}\omega_{\ssqbox{$b$}'}-\sum\lm_b\sum\lm_{\ssqbox{$b$}\in\Lambda}A_{ab}\omega_{\ssqbox{$b$}}\,.
\end{equation}

Clearly, the pseudo-coproduct \eqref{pseudo} is not symmetric with respect to the tensor components.
There is an alternative pseudo-coproduct $\sD^{\rm op}=P\sD P$ and a pseudo-$R$-matrix $\bf R$ defined by the following intertwining relation:
\begin{equation}\label{pseudo-R}
	{\bf R}\circ\sD(\fa)=\sD^{\rm op}(\fa)\circ {\bf R},\quad \forall\,\fa\in\myY(\fs\fl_{n+1})\,.
\end{equation}
This relation forces the pseudo-$R$-matrix to be \emph{strictly diagonal} in the crystal chain basis.
The pseudo-$R$-matrix cannot modify the eigenvalues of $h_k^a$ packed into the charge functions $\eta_{\lambda}^a(z)$.
As a rational function of $z$,  the charge function satisfies a rather restrictive property: it has only simple poles at those values of $\phi_{\Box}$, at which a cell $\Box$ can be added to or subtracted from a partition to obtain another partition \cite{Li:2020rij}.
In other words, the charge function of a crystal chain \eqref{pseudo}  uniquely determines the shape of each individual crystal $\lambda_{\ff}$ in the chain.
Since $\sD\left(h_k^a\right)=\sD^{\rm op}\left(h_k^a\right)$, the pseudo-$R$-matrix cannot modify these shapes and therefore must act diagonally.
This diagonal action is a mere renormalization of the basis in the tensor square \cite{Galakhov:2022uyu}:
\begin{equation}\label{diag_R}
	{\bf R}\,|\lambda_1\rangle_{u_1}\otimes|\lambda_2\rangle_{u_2}=\left(\prod\lm_c\prod_{\ssqbox{$a$}\in\lambda_1}\eta^c_{\varnothing_2}\left(\phi_{\ssqbox{$a$}}\right)\right)\left(\prod\lm_d\prod_{\ssqbox{$b$}\in\lambda_2}\eta^d_{\varnothing_1}\left(\phi_{\ssqbox{$b$}}\right)\right)^{-1}\left(\prod\lm_{\ssqbox{$a$}\in\lambda_1}\prod\lm_{\ssqbox{$b$}\in\lambda_2}\frac{\phi_{\ssqbox{$a$}}-\phi_{\ssqbox{$b$}}+\hbar A_{ab}}{\phi_{\ssqbox{$a$}}-\phi_{\ssqbox{$b$}}-\hbar A_{ab}}\right)|\lambda_1\rangle_{u_1}\otimes|\lambda_2\rangle_{u_2}\,,
\end{equation}
where by the product over $\sqbox{$a$}\in\lambda$ we we mean the product over all cells, irrespective of color.

Combining \eqref{pseudo-R} and \eqref{univ}, we obtain an expression for the $R$-matrix in terms of the pseudo-$R$-matrix and the ``universalizer'':
\begin{tcolorbox}
\begin{equation}\label{Gauss}
	R(u_1,u_2)=\left(P U_{u_2,u_1}P\right)^{-1}\cdot{\bf R}(u_1,u_2)\cdot U_{u_1,u_2}\,.
\end{equation}
\end{tcolorbox}
\noindent Let us note that the first matrix $\left(P U_{u_2,u_1}P\right)^{-1}$ on the r.h.s. of \eqref{Gauss} is unipotent lower-triangular; hence \eqref{Gauss} yields the Gauss (LDU) decomposition of the $R$-matrix.

In the literature \cite{drinfeld1990quasi,Pfeiffer2000Yangian,Terras2000,Pfeiffer2001QuantumAffine,Albert2000XYZ,Yang2004,Torrielli2023,Maillet2000}, one also performs a finer splitting of the diagonal pseudo-$R$-matrix into a product of two diagonal matrices, so that
\begin{equation}
	{\bf R}(u_1,u_2)=(PD_{u_2,u_1}P)^{-1}D_{u_1,u_2},\quad F_{u_1,u_2}:=D_{u_1,u_2}U_{u_1,u_2},\quad R(u_1,u_2)=\left(P F_{u_2,u_1}P\right)^{-1}\cdot F_{u_1,u_2}\,.
\end{equation}
Matrix $F$ is called a \emph{factorization (Drinfeld) twist}.

We have demonstrated that \eqref{univ} has an explicit solution for $\fa=h_0^a,h_1^a$; however, it is far from apparent why the solution we found also works for the cases $\fa={}^{\pm}e_0^a$ as well.
Moreover, there are counterexamples in more general shifted quiver Yangians \cite{Galakhov:2021xum} where where an ansatz such as \eqref{ansatz} fails.
We believe that the proof that the factorization twist (universalizer) $U$ exists can be carried out in a way similar to the proof that the $R$-matrix exists as a solution of the intertwining equations \eqref{intertw}.
First, one should work out this construction and check its self-consistency for fundamental representations.
Then one extends it to more general representations via the fusion procedure \cite{Cherednik1986,NazarovTarasov1998,KhoroshkinNazarovPapi2011,IsaevMolevOgievetsky2012,BazhanovFrassekLukowskiMeneghelliStaudacher2011}.
We will discuss a physical motivation for the splitting in \eqref{Gauss} and the meaning of these elements in Sec.~\ref{sec:R-mat_Berry}.
Alternatively, we can  reverse the logic. 
As we will discuss in the next subsection, the YBE provides the RLL relations, from which the Yangian algebra relations are traditionally deduced.
Relations such as \eqref{intertw} and \eqref{univ} also follow from the YBE.
Thus, if a solution to the YBE for representations of the Yangian exists, the procedure \eqref{Gauss} will reconstruct it from the known matrix elements of $\myY(\fg)$.


\subsection{Diagrammatic relations}\label{sec:diag}

Now we would like to digress a little to recall that the Yangian $\myY_{\hbar}(\fg)$ is the Drinfeld quantization of the standard Lie bialgebra structure on $\fg$ \cite{drinfeld1990quasi,Tolstoy2003}.
This quantization is performed using the classical $R$-matrix \cite{Belavin1982}.
Another canonical, and in some respects simpler, representative of such a quantization is the quantum group $U_q(\fg)$.
Many properties of $\myY_{\hbar}(\fg)$ and $U_q(\fg)$ related to the Hopf structure are alike.
The consistency relations induced by the Hopf structure of $U_q(\fg)$ admit a nice diagrammatic representation in terms of ribbon graphs \cite{RT1,RT2,1112.2654,Mironov:2015aia,Mironov:2015qma,Anokhina:2024lbn,Galakhov:2014sha} and underlie 3d TQFTs such as Chern-Simons theory \cite{Witten:1988hf}.
We would like to adopt a similar notation for our case.

A structure closely analogous to $U$ that appears for $U_q(\fg)$ is the $q$-Clebsch-Gordan coefficients, which map vectors of a tensor product explicitly to vectors in isotypic irreducible components.
This similarity allows one to compare $q$-Racah coefficients with Drinfeld associators \cite{Lanina:2026kew}.\footnote{
In the 2d CFT counterpart of quantum groups, Clebsch-Gordan coefficients appear in the standard way in OPE coefficients.
The latter are related to Baxter operators that enter the Hecke correspondence and the analytic Langlands program \cite{Levin:2003ab,Ambrosino:2025ab}.
In our language, the operators $U$ are also induced by a \emph{dual} Hecke correspondence.
By construction, the analytic Hecke correspondence shifts the Chern classes of bundles at a point of the curve. 
For example, it shifts the topological number of Yang--Mills instantons, i.e., second Chern class of self-dual bundles.
The quiver description we work with here can be obtained in certain cases by a nonlinear Fourier transformation, known as the Atiyah-Drinfeld-Hitchin-Manin (ADHM) construction \cite{Dorey:2002ik}.
}
However, unlike in $U_q(\fg)$, tensor products of $\myY(\fg)$ admit no isotypic decomposition into irreducibles.
Instead, $U$ maps tensor product vectors to vectors of the crystal chain representation.
Graphically, $q$-Clebsh-Gordan coefficients are depicted as 3-valent fusion/fission vertices.
Here we depict the $k$-crystal chain representation as $k$ closely spaced lines.
Thus, the $U$-element is depicted as the fusion of two lines into a dashed ribbon, and the $U^{-1}$-element as the corresponding fission:
\begin{equation}\label{U_diag}
	U=\begin{array}{c}
		\begin{tikzpicture}
			\draw[line width=1mm] (-0.7,-1) to[out=90,in=270] (-0.1,0) -- (-0.1,0.7) (0.7,-1) to[out=90,in=270] (0.1,0) -- (0.1,0.7);
			\draw[dashed, white, line width =0.5mm] (-0.1,0) -- (-0.1,0.7) (0.1,0) -- (0.1,0.7);
		\end{tikzpicture}
	\end{array},\quad U^{-1}=\begin{array}{c}
	\begin{tikzpicture}[yscale=-1]
		\draw[line width=1mm] (-0.7,-1) to[out=90,in=270] (-0.1,0) -- (-0.1,0.7) (0.7,-1) to[out=90,in=270] (0.1,0) -- (0.1,0.7);
		\draw[dashed, white, line width =0.5mm] (-0.1,0) -- (-0.1,0.7) (0.1,0) -- (0.1,0.7);
	\end{tikzpicture}
	\end{array}\,.
\end{equation}
For the $R$-matrix, we have the following depiction:
\begin{equation}
	R=\begin{array}{c}
		\begin{tikzpicture}[scale=0.7]
			\draw[line width=1mm] (-1,-1) to[out=90,in=270] (1,1) (1,-1) to[out=90,in=270] (-1,1);
		\end{tikzpicture}
	\end{array} =\begin{array}{c}
	\begin{tikzpicture}
		\draw[line width=1mm] (-0.7,-1) to[out=90,in=270] (-0.2,-0.5) to[out=90,in=270] (0.2,0.5) to[out=90,in=270] (0.7,1) (0.7,-1) to[out=90,in=270] (0.2,-0.5) to[out=90,in=270] (-0.2,0.5) to[out=90,in=270] (-0.7,1);
		\draw[dashed, white, line width =0.5mm] (-0.2,-0.5) to[out=90,in=270] (0.2,0.5) (0.2,-0.5) to[out=90,in=270] (-0.2,0.5);
	\end{tikzpicture}
	\end{array}\,,
\end{equation}
where the intersection of dashed lines represents the diagonal pseudo-$R$-matrix.

The Yangian algebra relations follow from the RLL relations, which, in turn, follow from the YBE.
The Yangian relations are obtained by specifying vectors and spectral parameters on one of the strands entering the $R$-matrix crossing.
Graphically, the $R$-matrix becomes a defect on a single strand -- an operator acting in the corresponding representation.
For example, for the evaluation representations, we obtain:
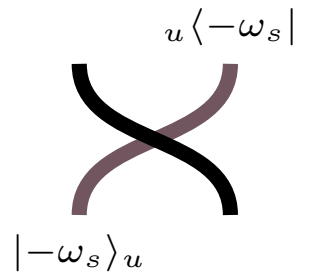
\begin{equation}
	\begin{aligned}
	&\begin{array}{c}
		\begin{tikzpicture}
			\draw[color1, line width =1mm] (-0.5,-0.5) to[out=90,in=270] (0.5,0.5);
			\draw[line width =1mm]  (0.5,-0.5) to[out=90,in=270] (-0.5,0.5);
			\node[below] at (-0.5,-0.5) {$\scriptstyle |-\omega_s\rangle_u$};
			\node[above] at (0.5,0.5) {$\scriptstyle {}_u\langle-\omega_s|$};
		\end{tikzpicture}
	\end{array}=h^s(u)=1+2\hbar\sum\lm_{k=0}\frac{h_k^s}{u^{k+1}}\,,\\
	& \begin{array}{c}
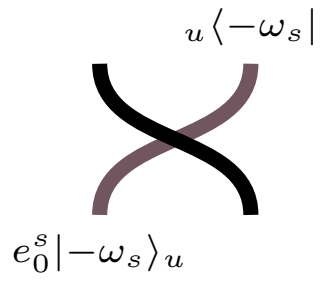
	\begin{tikzpicture}
		\draw[color1, line width =1mm] (-0.5,-0.5) to[out=90,in=270] (0.5,0.5);
		\draw[line width =1mm]  (0.5,-0.5) to[out=90,in=270] (-0.5,0.5);
		\node[below] at (-0.5,-0.5) {$\scriptstyle e_0^s |-\omega_s\rangle_u$};
		\node[above] at (0.5,0.5) {$\scriptstyle {}_u\langle-\omega_s|$};
	\end{tikzpicture}
	\end{array}=h^s(u)\sum\lm_{k=0}^{\infty}\frac{e_k^s}{u^{k+1}},\; \begin{array}{c}
	
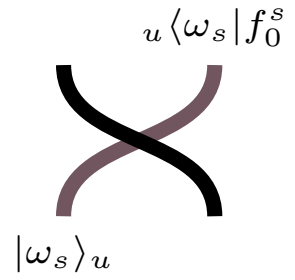
\begin{tikzpicture}
		\draw[color1, line width =1mm] (-0.5,-0.5) to[out=90,in=270] (0.5,0.5);
		\draw[line width =1mm]  (0.5,-0.5) to[out=90,in=270] (-0.5,0.5);
		\node[below] at (-0.5,-0.5) {$\scriptstyle |\omega_s\rangle_u$};
		\node[above] at (0.5,0.5) {$\scriptstyle {}_u\langle\omega_s|f_0^s $};
	\end{tikzpicture}
	\end{array}=\sum\lm_{k=0}^{\infty}\frac{f_k^s}{u^{k+1}}\left[h^s(u)\right]^{-1}\,,
	\end{aligned}
\end{equation}
where $|\pm \omega_s\rangle$ is the highest/lowest weight vector corresponding to the evaluation module with lowest weight $-\omega_s$.
Specifying matrix elements in the YBE then leads to relations among the Yangian generators.
Some details and examples of such a derivation can be found in \cite[App.~B]{Litvinov:2020zeq}.

In this language, the Yangian coproduct is mimicked by considering parallel strands crossed by a single control strand, as in the transfer matrix construction \eqref{transfer}.
Relations such as the intertwining relation \eqref{intertw} become apparent: it suffices to fix one of the strands as the control strand.
Similarly, the defining relations for the $q$-Clebsh-Gordan coefficients in $U_q(\fg)$, which allow one to pull a 3-valent vertex across another strand have an analog in $\myY_{\hbar}(\fg)$.
This is a graphical depiction of \eqref{main_alg}:
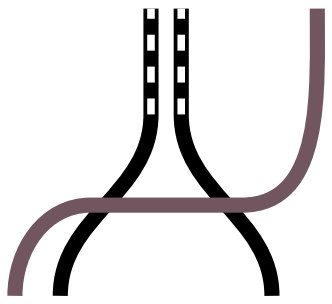
\begin{equation}\label{CGanalog}
	\begin{array}{c}
		\begin{tikzpicture}
			\draw[line width=1mm] (-0.7,-1.2) to[out=90,in=270] (-0.1,0) -- (-0.1,0.7) (0.7,-1.2) to[out=90,in=270] (0.1,0) -- (0.1,0.7);
			\draw[dashed, white, line width =0.5mm] (-0.1,0) -- (-0.1,0.7) (0.1,0) -- (0.1,0.7);
			\draw[color1, line width = 1mm] (-1,-1.2) to[out=90,in=180] (-0.5,-0.6) -- (0.5,-0.6) to[out=0,in=270] (1,0.7);
		\end{tikzpicture}
	\end{array}=\begin{array}{c}
	
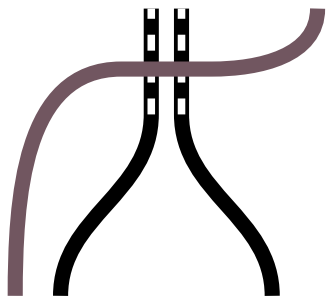
\begin{tikzpicture}
		\draw[line width=1mm] (-0.7,-1.2) to[out=90,in=270] (-0.1,0) -- (-0.1,0.7) (0.7,-1.2) to[out=90,in=270] (0.1,0) -- (0.1,0.7);
		\draw[dashed, white, line width =0.5mm] (-0.1,0) -- (-0.1,0.7) (0.1,0) -- (0.1,0.7);
		\draw[color1, line width = 1mm] (-1,-1.2) to[out=90,in=180] (-0.3,0.3) -- (0.3,0.3) to[out=0,in=270] (1,0.7);
	\end{tikzpicture}
	\end{array}\,.
\end{equation}

We will also need a relation that is analogous to the definition of the $q$-Racah coefficient.
In particular, consider ``cables'' -- a collection of strands (for cables in $U_q(\fg)$, see \cite{Anokhina:2013ica,cooper2012categorification,2013arXiv1309.3796W}). 
It does not matter in what order we fuse the strands into a single cable.
In the case of $U_q(\fg)$ the resulting isotypic bases depend explicitly on the cable fusion order and are related by the very $q$-Racah transition matrices.
As we mentioned, there is no isotypic decomposition theorem for $\myY(\fg)$.
Thus all the vectors are mapped to the crystal chains, and different orders of fusion are simply equivalent:
\begin{equation}\label{regluing}
	\begin{array}{c}
		\begin{tikzpicture}[yscale=0.8]
			\draw[line width = 1mm] (-1,-2) to[out=90,in=270] (-0.6,-1) to[out=90,in=270] (-0.2,0) -- (-0.2,0.5);
			\draw[line width = 1mm] (0,-2) to[out=90,in=270] (-0.4,-1) to[out=90,in=270] (0,0) -- (0,0.5);
			\draw[line width = 1mm] (1,-2) to[out=90,in=270] (0.2,0) -- (0.2,0.5);
			\draw[dashed, white, line width =0.5mm] (-0.6,-1) to[out=90,in=270] (-0.2,0) -- (-0.2,0.5) (-0.4,-1) to[out=90,in=270] (0,0) -- (0,0.5) (0.2,0) -- (0.2,0.5);
		\end{tikzpicture}
	\end{array}=\begin{array}{c}
	\begin{tikzpicture}[yscale=0.8, xscale=-1]
		\draw[line width = 1mm] (-1,-2) to[out=90,in=270] (-0.6,-1) to[out=90,in=270] (-0.2,0) -- (-0.2,0.5);
		\draw[line width = 1mm] (0,-2) to[out=90,in=270] (-0.4,-1) to[out=90,in=270] (0,0) -- (0,0.5);
		\draw[line width = 1mm] (1,-2) to[out=90,in=270] (0.2,0) -- (0.2,0.5);
		\draw[dashed, white, line width =0.5mm] (-0.6,-1) to[out=90,in=270] (-0.2,0) -- (-0.2,0.5) (-0.4,-1) to[out=90,in=270] (0,0) -- (0,0.5) (0.2,0) -- (0.2,0.5);
	\end{tikzpicture}
	\end{array}, \quad\begin{array}{c}
	\begin{tikzpicture}[yscale=-0.8]
		\draw[line width = 1mm] (-1,-2) to[out=90,in=270] (-0.6,-1) to[out=90,in=270] (-0.2,0) -- (-0.2,0.5);
		\draw[line width = 1mm] (0,-2) to[out=90,in=270] (-0.4,-1) to[out=90,in=270] (0,0) -- (0,0.5);
		\draw[line width = 1mm] (1,-2) to[out=90,in=270] (0.2,0) -- (0.2,0.5);
		\draw[dashed, white, line width =0.5mm] (-0.6,-1) to[out=90,in=270] (-0.2,0) -- (-0.2,0.5) (-0.4,-1) to[out=90,in=270] (0,0) -- (0,0.5) (0.2,0) -- (0.2,0.5);
	\end{tikzpicture}
	\end{array}=\begin{array}{c}
	\begin{tikzpicture}[yscale=-0.8, xscale=-1]
		\draw[line width = 1mm] (-1,-2) to[out=90,in=270] (-0.6,-1) to[out=90,in=270] (-0.2,0) -- (-0.2,0.5);
		\draw[line width = 1mm] (0,-2) to[out=90,in=270] (-0.4,-1) to[out=90,in=270] (0,0) -- (0,0.5);
		\draw[line width = 1mm] (1,-2) to[out=90,in=270] (0.2,0) -- (0.2,0.5);
		\draw[dashed, white, line width =0.5mm] (-0.6,-1) to[out=90,in=270] (-0.2,0) -- (-0.2,0.5) (-0.4,-1) to[out=90,in=270] (0,0) -- (0,0.5) (0.2,0) -- (0.2,0.5);
	\end{tikzpicture}
	\end{array}\,.
\end{equation}


\section{Quiver side}\label{sec:quiver}

\subsection{Quiver SQM}\label{sec:QSQM}

We begin this section with a brief review of quiver supersymmetric quantum mechanics (SQM). 
Our goal is only to recall the ingredients needed below, not to give a complete treatment; for further details, we refer the reader to \cite{Li:2020rij,Galakhov:2020vyb,Li:2023zub,Galakhov:2024bzs}.
We work with $\CN=4$ SQM, which is canonically obtained from $\CN=1$ 4d SYM theory by dimensional reduction.
Quiver arrows correspond to chiral (Higgs) fields, which reduce trivially.
Quiver nodes correspond to gauge multiplets containing the vector fields $A_{\mu}$.
During the reduction, the spatial components $A_{1,2,3}$ become scalars: $A_1$ and $A_2$ combine into a complex scalar $\Phi=A_1+\I A_2$, while $A_3$ becomes a real scalar $X=A_3$.
Framing nodes differ from gauge nodes in that they carry no dynamics; there is no corresponding dynamical term in the Lagrangian.
However, the framing fields are allowed to acquire vevs.
In particular, for a $U(1)$ framing node $\ff$, $\langle\Phi\rangle=\left(u_{\ff}\right)$, where $u_{\ff}$ is the complex spectral parameter associated with $\ff$.
The corresponding real parameters $x_{\ff}$, defined by $\langle X\rangle=x_{\ff}$, will be discussed in detail below.

In this setting, wave functions $\Psi$ correspond, to first order, to Gaussian fluctuations around classical vacua, i.e., fixed points on the quiver representation moduli space on the Higgs branch.
Here we assume that all stability (Fayet-Iliopoulos) parameters lie in the cyclic chamber.
In this case, the fixed points are in one-to-one correspondence with crystal chains.
First, let us describe the construction for a single framing node with parameters $(u_{\ff},x_{\ff})$.
The description of the vacuum includes explicit values of the quiver morphisms, i.e., of the maps between the vector spaces $V_a$ associated with the quiver nodes -- the Higgs vevs.
To do so, we describe $V_a$ as the vector space spanned by formal vectors associated with cells of color $a$:
\begin{equation}
	V_a={\rm Span}\left(|\sqbox{$a$}\rangle\right)\,.
\end{equation}
The scalars in the gauge multiplets then acquire the following eigenvalues:
\begin{equation}
	\langle\Phi_a\rangle|\sqbox{$a$}\rangle=\phi_{\ssqbox{$a$}}|\sqbox{$a$}\rangle,\quad \langle X_a\rangle|\sqbox{$a$}\rangle=x_{\ff}|\sqbox{$a$}\rangle\,.
\end{equation}
Matrix elements of the Higgs fields $H=A,B,C$ acquire nonzero vevs if the cells are connected by an edge parallel to the corresponding axis $A$, $B$, or $C$ (see Fig.~\ref{fig:parti}(b)); hence the names of the axes:
\begin{equation}
	\langle \sqbox{$b$}'|H|\sqbox{$a$}\rangle= \left\{\begin{array}{ll}
		1 &,\mbox{ if there is an edge }\sqbox{$a$}\to\sqbox{$b$}'\,,\\
		0 &,\mbox{ otherwise.}
	\end{array}\right.
\end{equation}

For crystal chains, we construct the fixed points and define the vevs of the quiver morphisms as if the vector spaces associated with the nodes were directly tensor products of the spaces associated with each framing node:
\begin{equation}
	V_a=\bigotimes\lm_{\ff}V_a^{(\ff)}\,.
\end{equation}


\subsection{Projected Berry connection and tunneling} \label{sec:Berry}

For the systems discussed previously, a natural question is how to construct a projected Berry connection on a highly degenerate ground-state subspace of the Hilbert space.
Geometrically, this projected Berry connection is realized as a Gauss--Manin connection \cite{manin1958algebraic,Moore:2017byz,Dedushenko:2022pem,Ferrari:2023hza}.
This construction is expected to apply generally to theories whose Hamiltonian has the form $\sH=\frac{1}{2}\left\{Q,Q^{\dagger}\right\}$ for some supercharge $Q$ and its conjugate $Q^{\dagger}$.
In our case, $Q$ and $Q^{\dagger}$ generate a subalgebra of the original $\mathcal{N}=4$ 1d SUSY algebra.

In our case, after replacing the fermion variables by the associated forms in the cotangent bundle, the supercharge takes the form of a twisted de Rham-Dolbeault differential \cite{Galakhov:2020vyb}.
The twisting combines corrections arising from the equivariant vector field $V$, from the superpotential, and from conjugation by the exponential of the Morse height function $h$:
\begin{equation}\label{super}
	Q=e^{-h}\left(d_X+\p_{\Phi, H}+\iota_{\bar V}+dW\wedge\right)e^h\,. 
\end{equation}
The height function is given explicitly by
\begin{equation}\label{height}
	h=\sum\lm_{a\in {\rm nodes}}\Tr\, X_a\left(\zeta_a-\sum\lm_{b}\sum\lm_{\alpha\in{\rm arrow}(b\to a)}H_{\alpha}H_{\alpha}^{\dagger}+\sum\lm_c\sum\lm_{\beta\in{\rm arrow}(a\to c)}H^{\dagger}_{\beta}H_{\beta}\right)\,. 
\end{equation}
The sum in \eqref{height} runs over all quiver nodes, both gauge and framing.
For the framing nodes, $X$ is replaced by its vacuum expectation value $x_{\ff}\cdot\bbone$, and the Fayet--Iliopoulos parameters $\zeta$ are set to zero.

From \eqref{super} it is transparent, if we modify the parameters $x_{\ff}$, the supercharge is conjugated by an operator $\CO$,
\begin{equation}
	Q\mapsto e^{-\CO}Qe^{\CO},\quad Q^{\dagger}\mapsto e^{\CO}Q^{\dagger}e^{-\CO}\,.
\end{equation}

We now assume that the parameters $x_{\ff}$ depend slowly on time, varying from some value in the far past to a new value in the far future, so that the supercharge becomes time-dependent:
\begin{equation}
	Q(t)=e^{-\CO(t)}Qe^{\CO(t)},\quad Q^{\dagger}(t)=e^{\CO(t)}Q^{\dagger}e^{-\CO(t)}\,.
\end{equation}
If we define the time-dependent Hamiltonian $\sH(t)=\left\{Q(t),Q^{\dagger}(t)\right\}$, it turns out that $\sH(t)$ preserves neither $Q(t)$ nor $Q^{\dagger}(t)$.
We therefore add a correction \cite{Galakhov:2025fhd},
\begin{equation}\label{Hamiltonian}
	\sH(t)=\left\{Q(t),Q^{\dagger}(t)\right\}-\I\p_t\CO(t)\,,
\end{equation}
chosen so that $\sH(t)$ preserves $Q(t)$, i.e., $\left[\I\p_t-H(t),Q(t)\right]=0$.
Note that this Hamiltonian is no longer Hermitian: the correction required to preserve $Q(t)$ is not Hermitian.

In the far past and the far future, $Q(t)$ is constant, and, by the Hodge theorem, the ground states correspond to its cohomology.
The projected Berry connection is then written as
\begin{equation}\label{Berry}
	\begin{aligned}
		& \sU:={\rm Pexp}\int\lm_{-\infty}^{+\infty} (-\I H(t))\,dt\,,\\
		& \sU|\Psi_v(-\infty))\rangle=|\Psi_v(+\infty)\rangle+\sum\lm_{\iota\in{\rm instantons}(v\to v')}\CA(\iota)|\Psi_v(+\infty)\rangle+Q(+\infty)\times \left(\ldots\right)\,.
	\end{aligned}
\end{equation}
The tunneling instantons summed over in \eqref{Berry} are directed flows in the direction of increasing $h$, while anti-instantons flow in the opposite direction.
This makes the Berry connection matrix upper-triangular: the matrix element $\langle v'|\sU|v\rangle$ between vacua $v$ and $v'$ can be nonzero only if $h(v) \geq h(v')$ (so that a tunneling instanton has a chance to flow in that direction).
This is a direct consequence of the non-Hermitian Hamiltonian \eqref{Hamiltonian} introduced above.
The tunneling amplitudes $\CA(\iota)$ are given by the corresponding one-loop superdeterminants in the instanton background.

It is important to note that the projected Berry transport $\sU$ on $Q(t)$-cohomology is locally flat.
This is a direct consequence of supersymmetry \cite[Sec.~3.1.2]{Galakhov:2025fhd}.

In quiver SQM, instantons for the bosonic fields are obtained by setting the supersymmetry variations of the fermion fields to zero and performing a Wick rotation $t\to -\I\tau$:
\begin{subequations}
	\begin{equation}\label{inst_smp}
		\p_{\tau}H_{a\to b}=X_bH_{a\to b}-H_{a\to b}X_a\,,
	\end{equation}
	\begin{equation}\label{inst_hard}
		\p_{\tau}X_{a}=\zeta_a-\sum\lm_{b}\sum\lm_{\alpha\in{\rm arrow}(b\to a)}H_{\alpha}H_{\alpha}^{\dagger}+\sum\lm_c\sum\lm_{\beta\in{\rm arrow}(a\to c)}H^{\dagger}_{\beta}H_{\beta}\,,
	\end{equation}
	\begin{equation}\label{inst_equiv}
		\p_{\tau}\Phi=0,\quad \Phi_bH_{a\to b}-H_{a\to b}\Phi_a=\epsilon_H\,H_{a\to b}\,.
	\end{equation}
\end{subequations}
Here $H_{a\to b}$ is any Higgs field corresponding to the arrow $a\to b$, and $\epsilon_H$ is the corresponding equivariant weight.
There is also a constraint analogous to the Gauss law, which we do not consider here.
Unfortunately, these equations admit no nonstationary solution, since \eqref{inst_equiv} contains no dynamical part with $\tau$-derivatives.
In \cite{Galakhov:2025fhd} it was argued that the flow part, \eqref{inst_smp} and \eqref{inst_hard}, dominates, so that instantons may be considered with the equivariant part omitted; note, however, that the one-loop determinant in the instanton background depends explicitly on the equivariant weights.
The equations remain complicated; however, \eqref{inst_smp} can be solved explicitly,
\begin{equation}\label{homo_eq}
	H_{a\to b}(\tau_1)=T_b\times H_{a\to b}(\tau_0)\times T_a^{-1},\quad T_a={\rm Pexp}\int\lm_{\tau_0}^{\tau_1}X_a(\tau')\,d\tau'\,.
\end{equation}
This equation states that the family $\{T_a\}$ is a homomorphism of quiver representations, i.e., that it intertwines the quiver morphisms $H_{a\to b}$ between the asymptotic states at $\tau_0\to-\infty$ and $\tau_1\to+\infty$, respectively \cite{Galakhov:2025fhd}.
This homomorphism yields a Hecke correspondence between quiver representations \cite{Nakajima1994,Galakhov:2020vyb}.

In general, the map given by the family $\{T_a\}$ is not an isomorphism.
The reason is that the exponents in the path-ordered exponentials of \eqref{homo_eq} are real, so some entries of the matrices $T_a$ are unbounded and may blow up along the instanton trajectory.

This effect was observed in \cite{Galakhov:2020vyb}.
For the elementary processes in which a cell leaves or merges with a crystal, the entries that blow up can be eliminated from \eqref{homo_eq}.
For example, suppose that a row of the matrix $T_a$ scales as $1/\Lambda$ as $\Lambda\to\infty$.
Then the inverse matrix $T_a^{-1}$ contains a row scaling as $\Lambda$, and its entries blow up.
Moving $T_a$ to the other side of \eqref{homo_eq} and taking the limit $\Lambda\to\infty$ effectively eliminates that row.
Thus, the final Hecke correspondence relations, once regularized, may in general be represented by non-rectangular matrices $T_a$.
Clearly, such a correspondence is not an isomorphism of quiver representations, but only a homomorphism.

The tunneling amplitude is constructed from the Berry transport of the effective tangent degrees of freedom and is expressed in terms of the corresponding equivariant weights, i.e., the flavor charges of the Higgs fields.
For details, see \cite{Galakhov:2020vyb,Galakhov:2024bzs,Galakhov:2025fhd}; here we recall only the expression that we will apply in what follows.

The equivariant weight $[\ldots]$ is computed separately for each matrix element of the field $H$ by the following formula:
\begin{equation}
	\left[\left(H_{a\to b}\right)_{\ssqbox{$b$}',\ssqbox{$a$}}\right]=\phi_{\ssqbox{$b$}'}-\phi_{\ssqbox{$a$}}-{\bf f}_H\,,
\end{equation}
where ${\bf f}_H$ is the weight, i.e., the flavor charge, of the field $H$ (see \eqref{fcharges}).
For a fixed point, the tangent space $\sV$ is spanned by the chiral Higgs fields:
\begin{equation}
	\sV=\bigoplus_i \IC\,H_i\,.
\end{equation}
The localized Euler class of the tangent space at the fixed point, which enters, for example, the Atiyah--Bott localization formula, is then simply a product of weights:
\begin{equation}
	{\bf eu}(\sV)=:[\sV]=\prod\lm_{i}[H_i]\,.
\end{equation}

Geometrically, an amplitude associated with an instanton trajectory describing the Hecke correspondence can then be computed as an equivariant integral over the Hecke correspondence surface $\Sigma$.
In particular, suppose that such a correspondence exists between two fixed points, and denote the corresponding tangent spaces by $\sV^{(1)}$ and $\sV^{(2)}$.
The Hecke correspondence extends to the tangent spaces in terms of linear relations.
These relations allow one to split the spaces,
\begin{equation}
	\sV^{(1)}=\sV^{(1)}_{\rm zero}\oplus\sV^{(1)}_{\rm eq}\oplus \sV^{(1)}_{\rm arb},\quad \sV^{(2)}=\sV^{(2)}_{\rm zero}\oplus\sV^{(2)}_{\rm eq}\oplus \sV^{(2)}_{\rm arb}\,.
\end{equation} 
so that the Hecke correspondence equations take the following form:
\begin{equation}\label{decomp}
	\sV^{(1)}_{\rm zero}=\sV^{(2)}_{\rm zero}=0,\quad \sV^{(1)}_{\rm eq}=\sV^{(2)}_{\rm eq},\quad 0\cdot \sV^{(1)}_{\rm arb}=0,\quad 0\cdot\sV^{(2)}_{\rm arb}=0\,.
\end{equation}
Here the subscript ``zero'' indicates that the equations for $\Sigma$ force those degrees of freedom to vanish.
The subscript ``eq'' labels the degrees of freedom that are \emph{equated}.
Finally, the subscript ``arb'' indicates that the corresponding degrees of freedom are unconstrained by the equations for $\Sigma$ and can be \emph{arbitrary}.

The amplitude expression is then
\begin{equation}\label{amplitude}
	\CA=\frac{{\bf eu}(\CV^{(1)})}{{\bf eu}(\Sigma)}=\frac{\left[\sV^{(1)}_{\rm zero}\oplus\sV^{(1)}_{\rm eq}\oplus \sV^{(1)}_{\rm arb}\right]}{\left[\sV^{(1)}_{\rm eq}\oplus \sV^{(1)}_{\rm arb}\oplus \sV^{(2)}_{\rm arb}\right]}=\frac{\left[\sV^{(1)}_{\rm zero}\right]}{\left[\sV_{\rm arb}^{(2)}\right]}\,.
\end{equation}


\subsection{\texorpdfstring{$R$}{R}-matrix reconstruction}\label{sec:R-mat_Berry}

\begin{figure}[ht!]
	\centering
	\begin{tikzpicture}[scale=0.35]
		\tikzset{box/.style ={fill=white!40!color8, thick}}
		\draw[-stealth] (0,0) -- (0., 3.23607);
		\draw[-stealth] (0,0) -- (1.81596, -2.09488);
		\draw[-stealth] (0,0) -- (17.8201, 5.33698);
		\node[right] at (17.8201, 5.33698) {$\scriptstyle x$};
		\node[above] at (0., 3.23607) {$\scriptstyle {\rm Im}\,\phi$};
		\node[below right] at (1.81596, -2.09488) {$\scriptstyle {\rm Re}\,\phi$};
		\foreach \a/\b/\c/\d/\e/\f/\g/\h in {15.1641/5.64919/16.0551/5.91604/16.0551/6.72506/15.1641/6.45821, 16.0721/2.98372/16.9631/3.25057/16.9631/4.05958/16.0721/3.79273, 16.0721/3.79273/16.9631/4.05958/16.9631/4.8686/16.0721/4.60175, 16.98/1.12726/17.8711/1.39411/17.8711/2.20313/16.98/1.93628, 14.7101/3.74586/15.1641/3.22214/15.1641/4.03116/14.7101/4.55488, 14.7101/4.55488/15.1641/4.03116/15.1641/4.84018/14.7101/5.3639, 14.7101/5.3639/15.1641/4.84018/15.1641/5.64919/14.7101/6.17291, 14.7101/6.17291/15.1641/5.64919/15.1641/6.45821/14.7101/6.98193, 15.1641/3.22214/15.6181/2.69842/15.6181/3.50744/15.1641/4.03116, 15.1641/4.03116/15.6181/3.50744/15.6181/4.31646/15.1641/4.84018, 15.1641/4.84018/15.6181/4.31646/15.6181/5.12547/15.1641/5.64919, 15.6181/2.69842/16.0721/2.1747/16.0721/2.98372/15.6181/3.50744, 15.6181/3.50744/16.0721/2.98372/16.0721/3.79273/15.6181/4.31646, 15.6181/4.31646/16.0721/3.79273/16.0721/4.60175/15.6181/5.12547, 16.0721/2.1747/16.5261/1.65098/16.5261/2.46/16.0721/2.98372, 16.5261/1.65098/16.98/1.12726/16.98/1.93628/16.5261/2.46, 14.7101/6.98193/15.1641/6.45821/16.0551/6.72506/15.6011/7.24878, 15.1641/5.64919/15.6181/5.12547/16.5091/5.39232/16.0551/5.91604, 15.6181/5.12547/16.0721/4.60175/16.9631/4.8686/16.5091/5.39232, 16.0721/2.98372/16.5261/2.46/17.4171/2.72685/16.9631/3.25057, 16.5261/2.46/16.98/1.93628/17.8711/2.20313/17.4171/2.72685}
		{
			\draw[box, rounded corners=1] (\a,\b) -- (\c,\d) -- (\e,\f) -- (\g,\h) -- cycle;
		}
		\foreach \a/\b/\c/\d/\e/\f/\g/\h in {11.6001/5.39081/12.4911/5.65766/12.4911/6.46668/11.6001/6.19983, 11.6001/6.19983/12.4911/6.46668/12.4911/7.2757/11.6001/7.00885, 12.054/4.05808/12.9451/4.32493/12.9451/5.13394/12.054/4.86709, 12.508/1.91632/13.399/2.18317/13.399/2.99219/12.508/2.72534, 12.508/2.72534/13.399/2.99219/13.399/3.8012/12.508/3.53436, 11.1461/3.48748/11.6001/2.96376/11.6001/3.77278/11.1461/4.2965, 11.1461/4.2965/11.6001/3.77278/11.6001/4.5818/11.1461/5.10552, 11.1461/5.10552/11.6001/4.5818/11.6001/5.39081/11.1461/5.91453, 11.1461/5.91453/11.6001/5.39081/11.6001/6.19983/11.1461/6.72355, 11.1461/6.72355/11.6001/6.19983/11.6001/7.00885/11.1461/7.53257, 11.6001/2.96376/12.054/2.44004/12.054/3.24906/11.6001/3.77278, 11.6001/3.77278/12.054/3.24906/12.054/4.05808/11.6001/4.5818, 11.6001/4.5818/12.054/4.05808/12.054/4.86709/11.6001/5.39081, 12.054/2.44004/12.508/1.91632/12.508/2.72534/12.054/3.24906, 12.054/3.24906/12.508/2.72534/12.508/3.53436/12.054/4.05808, 11.1461/7.53257/11.6001/7.00885/12.4911/7.2757/12.0371/7.79942, 11.6001/5.39081/12.054/4.86709/12.9451/5.13394/12.4911/5.65766, 12.054/4.05808/12.508/3.53436/13.399/3.8012/12.9451/4.32493}
		{
			\draw[box, rounded corners=1] (\a,\b) -- (\c,\d) -- (\e,\f) -- (\g,\h) -- cycle;
		}
		\foreach \a/\b/\c/\d/\e/\f/\g/\h in {7.12805/2.94381/8.01906/3.21066/8.01906/4.01967/7.12805/3.75283, 8.49002/0.56363/9.38103/0.830479/9.38103/1.6395/8.49002/1.37265, 6.67406/2.65851/7.12805/2.13479/7.12805/2.94381/6.67406/3.46753, 6.67406/3.46753/7.12805/2.94381/7.12805/3.75283/6.67406/4.27655, 7.12805/2.13479/7.58204/1.61107/7.58204/2.42009/7.12805/2.94381, 7.58204/1.61107/8.03603/1.08735/8.03603/1.89637/7.58204/2.42009, 8.03603/1.08735/8.49002/0.56363/8.49002/1.37265/8.03603/1.89637, 6.67406/4.27655/7.12805/3.75283/8.01906/4.01967/7.56507/4.54339, 7.12805/2.94381/7.58204/2.42009/8.47305/2.68694/8.01906/3.21066, 7.58204/2.42009/8.03603/1.89637/8.92704/2.16322/8.47305/2.68694, 8.03603/1.89637/8.49002/1.37265/9.38103/1.6395/8.92704/2.16322}
		{
			\draw[box, rounded corners=1] (\a,\b) -- (\c,\d) -- (\e,\f) -- (\g,\h) -- cycle;
		}
		\foreach \a/\b/\c/\d/\e/\f/\g/\h in {4.47201/1.63799/5.36301/1.90484/5.36301/2.71385/4.47201/2.44701, 4.47201/2.44701/5.36301/2.71385/5.36301/3.52287/4.47201/3.25602, 4.926/0.305251/5.817/0.5721/5.817/1.38112/4.926/1.11427, 3.56403/1.87641/4.01802/1.35269/4.01802/2.16171/3.56403/2.68543, 3.56403/2.68543/4.01802/2.16171/4.01802/2.97073/3.56403/3.49445, 3.56403/3.49445/4.01802/2.97073/4.01802/3.77974/3.56403/4.30346, 4.01802/1.35269/4.47201/0.828972/4.47201/1.63799/4.01802/2.16171, 4.01802/2.16171/4.47201/1.63799/4.47201/2.44701/4.01802/2.97073, 4.01802/2.97073/4.47201/2.44701/4.47201/3.25602/4.01802/3.77974, 4.47201/0.828972/4.926/0.305251/4.926/1.11427/4.47201/1.63799, 3.56403/4.30346/4.01802/3.77974/4.90902/4.04659/4.45503/4.57031, 4.01802/3.77974/4.47201/3.25602/5.36301/3.52287/4.90902/4.04659, 4.47201/1.63799/4.926/1.11427/5.817/1.38112/5.36301/1.90484}
		{
			\draw[box, rounded corners=1] (\a,\b) -- (\c,\d) -- (\e,\f) -- (\g,\h) -- cycle;
		}
		\draw[dashed] (3.56403,4.30346) -- (3.56403,1.0674) -- (5.37999,-1.02749) (7.12805,5.37086) -- (7.12805,2.13479) -- (8.94401,0.0399094) (10.6921,6.43826) -- (10.6921,3.20219) -- (12.508,1.10731) (14.2561,7.50565) -- (14.2561,4.26958) -- (16.0721,2.1747);
		\node[above] at (4.45503, 4.57031) {$\scriptstyle \lambda_1$};
		\node[above] at (7.56507, 4.54339) {$\scriptstyle \lambda_2$};
		\node[above] at (12.0371, 7.79942) {$\scriptstyle \lambda_3$};
		\node[above] at (15.6011, 7.24878) {$\scriptstyle \lambda_4$};
		\node[right] at (6.74196, -2.59865) {$\scriptstyle (u_1,x_1)$};
		\node[right] at (10.306, -1.53125) {$\scriptstyle (u_2,x_2)$};
		\node[right] at (13.87, -0.463856) {$\scriptstyle (u_3,x_3)$};
		\node[right] at (17.434, 0.603539) {$\scriptstyle (u_4,x_4)$};
		\draw[-stealth] (6.74196, -2.59865) to[out=150, in=260] (3.56403, 1.8764);
		\draw[-stealth] (10.306, -1.53125) to[out=150, in=260] (6.67406, 2.65851);
		\draw[-stealth] (13.87, -0.463856) to[out=150, in=260] (11.1461, 3.48748);
		\draw[-stealth] (17.434, 0.603539) to[out=150, in=260] (14.7101, 3.74586);
	\end{tikzpicture}
	\caption{Cell arrangement of a crystal chain in the $(\phi,x)$-space}\label{fig:chain}
\end{figure}
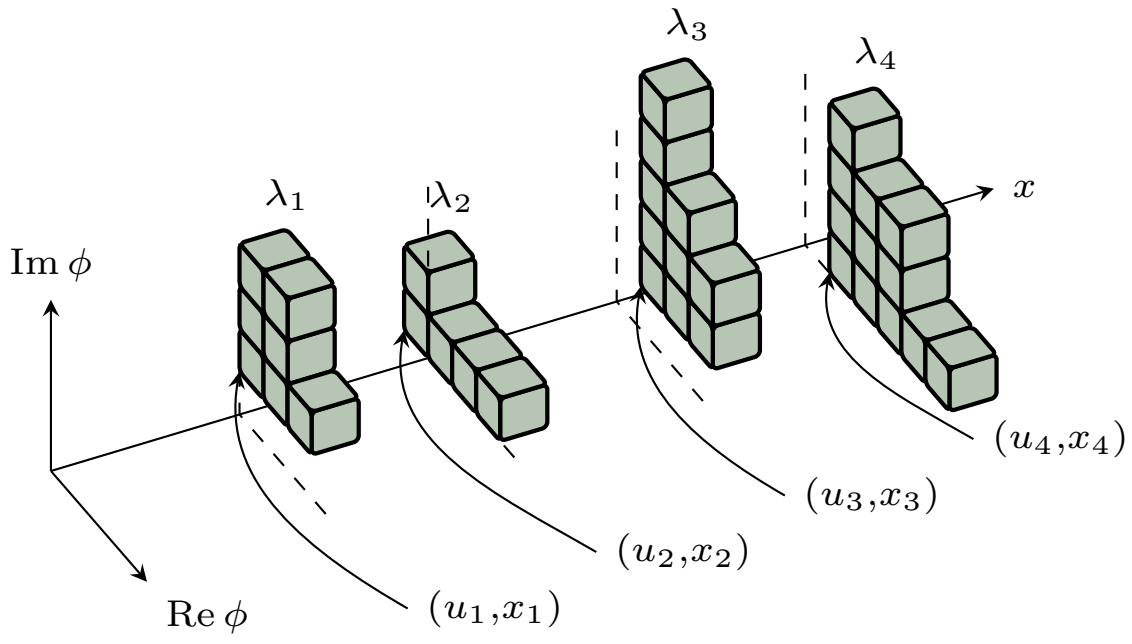

To visualize the crystal chains discussed above, we briefly recall the Higgs--Coulomb duality in quiver SQM \cite{Manschot:2013qma,Galakhov:2018lta,Beaujard:2021pii,Galakhov:2023aev}, which is the SQM counterpart of mirror symmetry in $\mathcal{N}=(2,2)$ 2D theories \cite{Hori:2000kt,Aganagic:2001uw,Hori:2013ika,Galakhov:2020upa}.
Although the organization of this duality for theories with a superpotential is generally an open question, it provides a useful visualization of the states discussed so far.
On the Coulomb branch, wave functions describe BPS molecules of dyons, and the dyon centers correspond to the eigenvalues of the scalars $\Phi$ and $X$ in the gauge multiplets.
It is therefore natural to visualize cells in crystal chains as dyons ``frozen'' in crystals.
Concretely, we embed the cells $\Box$ in the $(\phi,x)$-space with coordinates $(\phi_{\Box}\in\mathbb{C},x_{\ff}\in\mathbb{R})$ (see Fig.~\ref{fig:chain}).
In this arrangement, the crystals lie in isolated $\phi$-planes at fixed $x=x_{\ff}$, ordered along the $x$-axis.

The specific values of the stability parameters $\zeta_a$ are irrelevant, since the structure of states depends only on the chamber.
In the cyclic chamber, we can set all $\zeta_a=\zeta>0$.
In this setting, the value of the height function on a crystal chain reads:
\begin{equation}\label{height_chain}
	h(\Lambda)=\zeta\sum\lm_{\ff}x_{\ff}|\lambda_{\ff}|\,,
\end{equation}
where $|\lambda_{\ff}|$ is the number of cells in crystal $\lambda_{\ff}$.

We assume that in the asymptotic past and future, where the parameters $x_{\ff}$ are time-independent, the values $x_{\ff}$ in the corresponding crystal chains are widely separated.
Under this assumption, the height function \eqref{height_chain} induces an order on the crystal chain basis vectors: $h$ behaves as a positional value whose most significant digit is the occupation of the crystal at the largest $x_{\ff}$.
This is the standard mechanism by which superincreasing weights produce a lexicographic order, as in base-$B$ numeration (cf. \eqref{ord}).
In this setting, a tunneling instanton is depicted as a process in which a number of cells migrate from a crystal with smaller $x_{\ff}$ to a crystal with larger $x_{\ff}$ in the chain.
Since the weights are hierarchically separated, the leading contribution to $h$ comes from the crystal with the largest $x_{\ff}$ that changes; hence such a migration is the only way to increase $h$ at leading order, and no rearrangement of cells within the chain can increase $h$ otherwise.

\begin{figure}[ht!]
	\centering
	\begin{tikzpicture}
		\begin{scope}[scale=0.7]
			\draw[-stealth] (0,0) -- (13,0);
			\node[right] at (13,0) {$\scriptstyle x$};
			\draw[-stealth] (0,0) -- (0,3);
			\node[above] at (0,3) {$\scriptstyle {\rm Im}\,\phi$};
			\draw[-stealth] (0,0) -- (2.59808,-1.5);
			\node[below] at (2.59808,-1.5) {$\scriptstyle {\rm Re}\,\phi$};
			\draw (1.,0.) -- (3.59808,-1.5) -- (3.59808,1.5) -- (1.,3.) -- cycle;
			\draw (4.,0.)--(6.59808,-1.5)--(6.59808,1.5)--(4.,3.)--cycle;
			\draw (7.,0.)--(9.59808,-1.5)--(9.59808,1.5)--(7.,3.)--cycle;
			\draw (10.,0.)--(12.5981,-1.5)--(12.5981,1.5)--(10.,3.)--cycle;
			\foreach \a/\b/\c/\d/\e/\f/\g/\h in {1.43301/0.25/1.69282/0.1/1.69282/0.4/1.43301/0.55, 1.43301/0.55/1.69282/0.4/1.69282/0.7/1.43301/0.85, 1.43301/0.85/1.69282/0.7/1.69282/1./1.43301/1.15, 1.69282/0.1/1.95263/-0.05/1.95263/0.25/1.69282/0.4, 1.69282/0.4/1.95263/0.25/1.95263/0.55/1.69282/0.7, 1.69282/0.7/1.95263/0.55/1.95263/0.85/1.69282/1., 1.95263/-0.05/2.21244/-0.2/2.21244/0.1/1.95263/0.25}
			{\draw[fill=white!40!gray, rounded corners=2] (\a,\b) -- (\c,\d) -- (\e,\f) -- (\g,\h) -- cycle;}
			\foreach \a/\b/\c/\d/\e/\f/\g/\h in {4./1./4.25981/0.85/4.25981/1.15/4./1.3, 4./1.3/4.25981/1.15/4.25981/1.45/4./1.6, 4./1.6/4.25981/1.45/4.25981/1.75/4./1.9, 4./1.9/4.25981/1.75/4.25981/2.05/4./2.2, 4.25981/0.85/4.51962/0.7/4.51962/1./4.25981/1.15}
			{\draw[fill=color2, rounded corners=2] (\a,\b) -- (\c,\d) -- (\e,\f) -- (\g,\h) -- cycle;}
			\foreach \a/\b/\c/\d/\e/\f/\g/\h in {8.29904/-0.25/8.55885/-0.4/8.55885/-0.1/8.29904/0.05, 8.29904/0.05/8.55885/-0.1/8.55885/0.2/8.29904/0.35, 8.29904/0.35/8.55885/0.2/8.55885/0.5/8.29904/0.65, 8.29904/0.65/8.55885/0.5/8.55885/0.8/8.29904/0.95, 8.29904/0.95/8.55885/0.8/8.55885/1.1/8.29904/1.25, 8.55885/-0.4/8.81865/-0.55/8.81865/-0.25/8.55885/-0.1, 8.55885/-0.1/8.81865/-0.25/8.81865/0.05/8.55885/0.2, 8.55885/0.2/8.81865/0.05/8.81865/0.35/8.55885/0.5, 8.81865/-0.55/9.07846/-0.7/9.07846/-0.4/8.81865/-0.25, 8.81865/-0.25/9.07846/-0.4/9.07846/-0.1/8.81865/0.05}
			{\draw[fill=color4, rounded corners=2] (\a,\b) -- (\c,\d) -- (\e,\f) -- (\g,\h) -- cycle;}
			\foreach \a/\b/\c/\d/\e/\f/\g/\h in {10.433/0.75/10.6928/0.6/10.6928/0.9/10.433/1.05, 10.433/1.05/10.6928/0.9/10.6928/1.2/10.433/1.35, 10.433/1.35/10.6928/1.2/10.6928/1.5/10.433/1.65, 10.433/1.65/10.6928/1.5/10.6928/1.8/10.433/1.95, 10.6928/0.6/10.9526/0.45/10.9526/0.75/10.6928/0.9, 10.6928/0.9/10.9526/0.75/10.9526/1.05/10.6928/1.2, 10.6928/1.2/10.9526/1.05/10.9526/1.35/10.6928/1.5, 10.9526/0.45/11.2124/0.3/11.2124/0.6/10.9526/0.75, 10.9526/0.75/11.2124/0.6/11.2124/0.9/10.9526/1.05, 10.9526/1.05/11.2124/0.9/11.2124/1.2/10.9526/1.35, 11.2124/0.3/11.4722/0.15/11.4722/0.45/11.2124/0.6, 11.4722/0.15/11.7321/0./11.7321/0.3/11.4722/0.45}
			{\draw[fill=white!40!gray, rounded corners=2] (\a,\b) -- (\c,\d) -- (\e,\f) -- (\g,\h) -- cycle;}
			\node[below] at (3.59808, -1.5) {$\scriptstyle x_1$};
			\node[color1, below] at (6.59808, -1.5) {$\scriptstyle x_2$};
			\node[color3, below] at (9.59808, -1.5) {$\scriptstyle x_3$};
			\node[below] at (12.5981, -1.5) {$\scriptstyle x_4$};
			\draw[color5, thick, postaction=decorate, decoration={markings, mark= at position 0.7 with {\arrow{stealth}}}] (4.38971, 1.225) to[out=0,in=180] node[pos=0.4,below left] {\small tunnel} (8.94856, 0.125);
			\draw[fill=color6, rounded corners=2] (4.25981, 1.15) -- (4.51962, 1.) -- (4.51962, 1.3) -- (4.25981, 1.45) -- cycle;
			\draw[fill=color6, rounded corners=2] (8.81865, 0.05) -- (8.81865, 0.35) -- (9.07846, 0.2) -- (9.07846, -0.1) -- cycle;
			\draw[-stealth] (1., 0.) -- (1.43301, 0.25) node[pos=0.5,above] {$\scriptstyle u_1$};
			\draw[-stealth] (4., 0.) -- (4., 1.) node[pos=0.5,right] {$\scriptstyle\color{color1} u_2$};
			\draw[-stealth] (7., 0.) -- (8.29904, -0.25) node[pos=0.9,below] {$\scriptstyle\color{color3} u_3$};
			\draw[-stealth] (10., 0.) -- (10.433, 0.75) node[pos=0.5,right] {$\scriptstyle u_4$};
		\end{scope}
		\begin{scope}[shift={(0,3.5)}]
		\begin{scope}[scale=0.7]
			\draw[-stealth] (0,0) -- (13,0);
			\node[right] at (13,0) {$\scriptstyle x$};
			\draw[-stealth] (0,0) -- (0,3);
			\node[above] at (0,3) {$\scriptstyle {\rm Im}\,\phi$};
			\draw[-stealth] (0,0) -- (2.59808,-1.5);
			\node[below] at (2.59808,-1.5) {$\scriptstyle {\rm Re}\,\phi$};
			\draw (1.,0.) -- (3.59808,-1.5) -- (3.59808,1.5) -- (1.,3.) -- cycle;
			\draw (5.5,0.)--(8.09808,-1.5)--(8.09808,1.5)--(5.5,3.)--cycle;
			\draw (10.,0.)--(12.5981,-1.5)--(12.5981,1.5)--(10.,3.)--cycle;
			\foreach \a/\b/\c/\d/\e/\f/\g/\h in {1.43301/0.25/1.69282/0.1/1.69282/0.4/1.43301/0.55, 1.43301/0.55/1.69282/0.4/1.69282/0.7/1.43301/0.85, 1.43301/0.85/1.69282/0.7/1.69282/1./1.43301/1.15, 1.69282/0.1/1.95263/-0.05/1.95263/0.25/1.69282/0.4, 1.69282/0.4/1.95263/0.25/1.95263/0.55/1.69282/0.7, 1.69282/0.7/1.95263/0.55/1.95263/0.85/1.69282/1., 1.95263/-0.05/2.21244/-0.2/2.21244/0.1/1.95263/0.25}
			{\draw[fill=white!40!gray, rounded corners=2] (\a,\b) -- (\c,\d) -- (\e,\f) -- (\g,\h) -- cycle;}
			\foreach \a/\b/\c/\d/\e/\f/\g/\h in {5.5/1./5.75981/0.85/5.75981/1.15/5.5/1.3, 5.5/1.3/5.75981/1.15/5.75981/1.45/5.5/1.6, 5.5/1.6/5.75981/1.45/5.75981/1.75/5.5/1.9, 5.5/1.9/5.75981/1.75/5.75981/2.05/5.5/2.2, 5.75981/0.85/6.01962/0.7/6.01962/1./5.75981/1.15}
			{\draw[fill=color2, rounded corners=2] (\a,\b) -- (\c,\d) -- (\e,\f) -- (\g,\h) -- cycle;}
			\foreach \a/\b/\c/\d/\e/\f/\g/\h in {6.79904/-0.25/7.05885/-0.4/7.05885/-0.1/6.79904/0.05, 6.79904/0.05/7.05885/-0.1/7.05885/0.2/6.79904/0.35, 6.79904/0.35/7.05885/0.2/7.05885/0.5/6.79904/0.65, 6.79904/0.65/7.05885/0.5/7.05885/0.8/6.79904/0.95, 6.79904/0.95/7.05885/0.8/7.05885/1.1/6.79904/1.25, 7.05885/-0.4/7.31865/-0.55/7.31865/-0.25/7.05885/-0.1, 7.05885/-0.1/7.31865/-0.25/7.31865/0.05/7.05885/0.2, 7.05885/0.2/7.31865/0.05/7.31865/0.35/7.05885/0.5, 7.31865/-0.55/7.57846/-0.7/7.57846/-0.4/7.31865/-0.25, 7.31865/-0.25/7.57846/-0.4/7.57846/-0.1/7.31865/0.05}
			{\draw[fill=color4, rounded corners=2] (\a,\b) -- (\c,\d) -- (\e,\f) -- (\g,\h) -- cycle;}
			\foreach \a/\b/\c/\d/\e/\f/\g/\h in {10.433/0.75/10.6928/0.6/10.6928/0.9/10.433/1.05, 10.433/1.05/10.6928/0.9/10.6928/1.2/10.433/1.35, 10.433/1.35/10.6928/1.2/10.6928/1.5/10.433/1.65, 10.433/1.65/10.6928/1.5/10.6928/1.8/10.433/1.95, 10.6928/0.6/10.9526/0.45/10.9526/0.75/10.6928/0.9, 10.6928/0.9/10.9526/0.75/10.9526/1.05/10.6928/1.2, 10.6928/1.2/10.9526/1.05/10.9526/1.35/10.6928/1.5, 10.9526/0.45/11.2124/0.3/11.2124/0.6/10.9526/0.75, 10.9526/0.75/11.2124/0.6/11.2124/0.9/10.9526/1.05, 10.9526/1.05/11.2124/0.9/11.2124/1.2/10.9526/1.35, 11.2124/0.3/11.4722/0.15/11.4722/0.45/11.2124/0.6, 11.4722/0.15/11.7321/0./11.7321/0.3/11.4722/0.45}
			{\draw[fill=white!40!gray, rounded corners=2] (\a,\b) -- (\c,\d) -- (\e,\f) -- (\g,\h) -- cycle;}
			\node[below] at (3.59808, -1.5) {$\scriptstyle x_1$};
			\node[below] at (8.09808, -1.5) {$\scriptstyle {\color{color1} x_2}={\color{color3} x_3}$};
			\node[below] at (12.5981, -1.5) {$\scriptstyle x_4$};
			\draw[-stealth] (1., 0.) -- (1.43301, 0.25) node[pos=0.5,above] {$\scriptstyle u_1$};
			\draw[-stealth] (5.5, 0.) -- (5.5, 1.) node[pos=0.5,right] {$\scriptstyle\color{color1} u_2$};
			\draw[-stealth] (5.5, 0.) -- (6.79904, -0.25) node[pos=0.9,below] {$\scriptstyle\color{color3} u_3$};
			\draw[-stealth] (10., 0.) -- (10.433, 0.75) node[pos=0.5,right] {$\scriptstyle u_4$};
		\end{scope}
		\end{scope}
		\begin{scope}[shift={(0,7)}]
			\begin{scope}[scale=0.7]
				\draw[-stealth] (0,0) -- (13,0);
				\node[right] at (13,0) {$\scriptstyle x$};
				\draw[-stealth] (0,0) -- (0,3);
				\node[above] at (0,3) {$\scriptstyle {\rm Im}\,\phi$};
				\draw[-stealth] (0,0) -- (2.59808,-1.5);
				\node[below] at (2.59808,-1.5) {$\scriptstyle {\rm Re}\,\phi$};
				\draw (1.,0.) -- (3.59808,-1.5) -- (3.59808,1.5) -- (1.,3.) -- cycle;
				\draw (4.,0.)--(6.59808,-1.5)--(6.59808,1.5)--(4.,3.)--cycle;
				\draw (7.,0.)--(9.59808,-1.5)--(9.59808,1.5)--(7.,3.)--cycle;
				\draw (10.,0.)--(12.5981,-1.5)--(12.5981,1.5)--(10.,3.)--cycle;
				\foreach \a/\b/\c/\d/\e/\f/\g/\h in {1.43301/0.25/1.69282/0.1/1.69282/0.4/1.43301/0.55, 1.43301/0.55/1.69282/0.4/1.69282/0.7/1.43301/0.85, 1.43301/0.85/1.69282/0.7/1.69282/1./1.43301/1.15, 1.69282/0.1/1.95263/-0.05/1.95263/0.25/1.69282/0.4, 1.69282/0.4/1.95263/0.25/1.95263/0.55/1.69282/0.7, 1.69282/0.7/1.95263/0.55/1.95263/0.85/1.69282/1., 1.95263/-0.05/2.21244/-0.2/2.21244/0.1/1.95263/0.25}
				{\draw[fill=white!40!gray, rounded corners=2] (\a,\b) -- (\c,\d) -- (\e,\f) -- (\g,\h) -- cycle;}
				\foreach \a/\b/\c/\d/\e/\f/\g/\h in {7./1./7.25981/0.85/7.25981/1.15/7./1.3, 7./1.3/7.25981/1.15/7.25981/1.45/7./1.6, 7./1.6/7.25981/1.45/7.25981/1.75/7./1.9, 7./1.9/7.25981/1.75/7.25981/2.05/7./2.2, 7.25981/0.85/7.51962/0.7/7.51962/1./7.25981/1.15}
				{\draw[fill=color2, rounded corners=2] (\a,\b) -- (\c,\d) -- (\e,\f) -- (\g,\h) -- cycle;}
				\foreach \a/\b/\c/\d/\e/\f/\g/\h in {5.29904/-0.25/5.55885/-0.4/5.55885/-0.1/5.29904/0.05, 5.29904/0.05/5.55885/-0.1/5.55885/0.2/5.29904/0.35, 5.29904/0.35/5.55885/0.2/5.55885/0.5/5.29904/0.65, 5.29904/0.65/5.55885/0.5/5.55885/0.8/5.29904/0.95, 5.29904/0.95/5.55885/0.8/5.55885/1.1/5.29904/1.25, 5.55885/-0.4/5.81865/-0.55/5.81865/-0.25/5.55885/-0.1, 5.55885/-0.1/5.81865/-0.25/5.81865/0.05/5.55885/0.2, 5.55885/0.2/5.81865/0.05/5.81865/0.35/5.55885/0.5, 5.81865/-0.55/6.07846/-0.7/6.07846/-0.4/5.81865/-0.25, 5.81865/-0.25/6.07846/-0.4/6.07846/-0.1/5.81865/0.05}
				{\draw[fill=color4, rounded corners=2] (\a,\b) -- (\c,\d) -- (\e,\f) -- (\g,\h) -- cycle;}
				\foreach \a/\b/\c/\d/\e/\f/\g/\h in {10.433/0.75/10.6928/0.6/10.6928/0.9/10.433/1.05, 10.433/1.05/10.6928/0.9/10.6928/1.2/10.433/1.35, 10.433/1.35/10.6928/1.2/10.6928/1.5/10.433/1.65, 10.433/1.65/10.6928/1.5/10.6928/1.8/10.433/1.95, 10.6928/0.6/10.9526/0.45/10.9526/0.75/10.6928/0.9, 10.6928/0.9/10.9526/0.75/10.9526/1.05/10.6928/1.2, 10.6928/1.2/10.9526/1.05/10.9526/1.35/10.6928/1.5, 10.9526/0.45/11.2124/0.3/11.2124/0.6/10.9526/0.75, 10.9526/0.75/11.2124/0.6/11.2124/0.9/10.9526/1.05, 10.9526/1.05/11.2124/0.9/11.2124/1.2/10.9526/1.35, 11.2124/0.3/11.4722/0.15/11.4722/0.45/11.2124/0.6, 11.4722/0.15/11.7321/0./11.7321/0.3/11.4722/0.45}
				{\draw[fill=white!40!gray, rounded corners=2] (\a,\b) -- (\c,\d) -- (\e,\f) -- (\g,\h) -- cycle;}
				\node[below] at (3.59808, -1.5) {$\scriptstyle x_1$};
				\node[color3, below] at (6.59808, -1.5) {$\scriptstyle x_3$};
				\node[color1, below] at (9.59808, -1.5) {$\scriptstyle x_2$};
				\node[below] at (12.5981, -1.5) {$\scriptstyle x_4$};
				\draw[color5, thick, postaction=decorate, decoration={markings, mark= at position 0.7 with {\arrow{stealth}}}] (5.94856, 0.125) to[out=0,in=180] node[pos=0.8,above left] {\small tunnel} (7.38971, 1.225);
				\draw[fill=color6, rounded corners=2] (7.25981, 1.15) -- (7.25981, 1.45) -- (7.51962, 1.3) -- (7.51962, 1.) -- cycle;
				\draw[fill=color6, rounded corners=2] (5.81865, 0.05) -- (5.81865, 0.35) -- (6.07846, 0.2) -- (6.07846, -0.1) -- cycle;
				\draw[-stealth] (1., 0.) -- (1.43301, 0.25) node[pos=0.5,above] {$\scriptstyle u_1$};
				\draw[-stealth] (7., 0.) -- (7., 1.) node[pos=0.5,right] {$\scriptstyle\color{color1} u_2$};
				\draw[-stealth] (4., 0.) -- (5.29904, -0.25) node[pos=0.9,below] {$\scriptstyle\color{color3} u_3$};
				\draw[-stealth] (10., 0.) -- (10.433, 0.75) node[pos=0.5,right] {$\scriptstyle u_4$};
			\end{scope}
		\end{scope}
		\draw[line width = 1mm] (-4,-1) -- (-4,9) (-1,-1) -- (-1,9);
		\draw[color1, line width = 1mm] (-3.5,-1) to[out=90,in=270] (-2.7,2) to[out=90,in=270] (-2.3,6) to[out=90,in=270] (-1.5,9);
		\draw[color3, line width = 1mm] (-1.5,-1) to[out=90,in=270] (-2.3,2) to[out=90,in=270] (-2.7,6) to[out=90,in=270] (-3.5,9);
		\draw[line width = 0.5mm,white,dashed] (-2.7,2) to[out=90,in=270] (-2.3,6);
		\draw[line width = 0.5mm,white,dashed] (-2.3,2) to[out=90,in=270] (-2.7,6);
		\node[right,rotate=-90] at (-4,-1) {$\scriptstyle (x_1(t),u_1)$};
		\node[color1,right,rotate=-90] at (-3.5,-1) {$\scriptstyle (x_2(t),u_2)$};
		\node[color3,right,rotate=-90] at (-1.5,-1) {$\scriptstyle (x_3(t),u_3)$};
		\node[right,rotate=-90] at (-1,-1) {$\scriptstyle (x_4(t),u_4)$};
		\draw[rounded corners=5] (-4.2,3.1) -- (-4.2,4.9) -- (-0.8,4.9) -- (-0.8,3.1) -- cycle;
		\draw[rounded corners=5] (-4.2,-0.9) -- (-4.2,2.9) -- (-0.8,2.9) -- (-0.8,-0.9) -- cycle;
		\draw[rounded corners=5] (-4.2,8.9) -- (-4.2,5.1) -- (-0.8,5.1) -- (-0.8,8.9) -- cycle;
		\node[left] at (-4.2,4) {(\ref{en_c}) ${\bf R}(u_{12})$};
		\node[left] at (-4.2,1) {(\ref{en_b}) $U(u_{12})$};
		\node[left] at (-4.2,7) {(\ref{en_d}) $L(u_{12})$};
	\end{tikzpicture}
	\caption{A depiction of the $R$-matrix as a Berry transport process.}\label{fig:R_phys}
\end{figure}
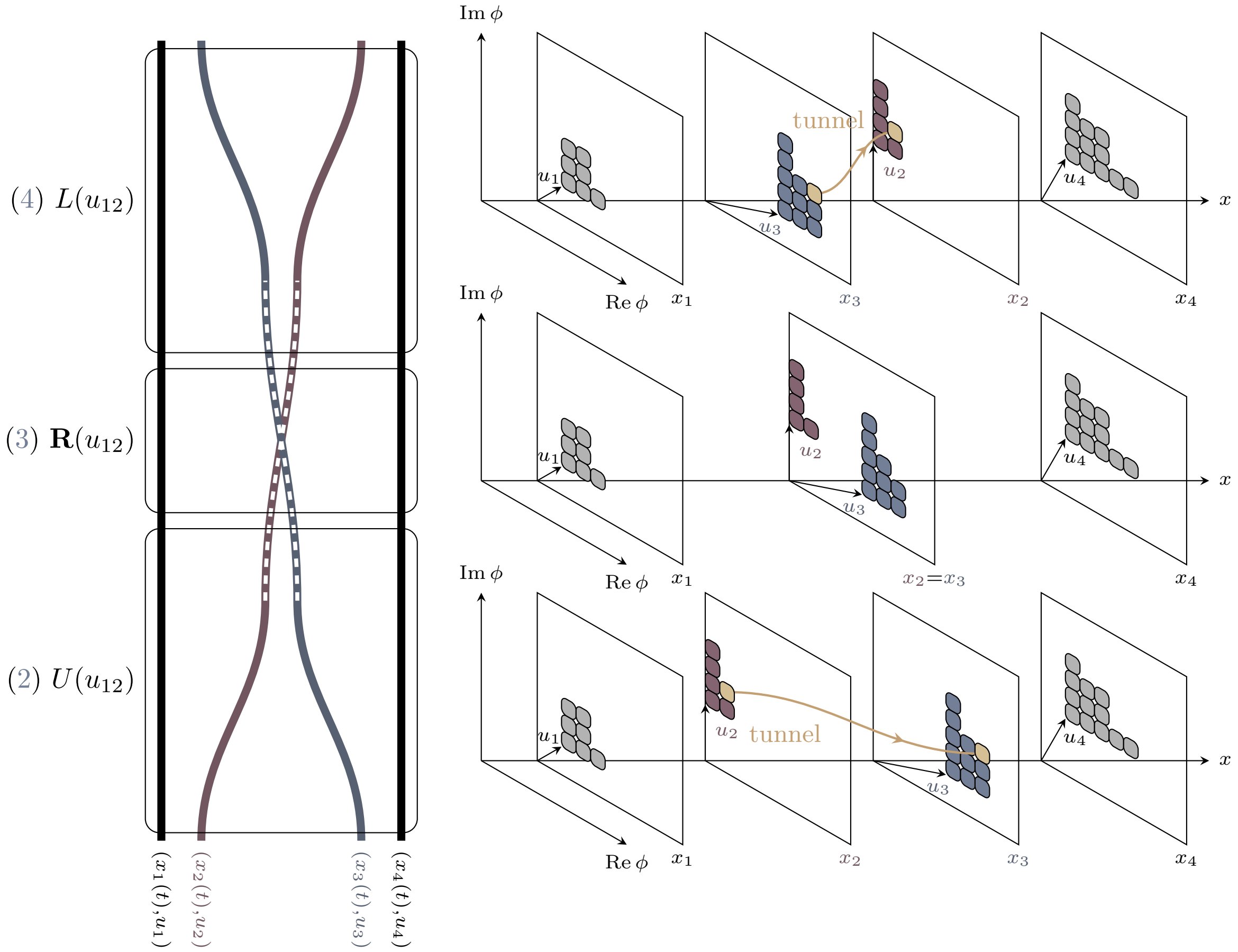

We depict the $R$-matrix in Fig.~\ref{fig:R_phys} as a process in which two nearby ordered values of $x_{\ff}$, say $x_2$ and $x_3$, are permuted.
In this language, the pictures of $R$-matrices adopted in Secs.~\ref{sec:spin_chains} and \ref{sec:diag} depict literally worldlines corresponding to the trajectories $x_{\ff}(t)$, with $x$ along the horizontal axis and $t$ along the vertical axis (see Fig.~\ref{fig:R_phys}).
The $R$-matrix transition then consists of five phases:
\begin{enumerate}
	\item \label{en_a} At $t=-\infty$, the parameters $x_{\ff}$ are widely separated.
	Tunneling is suppressed by the instanton factor $e^{-\ldots}$.
	\item \label{en_b} The values $x_2$ and $x_3$ approach each other, with $x_2<x_3$.
	Tunneling becomes possible, and the corresponding amplitudes, given by the formula in \eqref{Berry}, produce the \emph{upper-triangular} matrix $U(u_{12})$ for the factorization twist.
	Further jumps, such as those from $x_2$ to $x_4$, remain suppressed.
	\item \label{en_c} At $x_2=x_3$, tunneling is suppressed by the equivariant part.
	The two crystals lie in the same $\phi$-plane, and the system behaves as a chain multicrystal: the Hecke operators describe a representation of the Yangian algebra, which we called the pseudo-coproduct $\sD$ \eqref{pseudo}.
	The transition here is a mere redefinition of the basis, given by the diagonal pseudo-$R$-matrix \eqref{pseudo-R}.
	There is \emph{no tunneling} at this stage since the two configurations have the same $h$: a flow line must change $h$ monotonically, so it cannot connect critical points of equal height, even though distinct crystal configurations correspond to distinct isolated equivariant fixed points.
	\item \label{en_d} The inverse of phase \ref{en_b}: now $x_3<x_2$.
	This process is naturally described by the amplitude expression $L(u_{12})=\left(PU(-u_{12})P\right)^{-1}$, which is lower-triangular, since tunnelings are now directed from $x_3$ to $x_2$.
	\item \label{en_e} Final state.
	The parameters $x_{\ff}$ are again widely separated, and tunneling is again suppressed.
\end{enumerate}

In what follows, we will concentrate mostly on the factorization twist $U$ corresponding to the lower half of Fig.~\ref{fig:R_phys}.
Now our diagrammatic notation \eqref{U_diag} becomes clear: physically, we bring two crystals localized at $x_1$ and $x_2$ close to each other.
It is natural to \emph{filter} the amplitudes of $U=\sU$ in \eqref{Berry} by the number of cells transported in a jump.
This is the same filtration by the number $k$ of tunneling cells that we used when we classified the contributions $\Upsilon_k$ in \eqref{ansatz} as matrices with nonzero entries only above the $k^{\rm th}$ diagonal.

In \cite{Galakhov:2025fhd} it was argued that one-cell amplitudes can be evaluated relatively easily using ``mean field theory''.
One assumes that a single cell moves in the external field created by the immobile crystals, so the amplitude consists of three factors:
\begin{equation}\label{1-tunnel}
	\CA\Big[|\lambda_1+\sqbox{$a$}_1\rangle_{u_1}\otimes |\lambda_2\rangle_{u_2}\to|\lambda_1\rangle_{u_1}\otimes |\lambda_2+\sqbox{$a$}_2\rangle_{u_2}\Big]=\frac{-2\hbar}{\omega_1-\omega_2}\times\CA_{\lambda_1+\ssqbox{$a$}_1\to\lambda_1}\times\CA_{\lambda_2\to\lambda_2+\ssqbox{$a$}_2}\,.
\end{equation}
Here $-2\hbar$ is the equivariant weight of fields $C_a$.
The complex numbers $\omega_{1,2}$ are the cell contents of cells $\sqbox{$a$}_{1,2}$.
Their ratio describes the under-barrier tunneling amplitude for a single cell.
The amplitudes $\CA_{\lambda_1+\ssqbox{$a$}_1\to\lambda_1}$ and $\CA_{\lambda_2\to\lambda_2+\ssqbox{$a$}_2}$ correspond to a cell leaving one crystal and merging with the other, respectively.
They can be computed as the corresponding equivariant integrals over Hecke correspondences \cite{Galakhov:2024bzs}.
Expressions for these amplitudes coincide with the matrix coefficients of ${}^{\pm}e_0^a$ in \eqref{Y_rep} and \eqref{hysteresis}.
Thus the factorization twist reads
\begin{equation}\label{1-cell}
	\begin{aligned}
	&U|\lambda_1\rangle_{u_1}\otimes |\lambda_2\rangle_{u_2}=\sU|\lambda_1\rangle_{u_1}\otimes |\lambda_2\rangle_{u_2}=\\
	&=|\lambda_1\rangle_{u_1}\otimes |\lambda_2\rangle_{u_2}+\sum\lm_a\sum\lm_{\ssqbox{$a$}_1\in\lambda_1^-}\sum\lm_{\ssqbox{$a$}_2\in\lambda_2^+}\frac{-2\hbar}{\omega_1-\omega_2}\CA_{\lambda_1\to\lambda_1-\ssqbox{$a$}_1}\CA_{\lambda_2\to\lambda_2+\ssqbox{$a$}_2}|\lambda_1-\sqbox{$a$}_1\rangle_{u_1}\otimes |\lambda_2+\sqbox{$a$}_2\rangle_{u_2}+\ldots\,.
	\end{aligned}
\end{equation}
\noindent It explicitly satisfies \eqref{main_alg} explicitly up to higher-cell amplitudes.

Similarly, we expect that a generic $k$-cell transport amplitude takes the following form:
\begin{equation}\label{form-factor}
	\begin{aligned}
	&\CA\Big[|\lambda_1+\sqbox{$a$}_{1,1}+\sqbox{$b$}_{1,2}+\ldots\rangle_{u_1}\otimes |\lambda_2\rangle_{u_2}\to|\lambda_1\rangle_{u_1}\otimes |\lambda_2+\sqbox{$a$}_{2,1}+\sqbox{$b$}_{2,2}+\ldots\rangle_{u_2}\Big]=\\
	&=\CF(\omega_{1,1},\omega_{1,2},\ldots,\omega_{2,1},\omega_{2,2},\ldots)\times\CA_{\lambda_1+\ssqbox{$a$}_{1,1}+\ssqbox{$b$}_{1,2}+\ldots\to\lambda_1}\times\CA_{\lambda_2\to\lambda_2+\ssqbox{$a$}_{2,1}+\ssqbox{$b$}_{2,2}+\ldots}\,.
	\end{aligned}
\end{equation}
Here $\CF$ is a form factor for the coherent tunneling of $k$ cells, and the $\CA$ are amplitudes for cells leaving and merging with a crystal.
However, we are unable to present an expression as simple as \eqref{1-tunnel} for $\CF$ and $\CA$ in the general case.
The reason is that the cells interact with one another during the migration process.
We devote this paper to attempts to resolve these difficulties and construct higher-cell tunneling amplitudes.


\subsection{Attracting sets and stable envelopes}\label{sec:stab}

Stable envelopes were first defined in \cite{MO} (see also \cite{Aganagic:2021abc,okounkov2021inductive,okounkov2021nonabelian}) for Nakajima quiver varieties \cite{ginzburg2012lectures}.
The construction starts with the definition of an attracting set, a \emph{leaf}, associated with a fixed point (a classical vacuum) $\Lambda$ on the quiver variety.
The attracting set $\fL_{\Lambda}^{\{x\}}$ depends on the choice of a \emph{cocharacter}, real-valued eigenvalues $\{x\}$, belonging to a subtorus of the equivariant torus $T$.
Physically, the choice of $\{x\}$ corresponds to the choice of parameters $x_{\ff}$ associated with the framing nodes.
The attracting leaf $\fL_{\Lambda}^{\{x\}}$ is then defined as the set of points (Higgs fields) $H$ in the quiver variety $\fQ$ such that the flow pulls them toward $\Lambda$:
\begin{equation}
	\fL_{\Lambda}^{\{x\}}=\left\{H\in\fQ\,\bigg|\, \lim\lm_{t\to +\infty}e^{-X t}H=\Lambda \right\}\,,
\end{equation}
Alternatively, by reversing the flow direction, we can define $\fL_{\Lambda}^{\{x\}}$ as the trajectories of $H$-fields pushed out by the flow from the fixed point $\Lambda$.
Comparing this definition with \eqref{inst_smp}, we naturally observe that $\fL_{\Lambda}^{\{x\}}$ admits another definition as the collection of all instanton trajectories (flow lines) starting from $\Lambda$.
This definition is directly parallel to the definition of Lefschetz thimbles in symplectic geometry as the collection of all soliton trajectories (A-branes) in the corresponding Landau-Ginzburg model \cite{Gaiotto:2015aoa}.

We note that this definition omits equation \eqref{inst_hard}.
In principle, this is not surprising and fits the general framework of the Narasimhan--Seshadri--Hitchin--Kobayashi (NSHK) correspondence. 
Equation \eqref{inst_hard} is the real moment map equation.
We can eliminate the real moment map equation \cite{donaldson1983new,king1994moduli,Nakajima1994} by complexifying the gauge group and considering only stable configurations of the field $H$.

The stable envelopes $\operatorname{Stab}^{\{x\}}_{\Lambda}$ are defined as the unique equivariant cohomology classes in $H_T^*(\fQ)$ satisfying the normalization, support, and degree conditions.
However, we will avoid an explicit construction of the stable envelopes from first principles, since it is complicated; we therefore do not state all these constraints in full here.
Moreover, not all quiver varieties considered here are of Nakajima type.
For non-Nakajima quiver varieties, some of the constraints must be modified \cite{Ishtiaque:2023acr,Cao:2025yng,Cao:2026pir,Cao:2026vwf}.

Instead, we follow the physics route \cite{Dedushenko:2021mds,Bullimore:2021rnr}.
After localizing at the equivariant Euler classes of the fixed points, every equivariant cohomology class is a linear combination of the normalized fixed-point classes.
In practice, this means that the stable envelope decomposes as a linear combination of the ground state wave functions $\Psi_{\Lambda}$ associated with the fixed points, introduced in Sec.~\ref{sec:QSQM},
\begin{equation}
	{\rm Stab}_{\Lambda}^{\{x\}}=\sum\lm_{\Lambda'\geq \Lambda} {\rm Stab}_{\Lambda}^{\{x\}}\big|_{\Lambda'}\times \Psi_{\Lambda'}\,.
\end{equation}
The support condition for stable envelopes states that ${\rm Stab}_{\Lambda}^{\{x\}}$ is supported on the attracting set $\fA_{\Lambda}^{\{x\}}$.
In particular, the matrix ${\rm Stab}_{\Lambda}^{\{x\}}\big|_{\Lambda'}$ is upper-triangular, in other words ${\rm Stab}_{\Lambda}^{\{x\}}\big|_{\Lambda'}=0$ for $\Lambda'<\Lambda$, just as $U$ \eqref{ansatz} and $\sU$ \eqref{Berry}.
Here, as before, the order is induced by the Morse height function $h$ \eqref{height_chain}.

To avoid normalization ambiguities, we identify the matrix of transition amplitudes, i.e., the Berry transport matrix $U$, with the renormalized stable envelopes,
\begin{tcolorbox}
	\begin{equation}
		U_{ij}=\CA_{\Lambda_i\to\Lambda_j}=\frac{{\rm Stab}_{\Lambda_j}^{\{x\}}\big|_{\Lambda_i}}{{\rm Stab}_{\Lambda_i}^{\{x\}}\big|_{\Lambda_i}}\,.
	\end{equation}
\end{tcolorbox}
\noindent This relation is a dimensional reduction of the Janus interface partition function of 3D $\mathcal{N}=4$ theories \cite{Dedushenko:2021mds,Bullimore:2021rnr}.
In this identification, the matrix of localized stable envelopes plays the role of the factorization twist as originally designed in \cite{MO}.

To conclude this subsection, let us compare the approaches presented above in the simplest example, $T^*\mathbb{CP}^1$ following \cite{MO,Bullimore:2017lwu}.
$T^*\mathbb{CP}^1$ is a quiver variety of type $\mathsf{A}_1$; that is, it has a single gauge node (see Fig.~\ref{fig:quiver}) and one framing attachment.
The gauge node has dimension $1$, with scalar fields $(\phi,\xi)$; the framing node has dimension $2$, with scalar vevs $(u_a,x_a)$; and the spin parameter is $\ell=1$.

Assuming $\zeta>0$, the $D$-term constraint $[C,C^\dagger]+II^\dagger-J^\dagger J=\zeta$ has a simple solution in which the only field $I=(q_1,q_2)$ acquires a vev.
The $D$-term constraint then becomes $|q_1|^2+|q_2|^2=\zeta$, which describes a $3$-sphere $S^3$.
The gauge group is $U(1)\cong S^1$, so the base $\mathbb{CP}^1\cong S^3/S^1$ is represented as a Hopf bundle in this model, with projective coordinate $(q_1:q_2)$ in the base.
Denoting the components of the field $J$ by $J=(p_1,p_2)^T$, the $F$-term constraint reads $\partial_C W=IJ=p_1q_1+p_2q_2=0$.
Thus the fields $(p_1,p_2)$ correspond to the fiber of $\mathcal{O}(-2)\cong T^*\mathbb{CP}^1$.

Fixed points correspond geometrically to the poles of $\mathbb{CP}^1$: $(q_1:q_2)=(\sqrt{\zeta}:0)$ corresponds to the crystal chain $|\Box\rangle_{u_1}\otimes|\varnothing\rangle_{u_2}$, while $(q_1:q_2)=(0:\sqrt{\zeta})$ corresponds to $|\varnothing\rangle_{u_1}\otimes|\Box\rangle_{u_2}$.
The flow equations read
\begin{equation}
	\dot q_1(\tau)=e^{\xi(\tau)-x_1}q_1(\tau),\quad \dot q_2(\tau)=e^{\xi(\tau)-x_2}q_2(\tau),\quad \dot p_1(\tau)=e^{x_1-\xi(\tau)}p_1(\tau),\quad \dot p_2(\tau)=e^{x_2-\xi(\tau)}p_2(\tau)\,,
\end{equation}
where $\tau$ is the Euclidean time along the instanton trajectory.
Clearly, there are two chambers and two corresponding orderings of the basis vectors:
\begin{equation}
	x_1>x_2\;\Longleftrightarrow \; |\Box\rangle\otimes|\varnothing\rangle>|\varnothing\rangle\otimes|\Box\rangle,\quad x_1<x_2\;\Longleftrightarrow \; |\Box\rangle\otimes|\varnothing\rangle<|\varnothing\rangle\otimes|\Box\rangle\,.
\end{equation} 
Without loss of generality, we choose $x_1<x_2$, so that $x_1\leq\xi(\tau)\leq x_2$.
The instanton trajectories leaving the fixed point $q_1:q_2=0:1$ cover the fiber at that point:
\begin{equation}
	\fL_{0:1}=F_{0:1}=\{(q_1:q_2,p_1,p_2)=(0:1,0,z_3)\}\,.
\end{equation}
The instanton trajectories leaving the fixed point $q_1:q_2=1:0$ flow along the base $\mathbb{CP}^1$ toward the other fixed point $q_1:q_2=0:1$.
According to the rules of \cite{MO}, the attracting set in this case includes both $\mathbb{CP}^1$ and the attracting set at the other fixed point:
\begin{equation}
	\fL_{1:0}=\IC\IP^1\cup F_{0:1}=\{(q_1:q_2,p_1,p_2)=(z_1:z_2,0,0)\}\cup\{(q_1:q_2,p_1,p_2)=(0:1,0,z_3)\}\,.
\end{equation}
In this case, the stable envelopes are simply the Thom classes of the corresponding varieties:
\begin{equation}
	{\rm Stab}_{1:0}=[\IC\IP^1]+[F_{0:1}], \quad {\rm Stab}_{0:1}=[F_{0:1}]\,.
\end{equation}
On smooth varieties, the localization of Thom classes is given by the product of the equivariant weights of the normal degrees of freedom \cite{guillemin1999supersymmetry}.
Thus we obtain
\begin{equation}
	[\IC\IP^1]\big|_{1:0}=[p_2]=u_{12}-2\hbar,\quad [\IC\IP^1]\big|_{0:1}=[p_1]=u_{21}-2\hbar, \quad [F_{0:1}]\big|_{1:0}=0,\quad [F_{0:1}]\big|_{0:1}=u_{12}\,.
\end{equation}
For the localization of the stable envelopes, we obtain
\begin{equation}
	{\rm Stab}_{1:0}\big|_{1:0}=u_{12}-2\hbar,\quad {\rm Stab}_{1:0}\big|_{0:1}=-2\hbar,\quad {\rm Stab}_{0:1}\big|_{1:0}=0,\quad {\rm Stab}_{0:1}\big|_{0:1}=u_{12}\,. 
\end{equation}
The factorization twist matrix $U$ then coincides with the one computed from the amplitudes:
\begin{equation}
	U=\begin{pmatrix}
		1 & -\frac{2\hbar}{u_{12}}\\
		0 & 1\\
	\end{pmatrix}\,.
\end{equation}

\subsection{On shuffle algebras and Higgs-Coulomb duality}\label{sec:HCdu}

Close relations are known between affine Yangians and doubles of shuffle algebras \cite{Yang:2018pbw,Tsymbaliuk:2022bqx}, and, more broadly, between quantum toroidal algebras and the corresponding shuffle algebras \cite{Negut:2013cz,Negut:2019tto}.
Physically, we interpret these relations as a manifestation of the Higgs--Coulomb duality \cite{Denef:2002ru}.

In simpler terms, upon localization on either branch, the system acquires an effective connection.
In both cases, this connection is the magnetic-monopole vector-potential connection.
On the Coulomb branch, states are effectively represented by weakly interacting BPS dyons whose centers are labeled by the eigenvalues of the matrices $\Phi$ and $X$, so that these dyons float in 3d space.
On the Higgs branch, as discussed above, this is the Berry connection, and $\Phi$ and $X$ span the corresponding parameter space.

On the Coulomb branch, the eigenvalues $\phi$ and $x$ of the matrix fields $\Phi$ and $X$ represent the coordinates of the dyonic quasi-particles.
Further localization fixes the $x$'s to their vevs $\langle x\rangle$, while the $\phi$'s remain unconstrained degrees of freedom.
The wave functions are of Laughlin type for particles in the $\phi$-plane with a perpendicular external magnetic field; that is, they are holomorphic in $\phi$ up to a universal non-holomorphic prefactor, and symmetric in $\phi$.
The shuffle multiplication emerges naturally for the holomorphic wave functions of effective dyonic molecules fused into a single molecule:
\begin{equation}
	\Psi_1(\{\phi\})*\Psi_2(\{\tilde\phi\}):={\rm Sym}\left[\Psi_1(\{\phi\})\,\Psi_2(\{\tilde\phi\})\,\prod\lm_{p\in \Pi_{1}}\prod\lm_{q\in \Pi_{2}}\eta_{\fQ}(\phi_{p}-\tilde\phi_{q}|\langle x_p\rangle-\langle \tilde x_q\rangle)\right]\,.
\end{equation}
Here $\Pi_{1,2}$ denote the sets of elementary particles belonging to the first and second molecules, and the phases $\eta_{\fQ}$ arise after integration over the Higgs fields and the off-diagonal gauge fields in quiver $\fQ$.
The symmetrization runs over mutual shuffles of the sets $\{\phi\}$ and $\{\tilde\phi\}$.
If two quiver nodes $a$ and $b$ are connected by an arrow $a\to b$, then after integration over the corresponding elementary Higgs field $H$, the resulting phase is precisely the extra Landau-level-one wave function
\begin{equation}
	\eta_H(\phi_1,\phi_2)=(\phi_1-\phi_2-\mathbf{f}_H)^{\Theta(\langle x_1\rangle-\langle x_2\rangle)},
\end{equation}
where $\Theta$ is the Heaviside step function.
We assume that the entire magnetic flux is concentrated in one hemisphere; this convention explains the presence of the step function.
There is also a shift by the flavor charge in the $\phi$-plane.

As mentioned earlier, these results carry over naturally to the Higgs branch with a small correction.
When computing the phases, we do not assume that the dyons float freely in the $\phi$-plane; instead, they are confined to crystal nodes (see Fig.~\ref{fig:parti}(b)).

From this point of view, it is not surprising that the stable envelopes, when localized to fixed points, are expressed in terms of these phase functions $\eta$.
See, e.g., \cite[Thm.~5.2.1, Prop.~5.3.1]{Shenfeld:2013abc}, \cite[eq.~(48)]{Smirnov:2019abc}, \cite[Sec.~4.4.5]{Aganagic:2021abc}, and \cite[Sec.~2.3]{Bullimore:2017lwu}.
In geometric language, the fixed eigenvalues of $\Phi$ that appear in the crystal phases correspond to Chern roots \cite{Zhou:2021vcb,Ueda:2020csb}.

Similarly, in the language of tunneling amplitudes, the r.h.s. of the hysteresis relations \eqref{hysteresis} corresponds to a ratio of $\eta$-factors:
\begin{equation}
	\frac{\CA_{\lambda\to\lambda+\ssqbox{$b$}_2}\CA_{\lambda+\ssqbox{$b$}_2\to\lambda+\ssqbox{$a$}_1+\ssqbox{$b$}_1}}{\CA_{\lambda\to\lambda+\ssqbox{$a$}_1}\CA_{\lambda+\ssqbox{$a$}_1\to\lambda+\ssqbox{$a$}_1+\ssqbox{$b$}_2}}=\frac{\eta_{\fQ}(\phi_{\ssqbox{$a$}_1}-\phi_{\ssqbox{$b$}_2}|x_1>x_2)}{\eta_{\fQ}(\phi_{\ssqbox{$b$}_2}-\phi_{\ssqbox{$a$}_1}|x_2>x_1)}\,.
\end{equation}
Physically, this relation simply re-expresses a ratio of wave functions corresponding to the molecules $\sqbox{$a$}-\sqbox{$b$}$ and $\sqbox{$b$}-\sqbox{$a$}$ in terms of the $\eta$-phases as they appear in the effective description with permuted positions $x_1$ and $x_2$.
Note that the explicit values of $x_{1,2}$ do not contribute; only the relative order does.


\section{Tunnel fusion}\label{sec:fusion}

\subsection{Tunnel relations and wall-crossing subtleties}\label{sec:subtle}

We have associated $R$-matrices with Berry transport in Sec.~\ref{sec:R-mat_Berry}.
In this association, the framing parameters $x_{\ff}$ are varied along paths in their parameter space.
The pictorial representation of the $R$-matrices discussed in Sec.~\ref{sec:spin_chains} then consists literally of a set of trajectories $x_{\ff}(t)$.
Since the Berry transport is locally flat, the transported result depends only on the homotopy class of the path.
Consequently, the paths depicted on the l.h.s. and on the r.h.s. of the pictorial YBE \eqref{YBE} are homotopic, and the algebraic relation among the $R$-matrices follows from the flatness of the Berry transport.
Let us now reconsider relation \eqref{CGanalog} from this angle; it also follows from the flatness of the Berry connection.

To save space, we do not redraw Fig.~\ref{fig:R_phys} in full; instead, we denote the possible tunneling jumps by arrows:
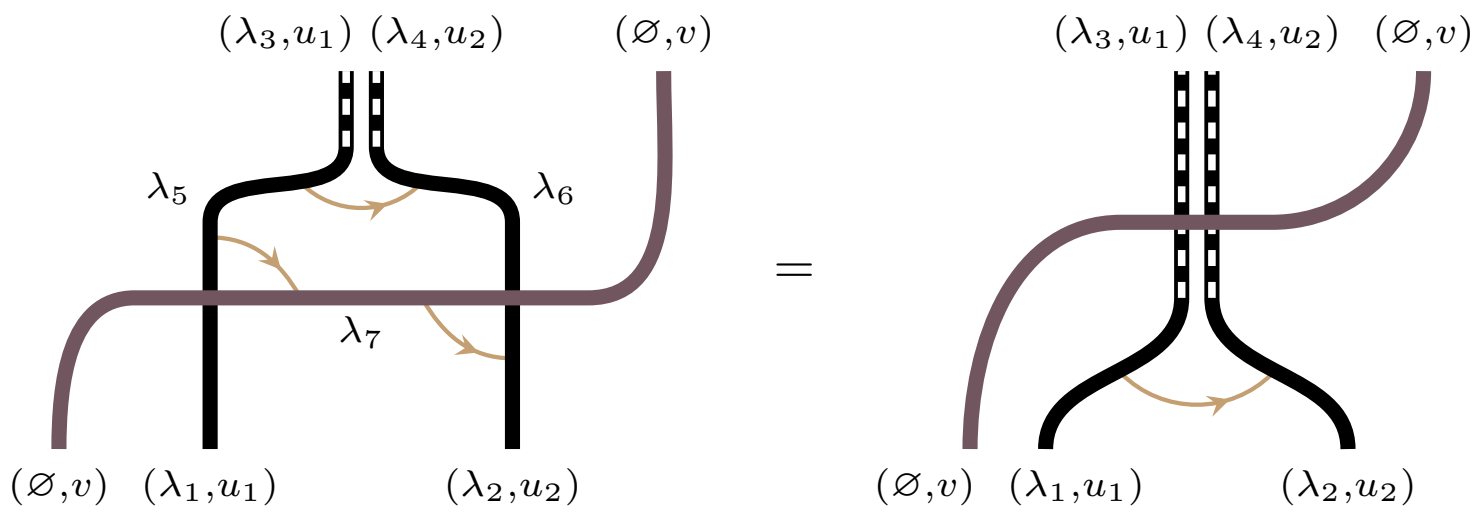
\begin{equation}\label{main_graph}
	\begin{array}{c}
		\begin{tikzpicture}
			\draw[thick, color5, postaction=decorate, decoration={markings, mark= at position 0.7 with {\arrow{stealth}}}] (-1,1.4) to[out=0,in=120] (-0.4,1);
			\draw[thick, color5, postaction=decorate, decoration={markings, mark= at position 0.7 with {\arrow{stealth}}}] (0.4,1) to[out=300,in=180] (1,0.6);
			\draw[thick, color5, postaction=decorate, decoration={markings, mark= at position 0.7 with {\arrow{stealth}}}] (-0.4,1.76) to[out=315,in=225] (0.4,1.76);
			\draw[line width = 1mm] (-1,0) -- (-1,1.5) to[out=90,in=270] (-0.1,2) -- (-0.1,2.5) (1,0) -- (1,1.5) to[out=90,in=270] (0.1,2) -- (0.1,2.5);
			\draw[line width = 0.5mm, dashed, white] (-0.1,2) -- (-0.1,2.5) (0.1,2) -- (0.1,2.5);
			\draw[line width = 1mm, color1] (-2,0) to[out=90,in=180] (-1.5,1) -- (1.5,1) to[out=0,in=270] (2,2.5);
			\node[below] at (-2,0) {$\scriptstyle (\varnothing,v)$};
			\node[above] at (2,2.5) {$\scriptstyle (\varnothing,v)$};
			\node[below] at (-1,0) {$\scriptstyle (\lambda_1,u_1)$};
			\node[below] at (1,0) {$\scriptstyle (\lambda_2,u_2)$};
			\node[above] at (-0.5,2.5) {$\scriptstyle (\lambda_3,u_1)$};
			\node[above] at (0.5,2.5) {$\scriptstyle (\lambda_4,u_2)$};
			\node[above left] at (-1,1.5) {$\scriptstyle \lambda_5$};
			\node[above right] at (1,1.5) {$\scriptstyle \lambda_6$};
			\node[below] at (0,1) {$\scriptstyle \lambda_7$};
		\end{tikzpicture}
	\end{array}=\begin{array}{c}
	\begin{tikzpicture}
		\draw[thick, color5, postaction=decorate, decoration={markings, mark= at position 0.7 with {\arrow{stealth}}}] (-0.5,0.5) to[out=315,in=225] (0.5,0.5);
		\draw[line width = 1mm] (-1,0) to[out=90,in=270] (-0.1,1) -- (-0.1,2.5) (1,0) to[out=90,in=270] (0.1,1) -- (0.1,2.5);
		\draw[line width = 0.5mm, dashed, white] (-0.1,1) -- (-0.1,2.5) (0.1,1) -- (0.1,2.5);
		\draw[line width = 1mm, color1] (-1.5,0) to[out=90,in=180] (-0.5,1.5) -- (0.5,1.5) to[out=0,in=270] (1.5,2.5);
		\node[below] at (-1.8,0) {$\scriptstyle (\varnothing,v)$};
		\node[above] at (1.5,2.5) {$\scriptstyle (\varnothing,v)$};
		\node[below] at (-0.8,0) {$\scriptstyle (\lambda_1,u_1)$};
		\node[below] at (1,0) {$\scriptstyle (\lambda_2,u_2)$};
		\node[above] at (-0.5,2.5) {$\scriptstyle (\lambda_3,u_1)$};
		\node[above] at (0.5,2.5) {$\scriptstyle (\lambda_4,u_2)$};
	\end{tikzpicture}
	\end{array}\,.
\end{equation}
Let us stress that, owing to the vacuum choice in the initial and final states of the control strand, there are no tunneling jumps before or after the crossing of the control strand with other strands.
However, tunneling jumps do occur in the intermediate channel.

We rewrite the graphical equation \eqref{main_graph} in the algebraic form:
\begin{equation}
	\sum\lm_{\lambda_5,\lambda_6,\lambda_7}\rho(\varnothing,\lambda_1)\CA_{(\lambda_1,\varnothing)\to(\lambda_5,\lambda_7)}\CA_{(\lambda_7,\lambda_2)\to(\varnothing,\lambda_6)}\rho(\varnothing,\lambda_6)\CA_{(\lambda_5,\lambda_6)\to (\lambda_3,\lambda_4)}=\CA_{(\lambda_1,\lambda_2)\to(\lambda_3,\lambda_4)}\rho(\varnothing,\lambda_3)\rho(\varnothing,\lambda_4)\,.
\end{equation}
Here the sum runs over all possible partitions labeling the internal lines.
The functions $\rho(\lambda_1,\lambda_2)$ are the eigenvalue functions of the diagonal operator $\mathbf{R}$ \eqref{diag_R}.
All expressions depend on the spectral parameters $u_1$, $u_2$, and $v$ in the corresponding way.
Here we omit this dependence for brevity.

Splitting the summation over $\lambda_7$ into the cases $\lambda_7=\varnothing$ and $\lambda_7\neq\varnothing$, we obtain the following relation:
\begin{tcolorbox}
\begin{equation}\label{main_amp}
\begin{aligned}
	&{\color{color7}Z_{(\lambda_1,\lambda_2)\to(\lambda_1,\lambda_2)}}\CA_{(\lambda_1,\lambda_2)\to(\lambda_3,\lambda_4)}=\sum\lm_{\lambda_5,\lambda_6}\rho(\varnothing,\lambda_1)\rho(\varnothing,\lambda_6){\color{color1}\CA^{\rm (cat)}_{(\lambda_1,\lambda_2)\to(\lambda_5,\lambda_6)}}\CA_{(\lambda_5,\lambda_6)\to (\lambda_3,\lambda_4)}\,,\\
	&\mbox{where } {\color{color7}Z}=\rho(\varnothing,\lambda_3)\rho(\varnothing,\lambda_4)-\rho(\varnothing,\lambda_1)\rho(\varnothing,\lambda_2),\;{\color{color1}\CA^{\rm (cat)}_{(\lambda_1,\lambda_2)\to(\lambda_5,\lambda_6)}}=\sum\lm_{\lambda_7\neq \varnothing}\CA_{(\lambda_1,\varnothing)\to(\lambda_5,\lambda_7)}\CA_{(\lambda_7,\lambda_2)\to(\varnothing,\lambda_6)}\,.
\end{aligned}
\end{equation}
\end{tcolorbox}
\noindent In this equation we separate out one amplitude $\CA^{\rm (cat)}$ and call it ``\emph{catalytic}.''
Physically, it describes a catalytic jump induced by the control strand: when the control strand approaches one of the other strands, it captures some cells from it and then donates them to the other strand when it reaches it later.
This process is rather restrictive.
By construction, we assume that at least one cell is transported by the catalytic process, and the control strand can accept only a limited number of extra cells, since by assumption it corresponds to some minuscule highest weight.
Thus $\CA^{\mathrm{(cat)}}$ is a finite sum over simplified amplitudes.
In practice, we can even glue the tunnel ends $(\lambda_5,\lambda_7)$ and $(\lambda_7,\lambda_2)$ over $\lambda_7$, so that $\CA^{\mathrm{(cat)}}$ is indeed an amplitude for cells migrating from the strand at $u_1$ to the strand at $u_2$.

Due to the well-established filtration of amplitudes by the number of transported cells, equation \eqref{main_amp} yields recursion relations that allow one to construct all $k$-cell amplitudes from simpler $(k-m)$-cell amplitudes and $m$-cell catalytic amplitudes for $m\geq 1$.
This is equation \eqref{main_alg} rewritten in terms of tunneling amplitudes.
We stress once again that the fact that amplitudes satisfy recursion relations \eqref{main_amp} follows from the flatness of the Berry connection, which is in turn a consequence of SUSY.

Equation \eqref{main_amp} suggests that a similar construction applies to the tunneling instanton trajectories themselves.
One could therefore construct an instanton trajectory describing the migration of $k$ cells as a ``path sum'' over \emph{glued} intermediate paths for $k-m$ and $m$ cells migrating in succession.
Taken at face value, this form would suggest that the $k$-cell amplitude is represented directly as a sum over $(k-m)$- and $m$-cell amplitudes.
The obstacle is that the l.h.s. of \eqref{main_amp} carries a nontrivial factor given by the pseudo-$R$-matrix eigenvalues.
As we will see in the examples, the individual $(k-m)$- and $m$-cell amplitudes are not divisible by this factor, so it can be canceled only by specific coherent sums of broken $(k-m)$--$m$ paths.
Alternatively, this obstruction can be rephrased as follows.
Combining amplitudes on the r.h.s. of \eqref{main_amp} gives a result that factorizes and contains the factor from the l.h.s., but the remaining factor, the form factor $\CF$ of \eqref{form-factor}, does not decompose canonically into broken $(k-m)$--$m$ amplitudes.

The issue described above shows that merely gluing $(k-m)$- and $m$-cell instanton paths is insufficient: the instantons, viewed as quasi-defects, experience wall-crossing phenomena.
These wall-crossing phenomena are a direct image of the wall-crossing phenomena for solitons in 2d $\CN=(2,2)$ theories \cite{Kontsevich:2008fj,Cecotti:2010Rtwist,Gaiotto:2011tf,Gaiotto:2012rg,Herbst:2008jq,Hori:2013ika,Gaiotto:2015aoa,Galakhov:2020upa,Brunner:2021ulc,Aganagic:2009OpenBPS,Khan:2020Categorical,Khan:2024yiy}, since our SQM is a dimensional reduction of that theory; the instantons are thus an image of the solitons of the 2d theory.
The wall-crossing relations can be quite involved, especially in our case, where one-loop determinants in the defect background contribute directly.
We therefore do not attempt to classify them here.
Instead, we consider several examples of these relations and verify explicitly that the tunneling amplitudes satisfy the recursion relations \eqref{main_amp}.


\subsection{\texorpdfstring{$\myY(\fs\fl_{n+1})$}{}, fundamental representation, emergent Weyl symmetry}\label{sec:sl(n+1)_fu}

We begin with a simple example, the evaluation representation of $\myY(\fs\fl_{n+1})$ with lowest weight $-\omega_1$ (highest weight $\omega_n$).
This implies that the framing is attached to the first node in Fig.~\ref{fig:quiver}, with parameter $r=1$.
In this case, the crystals (see Fig.~\ref{fig:parti}(b,c)) are simply strips of cells.
The total number of cells in the strip ranges from $0$ to $n$:
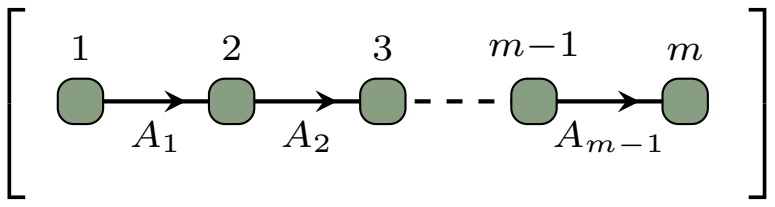
\begin{equation}
	|m\rangle=\left[\begin{array}{c}
		\begin{tikzpicture}
			\draw[thick, postaction=decorate, decoration={markings, mark= at position 0.7 with {\arrow{stealth}}}] (0,0) -- (1,0) node[pos=0.5, below] {$\scriptstyle A_1$};
			\draw[thick, postaction=decorate, decoration={markings, mark= at position 0.7 with {\arrow{stealth}}}] (1,0) -- (2,0) node[pos=0.5, below] {$\scriptstyle A_2$};
			\draw[thick, postaction=decorate, decoration={markings, mark= at position 0.7 with {\arrow{stealth}}}] (3,0) -- (4,0) node[pos=0.5, below] {$\scriptstyle A_{m-1}$};
			\draw[thick, dashed] (2,0) -- (3,0);
			\foreach \i in {0, 1, 2, 3, 4}
			{
				\begin{scope}[shift={(\i,0)}]
					\draw[fill=color8, rounded corners=3] (-0.15,-0.15) -- (-0.15,0.15) -- (0.15,0.15) -- (0.15,-0.15) -- cycle;
				\end{scope}
			}
			\node[above] at (0,0.15) {$\scriptstyle 1$};
			\node[above] at (1,0.15) {$\scriptstyle 2$};
			\node[above] at (2,0.15) {$\scriptstyle 3$};
			\node[above] at (3,0.15) {$\scriptstyle m-1$};
			\node[above] at (4,0.15) {$\scriptstyle m$};
		\end{tikzpicture}
	\end{array}\right],\quad I=(1-\delta_{m,0}),\quad m=0,1,\ldots,n\,.
\end{equation}
All the other fields are set to zero vevs.

Now consider a tensor product $|a\rangle_{u_1}\otimes |b\rangle_{u_2}$ assuming that $a>b$.
In the crystal chain there are two cells of each color $c=1,\ldots,b$ and one cell of each color $c=b+1,\ldots,a$.
It turns out that there are only two crystal chain configurations with these occupation numbers.
Indeed, a cell of color $c+2$ cannot be attached to a cell of color $c$ unless a cell of color $c+1$ lies in between.
These two configurations are:
\begin{equation}
	\begin{aligned}
	&|a\rangle_{u_1}\otimes|b\rangle_{u_2}=\left[\begin{array}{c}
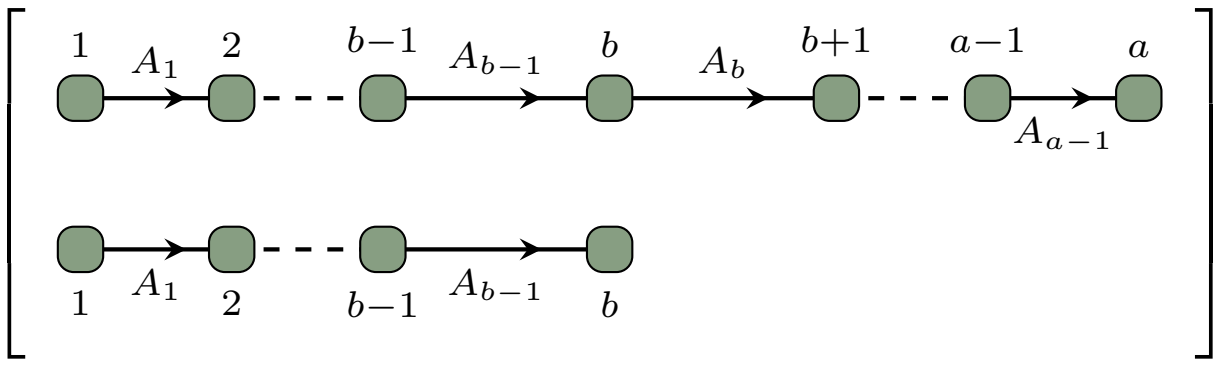
		\begin{tikzpicture}
			\draw[thick, postaction=decorate, decoration={markings, mark= at position 0.7 with {\arrow{stealth}}}] (0,0) -- (1,0) node[pos=0.5, above] {$\scriptstyle A_1$};
			\draw[thick, postaction=decorate, decoration={markings, mark= at position 0.7 with {\arrow{stealth}}}] (2,0) -- (3.5,0) node[pos=0.5, above] {$\scriptstyle A_{b-1}$};
			\draw[thick, postaction=decorate, decoration={markings, mark= at position 0.7 with {\arrow{stealth}}}] (3.5,0) -- (5,0) node[pos=0.5, above] {$\scriptstyle A_{b}$};
			\draw[thick, postaction=decorate, decoration={markings, mark= at position 0.7 with {\arrow{stealth}}}] (6,0) -- (7,0) node[pos=0.5, below] {$\scriptstyle A_{a-1}$};
			\draw[dashed, thick] (1,0) -- (2,0) (5,0) -- (6,0);
			\foreach \i in {0, 1, 2, 3.5, 5, 6, 7}
			{
				\begin{scope}[shift={(\i,0)}]
					\draw[fill=color8, rounded corners=3] (-0.15,-0.15) -- (-0.15,0.15) -- (0.15,0.15) -- (0.15,-0.15) -- cycle;
				\end{scope}
			}
			\node[above] at (0,0.15) {$\scriptstyle 1$};
			\node[above] at (1,0.15) {$\scriptstyle 2$};
			\node[above] at (2,0.15) {$\scriptstyle b-1$};
			\node[above] at (3.5,0.15) {$\scriptstyle b$};
			\node[above] at (5,0.15) {$\scriptstyle b+1$};
			\node[above] at (6,0.15) {$\scriptstyle a-1$};
			\node[above] at (7,0.15) {$\scriptstyle a$};
			\begin{scope}[shift={(0,-1)}]
				\draw[thick, postaction=decorate, decoration={markings, mark= at position 0.7 with {\arrow{stealth}}}] (0,0) -- (1,0) node[pos=0.5, below] {$\scriptstyle A_1$};
				\draw[thick, postaction=decorate, decoration={markings, mark= at position 0.7 with {\arrow{stealth}}}] (2,0) -- (3.5,0) node[pos=0.5, below] {$\scriptstyle A_{b-1}$};
				\draw[dashed, thick] (1,0) -- (2,0); 
				\foreach \i in {0, 1, 2, 3.5}
				{
					\begin{scope}[shift={(\i,0)}]
						\draw[fill=color8, rounded corners=3] (-0.15,-0.15) -- (-0.15,0.15) -- (0.15,0.15) -- (0.15,-0.15) -- cycle;
					\end{scope}
				}
				\node[below] at (0,-0.15) {$\scriptstyle 1$};
				\node[below] at (1,-0.15) {$\scriptstyle 2$};
				\node[below] at (2,-0.15) {$\scriptstyle b-1$};
				\node[below] at (3.5,-0.15) {$\scriptstyle b$};
			\end{scope}
		\end{tikzpicture}
	\end{array}\right]\,,\\
	&|b\rangle_{u_1}\otimes|a\rangle_{u_2}=\left[\begin{array}{c}
		
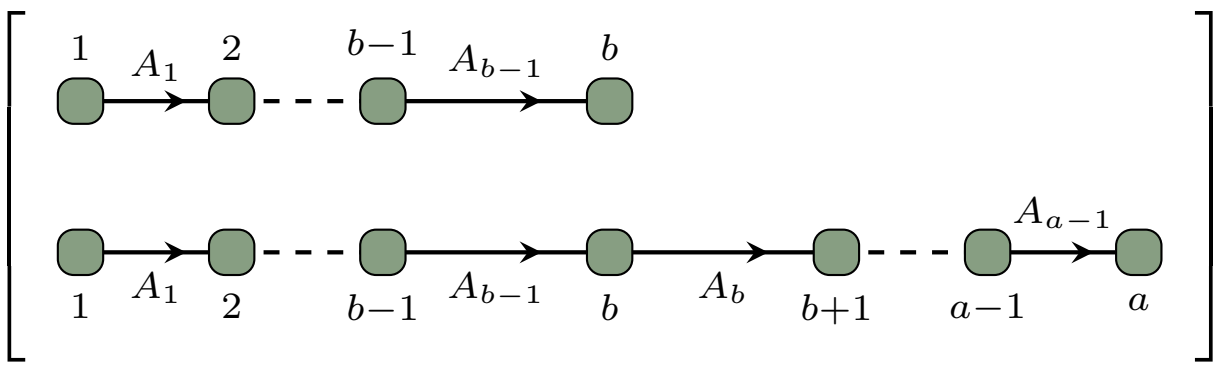
\begin{tikzpicture}
			\draw[thick, postaction=decorate, decoration={markings, mark= at position 0.7 with {\arrow{stealth}}}] (0,0) -- (1,0) node[pos=0.5, below] {$\scriptstyle A_1$};
			\draw[thick, postaction=decorate, decoration={markings, mark= at position 0.7 with {\arrow{stealth}}}] (2,0) -- (3.5,0) node[pos=0.5, below] {$\scriptstyle A_{b-1}$};
			\draw[thick, postaction=decorate, decoration={markings, mark= at position 0.7 with {\arrow{stealth}}}] (3.5,0) -- (5,0) node[pos=0.5, below] {$\scriptstyle A_{b}$};
			\draw[thick, postaction=decorate, decoration={markings, mark= at position 0.7 with {\arrow{stealth}}}] (6,0) -- (7,0) node[pos=0.5, above] {$\scriptstyle A_{a-1}$};
			\draw[dashed, thick] (1,0) -- (2,0) (5,0) -- (6,0);
			\foreach \i in {0, 1, 2, 3.5, 5, 6, 7}
			{
				\begin{scope}[shift={(\i,0)}]
					\draw[fill=color8, rounded corners=3] (-0.15,-0.15) -- (-0.15,0.15) -- (0.15,0.15) -- (0.15,-0.15) -- cycle;
				\end{scope}
			}
			\node[below] at (0,-0.15) {$\scriptstyle 1$};
			\node[below] at (1,-0.15) {$\scriptstyle 2$};
			\node[below] at (2,-0.15) {$\scriptstyle b-1$};
			\node[below] at (3.5,-0.15) {$\scriptstyle b$};
			\node[below] at (5,-0.15) {$\scriptstyle b+1$};
			\node[below] at (6,-0.15) {$\scriptstyle a-1$};
			\node[below] at (7,-0.15) {$\scriptstyle a$};
			\begin{scope}[shift={(0,1)}]
				\draw[thick, postaction=decorate, decoration={markings, mark= at position 0.7 with {\arrow{stealth}}}] (0,0) -- (1,0) node[pos=0.5, above] {$\scriptstyle A_1$};
				\draw[thick, postaction=decorate, decoration={markings, mark= at position 0.7 with {\arrow{stealth}}}] (2,0) -- (3.5,0) node[pos=0.5, above] {$\scriptstyle A_{b-1}$};
				\draw[dashed, thick] (1,0) -- (2,0); 
				\foreach \i in {0, 1, 2, 3.5}
				{
					\begin{scope}[shift={(\i,0)}]
						\draw[fill=color8, rounded corners=3] (-0.15,-0.15) -- (-0.15,0.15) -- (0.15,0.15) -- (0.15,-0.15) -- cycle;
					\end{scope}
				}
				\node[above] at (0,0.15) {$\scriptstyle 1$};
				\node[above] at (1,0.15) {$\scriptstyle 2$};
				\node[above] at (2,0.15) {$\scriptstyle b-1$};
				\node[above] at (3.5,0.15) {$\scriptstyle b$};
			\end{scope}
		\end{tikzpicture}
	\end{array}\right]\,.
	\end{aligned}
\end{equation}
This picture makes it clear that if an instanton jump occurs in this situation, the cells of colors $c=b+1,\ldots,a$ jump together, as a single quasi-cell.
In other words, although this is a genuine $(a-b)$-cell transition, there is no issue with the mutual ordering of cells along the flow discussed in Sec.~\ref{sec:subtle}.

We describe the instanton trajectory interpolating between $|a\rangle_{u_1}\otimes |b\rangle_{u_2}$ and $|b\rangle_{u_1}\otimes |a\rangle_{u_2}$ by the following field configuration:
\begin{equation}
	\begin{aligned}
		& I=\begin{pmatrix}
			1 & 0\\
			0 & 1
		\end{pmatrix},\quad J=0,\quad B_k=0,\quad C_k=0\,,\\
		& X_1=\begin{pmatrix}
			x_1 & 0\\
			0 & x_2
		\end{pmatrix},\quad\ldots,\quad X_b=\begin{pmatrix}
		x_1 & 0\\
		0 & x_2
		\end{pmatrix}, \quad X_{b+1}=({\color{color2}\xi}),\quad\ldots,\quad X_{a}=({\color{color2}\xi})\,,\\
		& \Phi_1=\begin{pmatrix}
			u_1 & 0\\
			0 & u_2
		\end{pmatrix},\; \ldots,\; \Phi_b=\begin{pmatrix}
			u_1-\hbar(b-1) & 0\\
			0 & u_2-\hbar(b-1)
		\end{pmatrix}, \; \Phi_{b+1}=({\color{color2}\phi}-\hbar b),\; \ldots,\; \Phi_{a}=({\color{color2}\phi}-\hbar(a-1))\,,\\
		& A_1=\begin{pmatrix}
			1 & 0\\
			0 & 1\\
		\end{pmatrix},\quad\ldots,\quad A_{b-1}=\begin{pmatrix}
		1 & 0\\
		0 & 1\\
		\end{pmatrix},\quad A_b=\begin{pmatrix}
		{\color{color2}q_1} & {\color{color2}q_2}
		\end{pmatrix},\quad A_{b+1}=(1),\quad \ldots,\quad A_{a-1}=(1)\,.
	\end{aligned}
\end{equation}
We define these fields in the framework of the NSHK correspondence, i.e., we treat them as representatives of orbits of the complexified gauge group.
The instanton is described by the following interpolation of the fields $q_1$, $q_2$, $\xi$, and $\phi$:
\begin{equation}
	(q_1:q_2,\xi,\phi):\quad |a\rangle_{u_1}\otimes |b\rangle_{u_2}=(1:0,x_1,u_1)\;\longrightarrow \;|b\rangle_{u_1}\otimes |a\rangle_{u_2}=(0:1,x_2,u_2)\,.
\end{equation}
This trajectory is equivalent to the standard instanton trajectory connecting the two poles of the $\mathbb{CP}^1$ model (see \cite[Sec.~5.1]{Galakhov:2025fhd}), which flows along a meridian on the $\mathbb{CP}^1$ sphere.
This model contains two Higgs fields $q_1$ and $q_2$, both charged under the $U(1)$ gauge group.
The fields $\phi$ and $\xi$ are the corresponding complex and real scalar fields, respectively, in the $U(1)$ gauge multiplet of $\CN=4$ SQM.

These instantons contribute directly to the 1-cell migrating amplitude \eqref{1-tunnel}.
Thus, there is no need to recalculate the amplitude explicitly; we simply adopt the expression in \eqref{1-tunnel}:
\begin{equation}
	\CA\left[|a\rangle_{u_1}\otimes |b\rangle_{u_2}\to |b\rangle_{u_1}\otimes |a\rangle_{u_2}\right]=-\frac{2\hbar}{u_{12}}\,.
\end{equation}
Thus, we obtain the following Berry transport, which gives the factorization twist on the algebra side:
\begin{equation}
	U\left(|a\rangle_{u_1}\otimes |b\rangle_{u_2}\right)=\sU\left(|a\rangle_{u_1}\otimes |b\rangle_{u_2}\right)=|a\rangle_{u_1}\otimes |b\rangle_{u_2}-\frac{2\hbar\,\theta_{a,b}}{u_{12}}|b\rangle_{u_1}\otimes |a\rangle_{u_2}\,,
\end{equation}
where
\begin{equation}
	\theta_{a,b}:=\left\{ \begin{array}{ll}
		1,& \mbox{if }a>b\,,\\
		0,& \mbox{if }a\le b\,.
	\end{array}\right.
\end{equation}

Also we find:
\begin{equation}
	{\bf R}\left(|n\rangle_{u_1}\otimes |m\rangle_{u_2}\right)=\left(\delta_{m,n}+\frac{2\hbar}{u_{12}}\theta_{m,n}-\frac{u_{12}}{\hbar}\right)|n\rangle_{u_1}\otimes |m\rangle_{u_2}\,.
\end{equation}

Combining these expressions, we obtain the canonical expression for the $R$-matrix \cite{slavnov2019algebraicbetheansatz}:
\begin{tcolorbox}
	\begin{equation}
		R(u_{12})=\left(PU(-u_{12})P\right)^{-1}{\bf R}(u_{12})\,U(u_{12})=-\frac{u_{12}}{\hbar}\cdot\bbone+P\,.
	\end{equation}
\end{tcolorbox}

We can rewrite the resulting factorization twist in the following form:
\begin{equation}
	U(u_{12})=\bbone-\frac{2\hbar}{u_{12}}\sum\lm_{\alpha\in\Delta_+}E_{-\alpha}\otimes E_{\alpha}\,.
\end{equation}
At first glance, this expression might seem somewhat \emph{surprising} as a tunneling amplitude.
First, the amplitude receives contributions with various numbers $k$ of migrating cells, all with the same ``coupling constant'' -- the inverse spectral parameter $u_{12}^{-1}$.
Second, these higher tunneling contributions are not arbitrary: they are organized into a sum that enumerates \emph{all} positive roots.

These are precisely the amplitudes that represent $\CA^{\mathrm{(cat)}}$ in \eqref{main_amp}, or $\Omega_+^a$ in \eqref{Omega}.
We would like to draw the reader's attention to the fact that this \emph{emergent} Weyl symmetry between different roots of the underlying Lie algebra $\fs\fl_{n+1}$ was \emph{not expected} from the QFT point of view \emph{a priori}.
By design, the quiver Yangian construction lands in Drinfeld's second (current) presentation, which singles out the Chevalley basis, while the more ``Weyl-symmetry-friendly'' first presentation (the J-presentation) remains hidden within this framework.
In particular, the Serre relations and their origin in the basic construction for generic quiver Yangians require a deeper discussion \cite{Li:2020rij,Negut:2022pka}.
Of course, this phenomenon is not novel in topological string theory.
D-branes, and more broadly M-branes, compactified on ADE McKay singularities \cite{McKay:1980} flow to a super-Yang-Mills theory with the full ADE gauge algebra \cite{Douglas:1996sw,Aspinwall:2004jr,Witten:1996Phase}.
However, we should remark that the Weyl symmetry in the quiver language is realized by reflection functors \cite{Bernstein:1973Coxeter,Nakajima:2003Reflection,Maffei:2005Quiver}, which act on the stability parameters of the quiver accordingly \cite{Lusztig:1976Finiteness,Galakhov:2026cwv}.
In our setup, the stability parameters remain fixed during the Berry transport.
 

\subsection{\texorpdfstring{$\myY(\fs\fl_{2})$}{}, non-evaluation representations, cascades} \label{sec:sl(2)_nonev}

In the previous section, the problem of the appearance of the factor in \eqref{factor} (the factor $Z$ in \eqref{main_amp}) did not arise.
The reason is that the $k$-cell jump was coherent, so the cell contents are correlated with one another.
The eigenvalue in \eqref{factor} therefore behaves as $\alpha(u_{12}+\hbar\beta)$.
This factor may resemble a one-loop contribution from a chiral multiplet.
Such a contribution may, in principle, appear in each summand on the r.h.s. of \eqref{factor} and \eqref{main_amp}.
To expose the problem, we need a situation in which such coherence is certainly absent, so that the factor on the l.h.s. of \eqref{factor}, or of \eqref{main_amp}, cannot be produced by the amplitudes on the r.h.s. separately and arises only when they are all combined.
The most suitable test case is therefore a tensor product of non-evaluation representations whose crystal nucleation centers are maximally separated, so that each factor depends on many independent spectral parameters.
Again, we can depict this setting as a crystal chain in which the individual crystal centers are split into two groups lying in two distinct $x$-planes (see Fig.~\ref{fig:R_phys}).
The factorization twist then proceeds as follows:
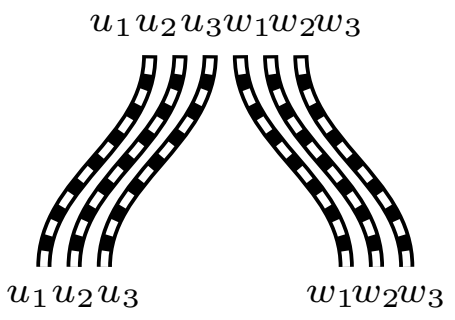
\begin{equation}\label{U_cab}
	U=\begin{array}{c}
		\begin{tikzpicture}[yscale=0.7]
			\draw[line width=1mm] (-1.2,-1) to[out=90,in=270] (-0.5,1) (-1,-1) to[out=90,in=270] (-0.3,1) (-0.8,-1) to[out=90,in=270] (-0.1,1) (0.8,-1) to[out=90,in=270] (0.1,1) (1,-1) to[out=90,in=270] (0.3,1) (1.2,-1) to[out=90,in=270] (0.5,1);
			\draw[white, dashed, line width=0.5mm] (-1.2,-1) to[out=90,in=270] (-0.5,1) (-1,-1) to[out=90,in=270] (-0.3,1) (-0.8,-1) to[out=90,in=270] (-0.1,1) (0.8,-1) to[out=90,in=270] (0.1,1) (1,-1) to[out=90,in=270] (0.3,1) (1.2,-1) to[out=90,in=270] (0.5,1);
			\node[below] at (-1.3,-1) {$\scriptstyle u_1$};
			\node[below] at (-1,-1) {$\scriptstyle u_2$};
			\node[below] at (-0.7,-1) {$\scriptstyle u_3$};
			\node[below] at (0.7,-1) {$\scriptstyle w_1$};
			\node[below] at (1,-1) {$\scriptstyle w_2$};
			\node[below] at (1.3,-1) {$\scriptstyle w_3$};
			\node[above] at (-0.75,1) {$\scriptstyle u_1$};
			\node[above] at (-0.45,1) {$\scriptstyle u_2$};
			\node[above] at (-0.15,1) {$\scriptstyle u_3$};
			\node[above] at (0.15,1) {$\scriptstyle w_1$};
			\node[above] at (0.45,1) {$\scriptstyle w_2$};
			\node[above] at (0.75,1) {$\scriptstyle w_3$};
		\end{tikzpicture}
	\end{array}\,.
\end{equation}
Here $u_i$ denote the spectral parameters of the first cluster and $w_i$ those of the second.

We work with the simplest case, in which all individual representations on the strands are the spin-$\sfrac{1}{2}$ representation of $\myY(\fs\fl_2)$.
The representations corresponding to the cables in \eqref{U_cab} are given by the pseudo-coproduct $\sD$ discussed in Sec.~\ref{sec:cry_chain}.
A similar setting was used in \cite{Yang:2024hsp}, where the non-evaluation representations were called ``multi-spin.''
In contrast to our setting, in \cite{Yang:2024hsp} the authors used evaluation spin-$\ell_i$ representations on the elementary strands; however, they did not exploit the dependence on the spectral parameters $u_i$ and $w_i$.

Thus, in our setting, a $k$-cell amplitude takes the following form.
A group of $k$ cells leaves the left cluster at the places with spectral parameters $u_{i_1},u_{i_2},\ldots,u_{i_k}$ and lands in the right cluster at the vacant positions with spectral parameters $w_{j_1},w_{j_2},\ldots,w_{j_k}$.
We assume that $i_1<i_2<\ldots<i_k$, and the same for $j_\beta$.
We define the corresponding amplitude as follows:
\begin{equation}
	\CB:=\CA\left[\scalebox{0.8}{$\ldots\otimes|{\color{color1}1}\rangle_{u_{i_1}}\otimes\ldots\otimes |{\color{color1}1}\rangle_{u_{i_2}}\otimes\ldots\otimes|{\color{color3}0}\rangle_{w_{j_1}}\otimes\ldots\otimes|{\color{color3}0}\rangle_{w_{j_2}}\otimes \ldots\to \ldots\otimes|{\color{color3}0}\rangle_{u_{i_1}}\otimes\ldots\otimes |{\color{color3}0}\rangle_{u_{i_2}}\otimes\ldots\otimes|{\color{color1}1}\rangle_{w_{j_1}}\otimes\ldots\otimes|{\color{color1}1}\rangle_{w_{j_2}}\otimes \ldots$}\right]\,.
\end{equation}
In this situation, we can still apply the mean-field logic and expect the amplitude to factorize.
Some factors are simple: they correspond to the phase factors generated by the cells that remain immobile during the process.
The remaining factor is a nontrivial form factor $\CF$ that accounts for the mutual interactions of the migrating cells but does not depend on the immobile ones.
This factorization takes the following form:
\begin{equation}\label{chain_factor}
	\CB=\prod\lm_{\alpha=1}^k\prod\lm_{\substack{s>i_{\alpha}\\
	s\neq i_{\gamma}}}\eta_{\lambda_{s}}\left(u_{i_\alpha}-u_s\right)\times\prod\lm_{\beta=1}^k\prod\lm_{\substack{p<j_{\beta}\\
	p\neq j_{\gamma}}}\eta_{\lambda_{p}}\left(w_{j_\beta}-w_p\right)^{-1}\times\CF^{(k)}(u_{i_1},u_{i_2},\ldots, u_{i_k};w_{j_1},w_{j_2},\ldots, w_{j_k})\,.
\end{equation}
Here $\lambda_p\in\{0,1\}$ labels the two elementary states, $|0\rangle$ and $|1\rangle$.
The corresponding charge functions read
\begin{equation}
	\eta_0(z)=\frac{z-2\hbar}{z},\quad \eta_1(z)=\frac{z+2\hbar}{z}\,.
\end{equation}

In our case, we obtain the following eigenvalue in \eqref{factor}:
\begin{equation}
	z=\sum\lm_{\beta=1}^k w_{j_\beta}-\sum\lm_{\alpha=1}^k u_{i_\alpha}\,.
\end{equation}
For $k\geq 2$, this combination does not match the one-loop equivariant factor of any chiral multiplet.
Thus, we conclude that factor $z$ cannot appear separately in each summand on the r.h.s. of \eqref{recursion}.
Equivalently, the corresponding factor $Z$ cannot appear separately in each summand on the r.h.s. of \eqref{main_amp}.
These factors appear on the r.h.s. of these relations only after the sum over paths has been performed.

In what follows, we focus on the case $k=2$, which is both simple and sufficiently representative.
Substituting the corresponding 1-cell amplitudes into the recursion relations \eqref{recursion} (resp., \eqref{main_amp}), we find that the combined amplitude on the right-hand side factorizes.
It contains the factor $z$ (resp., $Z$), and the remaining factors take the form of \eqref{chain_factor}, with the following form-factor:
\begin{equation}
	\CF^{(2)}(u_1,u_2;w_1,w_2)=-\frac{4 \hbar ^2 \left(-u_1 w_1-u_2 w_1-u_1 w_2-u_2 w_2+2 u_1 \hbar +2 u_2 \hbar +2 u_1 u_2-2 w_1 \hbar -2 w_2 \hbar +2 w_1 w_2+4 \hbar ^2\right)}{\left(u_1-w_1\right) \left(w_1-u_2\right) \left(u_1-w_2\right) \left(u_2-w_2\right)}\,.
\end{equation}
Looking at this expression, it is hard to reconstruct how it should be decomposed into a sum of separate amplitudes corresponding to different tunneling trajectories, what the space of those trajectories is, what the corresponding moduli spaces are, and so on.
To perform this decomposition in a way natural for this construction, we split the process corresponding to the factorization twist in this case into a collection of 1-cell processes, which we call ``\emph{cascades}.''
The idea is similar to that of the control strand performing an elementary transport, as discussed in Sec.~\ref{sec:subtle}.
Here we deform the trajectories of $x_{\ff}$ corresponding to this process by a homotopy, so that the gluing of two thick cables looks like a sequence of elementary processes.
For each cascade, we peel a single strand off the left cable and join it to the right cable.
We proceed with this strand migration one by one:
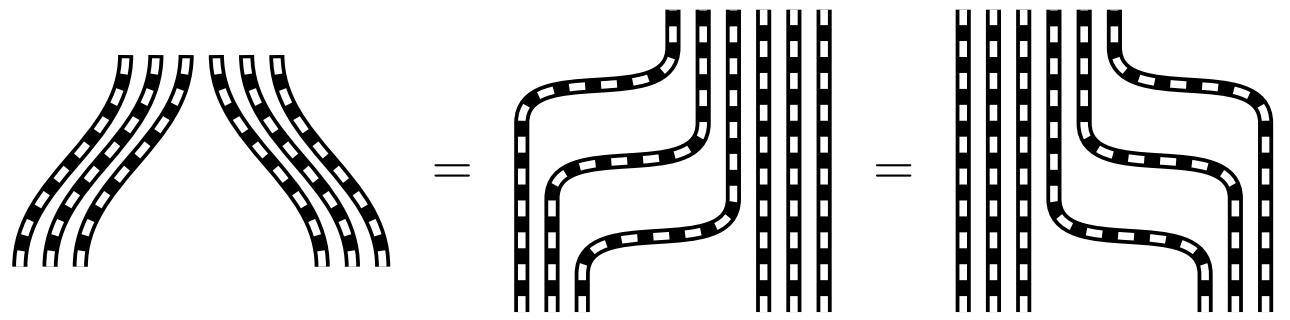
\begin{equation}\label{cascades}
	U=\begin{array}{c}
		\begin{tikzpicture}[yscale=0.7]
			\draw[line width=1mm] (-1.2,-1) to[out=90,in=270] (-0.5,1) (-1,-1) to[out=90,in=270] (-0.3,1) (-0.8,-1) to[out=90,in=270] (-0.1,1) (0.8,-1) to[out=90,in=270] (0.1,1) (1,-1) to[out=90,in=270] (0.3,1) (1.2,-1) to[out=90,in=270] (0.5,1);
			\draw[white, dashed, line width=0.5mm] (-1.2,-1) to[out=90,in=270] (-0.5,1) (-1,-1) to[out=90,in=270] (-0.3,1) (-0.8,-1) to[out=90,in=270] (-0.1,1) (0.8,-1) to[out=90,in=270] (0.1,1) (1,-1) to[out=90,in=270] (0.3,1) (1.2,-1) to[out=90,in=270] (0.5,1);
		\end{tikzpicture}
	\end{array}=\begin{array}{c}
	\begin{tikzpicture}
		\draw[line width=1mm] (-1,0) -- (-1,0.25) to[out=90,in=270] (0,0.75) -- (0,2) (-1.2,0) -- (-1.2,0.75) to[out=90,in=270] (-0.2,1.25) -- (-0.2,2) (-1.4,0) -- (-1.4,1.25) to[out=90,in=270] (-0.4,1.75) -- (-0.4,2) (0.2,0) -- (0.2,2) (0.4,0) -- (0.4,2) (0.6,0) -- (0.6,2);
		\draw[white, dashed, line width=0.5mm] (-1,0) -- (-1,0.25) to[out=90,in=270] (0,0.75) -- (0,2) (-1.2,0) -- (-1.2,0.75) to[out=90,in=270] (-0.2,1.25) -- (-0.2,2) (-1.4,0) -- (-1.4,1.25) to[out=90,in=270] (-0.4,1.75) -- (-0.4,2) (0.2,0) -- (0.2,2) (0.4,0) -- (0.4,2) (0.6,0) -- (0.6,2);
	\end{tikzpicture}
	\end{array}=\begin{array}{c}
	\begin{tikzpicture}[xscale=-1]
		\draw[line width=1mm] (-1,0) -- (-1,0.25) to[out=90,in=270] (0,0.75) -- (0,2) (-1.2,0) -- (-1.2,0.75) to[out=90,in=270] (-0.2,1.25) -- (-0.2,2) (-1.4,0) -- (-1.4,1.25) to[out=90,in=270] (-0.4,1.75) -- (-0.4,2) (0.2,0) -- (0.2,2) (0.4,0) -- (0.4,2) (0.6,0) -- (0.6,2);
		\draw[white, dashed, line width=0.5mm] (-1,0) -- (-1,0.25) to[out=90,in=270] (0,0.75) -- (0,2) (-1.2,0) -- (-1.2,0.75) to[out=90,in=270] (-0.2,1.25) -- (-0.2,2) (-1.4,0) -- (-1.4,1.25) to[out=90,in=270] (-0.4,1.75) -- (-0.4,2) (0.2,0) -- (0.2,2) (0.4,0) -- (0.4,2) (0.6,0) -- (0.6,2);
	\end{tikzpicture}
	\end{array}\,.
\end{equation}
From the algebraic point of view, this equivalence follows from \eqref{regluing}.

For the 2-cascade we obtain two more elementary 1-cell transports $\sU_1$ and $\sU_2$:
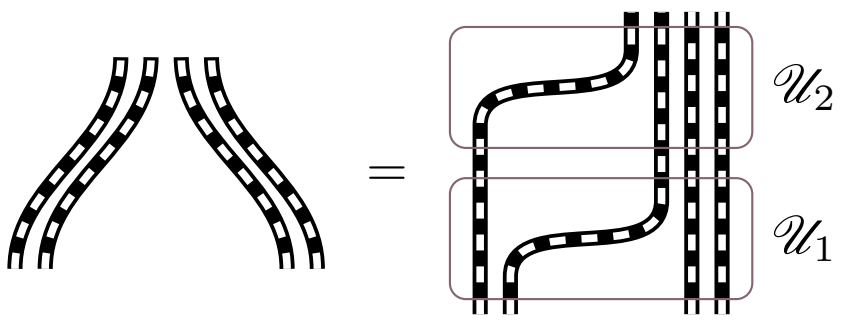
\begin{equation}
	\begin{array}{c}
		\begin{tikzpicture}[yscale=0.7]
			\draw[line width=1mm] (-1,-1) to[out=90,in=270] (-0.3,1) (-0.8,-1) to[out=90,in=270] (-0.1,1) (0.8,-1) to[out=90,in=270] (0.1,1) (1,-1) to[out=90,in=270] (0.3,1) ;
			\draw[white, dashed, line width=0.5mm] (-1,-1) to[out=90,in=270] (-0.3,1) (-0.8,-1) to[out=90,in=270] (-0.1,1) (0.8,-1) to[out=90,in=270] (0.1,1) (1,-1) to[out=90,in=270] (0.3,1);
		\end{tikzpicture}
	\end{array}=\begin{array}{c}
	\begin{tikzpicture}
		\draw[line width=1mm] (-1,0) --(-1,0.25) to[out=90,in=270] (0,0.75) -- (0,2) (-1.2,0) --(-1.2,1.25) to[out=90,in=270] (-0.2,1.75) -- (-0.2,2) (0.2,0) -- (0.2,2) (0.4,0) -- (0.4,2);
		\draw[white, dashed, line width=0.5mm] (-1,0) --(-1,0.25) to[out=90,in=270] (0,0.75) -- (0,2) (-1.2,0) --(-1.2,1.25) to[out=90,in=270] (-0.2,1.75) -- (-0.2,2) (0.2,0) -- (0.2,2) (0.4,0) -- (0.4,2);
		\draw[color2, rounded corners = 3] (-1.4,0.1) -- (-1.4,0.9) -- (0.6,0.9) -- (0.6,0.1) -- cycle;
		\node[right] at (0.6,0.5) {$\sU_1$};
		\draw[color2, rounded corners = 3] (-1.4,1.1) -- (-1.4,1.9) -- (0.6,1.9) -- (0.6,1.1) -- cycle;
		\node[right] at (0.6,1.5) {$\sU_2$};
	\end{tikzpicture}
	\end{array}\,.
\end{equation}
Since each process is 1-cell we compute transports easily from \eqref{1-cell}:
\begin{equation}
	\begin{aligned}
	&\left(\sU_1-\bbone\right)\left(|1\rangle_{u_1}\otimes |1\rangle_{u_2}\right)\otimes \left(|0\rangle_{w_1}\otimes |0\rangle_{w_2}\right)=\frac{-2\hbar}{u_2-w_1}\left(|1\rangle_{u_1}\otimes |0\rangle_{u_2}\right)\otimes \left(|1\rangle_{w_1}\otimes |0\rangle_{w_2}\right)+\\
	&\hspace{5cm}+\frac{-2\hbar}{u_2-w_2}\eta_0(w_2-w_1)\left(|1\rangle_{u_1}\otimes |0\rangle_{u_2}\right)\otimes \left(|0\rangle_{w_1}\otimes |1\rangle_{w_2}\right)\,,\\
	&\left(\sU_2-\bbone\right)\left(|1\rangle_{u_1}\otimes |0\rangle_{u_2}\right)\otimes \left(|0\rangle_{w_1}\otimes |1\rangle_{w_2}\right)=\frac{-2\hbar}{u_2-u_1}\left(|0\rangle_{u_1}\otimes |1\rangle_{u_2}\right)\otimes \left(|0\rangle_{w_1}\otimes |1\rangle_{w_2}\right)+\\
	&\hspace{5cm}+\frac{-2\hbar}{u_1-w_1}\eta_0(w_1-u_2)\left(|0\rangle_{u_1}\otimes |0\rangle_{u_2}\right)\otimes \left(|1\rangle_{w_1}\otimes |1\rangle_{w_2}\right)\,,\\
	&\left(\sU_2-\bbone\right)\left(|1\rangle_{u_1}\otimes |0\rangle_{u_2}\right)\otimes \left(|1\rangle_{w_1}\otimes |0\rangle_{w_2}\right)=\frac{-2\hbar}{u_2-u_1}\left(|0\rangle_{u_1}\otimes |1\rangle_{u_2}\right)\otimes \left(|1\rangle_{w_1}\otimes |0\rangle_{w_2}\right)+\\
	&\hspace{5cm}+\frac{-2\hbar}{u_1-w_2}\eta_0(w_2-u_2)\eta_1(w_2-w_1)\left(|0\rangle_{u_1}\otimes |0\rangle_{u_2}\right)\otimes \left(|1\rangle_{w_1}\otimes |1\rangle_{w_2}\right)\,.
	\end{aligned}
\end{equation}

Combining these expressions, we obtain the amplitude coinciding with the form-factor:
\begin{tcolorbox}
\begin{equation}
	\left({}_{u_1}\langle 0|\otimes {}_{u_2}\langle 0|\right)\otimes \left({}_{w_1}\langle 1|\otimes {}_{w_2}\langle 1|\right)\sU_2\sU_1\left(|1\rangle_{u_1}\otimes |1\rangle_{u_2}\right)\otimes \left(|0\rangle_{w_1}\otimes |0\rangle_{w_2}\right)=\CF^{(2)}(u_1,u_2;w_1,w_2)\,.
\end{equation}
\end{tcolorbox}
This approach allows the form-factor to be naturally decomposed into a sum of instanton paths:
\begin{equation}
	\hspace{-4mm}\raisebox{-7.5mm}{$\begin{array}{c}
	\begin{tikzpicture}
		\node{$\displaystyle \CF^{(2)}(u_1,u_2;w_1,w_2)=\frac{4\hbar^2\left(-w_1+w_2-2 \hbar \right) \left(-u_2+w_1-2 \hbar \right) }{\left(w_2-w_1\right) \left(u_1-w_1\right) \left(w_1-u_2\right) \left(u_2-w_2\right)}+\frac{4\hbar^2\left(-w_1+w_2+2 \hbar \right) \left(-u_2+w_2-2 \hbar \right) }{\left(w_2-w_1\right) \left(u_2-w_1\right) \left(u_1-w_2\right) \left(w_2-u_2\right)}\,.$};
		\foreach \x/\y/\h in {-4.8/1.4/-0.5} {
			\draw[ultra thin, fill] (\x,\h) to[out=270,in=180] (\x+0.3,\h-0.17) -- (0.5*\x+0.5*\y-0.3,\h-0.17) to[out=0,in=90] (0.5*\x+0.5*\y,\h-0.3) to[out=90,in=0] (0.5*\x+0.5*\y-0.3,\h-0.13) -- (\x+0.3,\h-0.13) to[out=180,in=270] (\x,\h);
			\draw[ultra thin, fill] (\y,\h) to[out=270,in=0] (\y-0.3,\h-0.17) -- (0.5*\x+0.5*\y+0.3,\h-0.17) to[out=180,in=90] (0.5*\x+0.5*\y,\h-0.3) to[out=90,in=180] (0.5*\x+0.5*\y+0.3,\h-0.13) -- (\y-0.3,\h-0.13) to[out=0,in=270] (\y,\h);
			\node[below] at (0.5*\x+0.5*\y,\h-0.3) {$\scriptstyle \CA_<\left[\scalebox{0.7}{$\begin{array}{c}
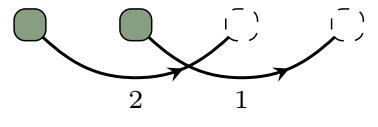
						\begin{tikzpicture}
							\draw[thick, postaction=decorate, decoration={markings, mark= at position 0.7 with {\arrow{stealth}}}] (1,0) to[out=315,in=180] (2,-0.5) to[out=0,in=225] (3,0);
							\draw[thick, postaction=decorate, decoration={markings, mark= at position 0.7 with {\arrow{stealth}}}] (0,0) to[out=315,in=180] (1,-0.5) to[out=0,in=225] (2,0);
							\node[below] at (2,-0.5) {$\scriptstyle 1$};
							\node[below] at (1,-0.5) {$\scriptstyle 2$};
							\begin{scope}
								\draw[fill=color8, rounded corners=3] (-0.15,-0.15) -- (-0.15,0.15) -- (0.15,0.15) -- (0.15,-0.15) -- cycle;
							\end{scope}
							\begin{scope}[shift={(1,0)}]
								\draw[fill=color8, rounded corners=3] (-0.15,-0.15) -- (-0.15,0.15) -- (0.15,0.15) -- (0.15,-0.15) -- cycle;
							\end{scope}
							\begin{scope}[shift={(2,0)}]
								\draw[dashed, fill=white, rounded corners=3] (-0.15,-0.15) -- (-0.15,0.15) -- (0.15,0.15) -- (0.15,-0.15) -- cycle;
							\end{scope}
							\begin{scope}[shift={(3,0)}]
								\draw[dashed, fill=white, rounded corners=3] (-0.15,-0.15) -- (-0.15,0.15) -- (0.15,0.15) -- (0.15,-0.15) -- cycle;
							\end{scope}
						\end{tikzpicture}
					\end{array}$}\right]$};
		}
		\foreach \x/\y/\h in {1.8/8/-0.5} {
			\draw[ultra thin, fill] (\x,\h) to[out=270,in=180] (\x+0.3,\h-0.17) -- (0.5*\x+0.5*\y-0.3,\h-0.17) to[out=0,in=90] (0.5*\x+0.5*\y,\h-0.3) to[out=90,in=0] (0.5*\x+0.5*\y-0.3,\h-0.13) -- (\x+0.3,\h-0.13) to[out=180,in=270] (\x,\h);
			\draw[ultra thin, fill] (\y,\h) to[out=270,in=0] (\y-0.3,\h-0.17) -- (0.5*\x+0.5*\y+0.3,\h-0.17) to[out=180,in=90] (0.5*\x+0.5*\y,\h-0.3) to[out=90,in=180] (0.5*\x+0.5*\y+0.3,\h-0.13) -- (\y-0.3,\h-0.13) to[out=0,in=270] (\y,\h);
			\node[below] at (0.5*\x+0.5*\y,\h-0.3) {$\scriptstyle \CA_<\left[\scalebox{0.7}{$\begin{array}{c}
						
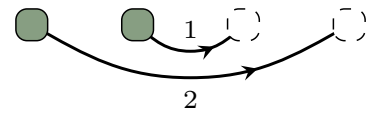
\begin{tikzpicture}
							\draw[thick, postaction=decorate, decoration={markings, mark= at position 0.7 with {\arrow{stealth}}}] (0,0) to[out=330,in=180] (1.5,-0.5) to[out=0,in=210] (3,0);
							\draw[thick, postaction=decorate, decoration={markings, mark= at position 0.7 with {\arrow{stealth}}}] (1,0) to[out=315,in=180] (1.5,-0.25) to[out=0,in=225] (2,0);
							\node[above] at (1.5,-0.25) {$\scriptstyle 1$};
							\node[below] at (1.5,-0.5) {$\scriptstyle 2$};
							\begin{scope}
								\draw[fill=color8, rounded corners=3] (-0.15,-0.15) -- (-0.15,0.15) -- (0.15,0.15) -- (0.15,-0.15) -- cycle;
							\end{scope}
							\begin{scope}[shift={(1,0)}]
								\draw[fill=color8, rounded corners=3] (-0.15,-0.15) -- (-0.15,0.15) -- (0.15,0.15) -- (0.15,-0.15) -- cycle;
							\end{scope}
							\begin{scope}[shift={(2,0)}]
								\draw[dashed, fill=white, rounded corners=3] (-0.15,-0.15) -- (-0.15,0.15) -- (0.15,0.15) -- (0.15,-0.15) -- cycle;
							\end{scope}
							\begin{scope}[shift={(3,0)}]
								\draw[dashed, fill=white, rounded corners=3] (-0.15,-0.15) -- (-0.15,0.15) -- (0.15,0.15) -- (0.15,-0.15) -- cycle;
							\end{scope}
						\end{tikzpicture}
					\end{array}$}\right]$};
		}
	\end{tikzpicture}
	\end{array}$}
\end{equation}
Here the jump labeled $1$ occurs before the jump labeled $2$.
This decomposition is not unique.
In fact, the equivalence of left $(<)$ and right $(>)$ cascades in \eqref{cascades} imposes the following relation on the amplitudes:
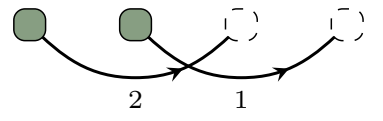
\begin{equation}\label{wc}
	\CA_{<}\left[\scalebox{0.7}{$\begin{array}{c}
			\begin{tikzpicture}
				\draw[thick, postaction=decorate, decoration={markings, mark= at position 0.7 with {\arrow{stealth}}}] (1,0) to[out=315,in=180] (2,-0.5) to[out=0,in=225] (3,0);
				\draw[thick, postaction=decorate, decoration={markings, mark= at position 0.7 with {\arrow{stealth}}}] (0,0) to[out=315,in=180] (1,-0.5) to[out=0,in=225] (2,0);
				\node[below] at (2,-0.5) {$\scriptstyle 1$};
				\node[below] at (1,-0.5) {$\scriptstyle 2$};
				\begin{scope}
					\draw[fill=color8, rounded corners=3] (-0.15,-0.15) -- (-0.15,0.15) -- (0.15,0.15) -- (0.15,-0.15) -- cycle;
				\end{scope}
				\begin{scope}[shift={(1,0)}]
					\draw[fill=color8, rounded corners=3] (-0.15,-0.15) -- (-0.15,0.15) -- (0.15,0.15) -- (0.15,-0.15) -- cycle;
				\end{scope}
				\begin{scope}[shift={(2,0)}]
					\draw[dashed, fill=white, rounded corners=3] (-0.15,-0.15) -- (-0.15,0.15) -- (0.15,0.15) -- (0.15,-0.15) -- cycle;
				\end{scope}
				\begin{scope}[shift={(3,0)}]
					\draw[dashed, fill=white, rounded corners=3] (-0.15,-0.15) -- (-0.15,0.15) -- (0.15,0.15) -- (0.15,-0.15) -- cycle;
				\end{scope}
			\end{tikzpicture}
		\end{array}$}\right]+\CA_{<}\left[\scalebox{0.7}{$\begin{array}{c}
		
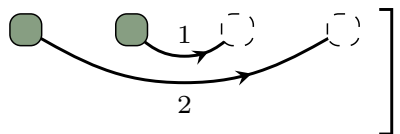
\begin{tikzpicture}
			\draw[thick, postaction=decorate, decoration={markings, mark= at position 0.7 with {\arrow{stealth}}}] (0,0) to[out=330,in=180] (1.5,-0.5) to[out=0,in=210] (3,0);
			\draw[thick, postaction=decorate, decoration={markings, mark= at position 0.7 with {\arrow{stealth}}}] (1,0) to[out=315,in=180] (1.5,-0.25) to[out=0,in=225] (2,0);
			\node[above] at (1.5,-0.25) {$\scriptstyle 1$};
			\node[below] at (1.5,-0.5) {$\scriptstyle 2$};
			\begin{scope}
				\draw[fill=color8, rounded corners=3] (-0.15,-0.15) -- (-0.15,0.15) -- (0.15,0.15) -- (0.15,-0.15) -- cycle;
			\end{scope}
			\begin{scope}[shift={(1,0)}]
				\draw[fill=color8, rounded corners=3] (-0.15,-0.15) -- (-0.15,0.15) -- (0.15,0.15) -- (0.15,-0.15) -- cycle;
			\end{scope}
			\begin{scope}[shift={(2,0)}]
				\draw[dashed, fill=white, rounded corners=3] (-0.15,-0.15) -- (-0.15,0.15) -- (0.15,0.15) -- (0.15,-0.15) -- cycle;
			\end{scope}
			\begin{scope}[shift={(3,0)}]
				\draw[dashed, fill=white, rounded corners=3] (-0.15,-0.15) -- (-0.15,0.15) -- (0.15,0.15) -- (0.15,-0.15) -- cycle;
			\end{scope}
		\end{tikzpicture}
		\end{array}$}\right]=\CA_{>}\left[\scalebox{0.7}{$\begin{array}{c}
		
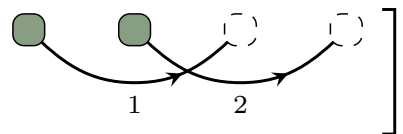
\begin{tikzpicture}
			\draw[thick, postaction=decorate, decoration={markings, mark= at position 0.7 with {\arrow{stealth}}}] (1,0) to[out=315,in=180] (2,-0.5) to[out=0,in=225] (3,0);
			\draw[thick, postaction=decorate, decoration={markings, mark= at position 0.7 with {\arrow{stealth}}}] (0,0) to[out=315,in=180] (1,-0.5) to[out=0,in=225] (2,0);
			\node[below] at (2,-0.5) {$\scriptstyle 2$};
			\node[below] at (1,-0.5) {$\scriptstyle 1$};
			\begin{scope}
				\draw[fill=color8, rounded corners=3] (-0.15,-0.15) -- (-0.15,0.15) -- (0.15,0.15) -- (0.15,-0.15) -- cycle;
			\end{scope}
			\begin{scope}[shift={(1,0)}]
				\draw[fill=color8, rounded corners=3] (-0.15,-0.15) -- (-0.15,0.15) -- (0.15,0.15) -- (0.15,-0.15) -- cycle;
			\end{scope}
			\begin{scope}[shift={(2,0)}]
				\draw[dashed, fill=white, rounded corners=3] (-0.15,-0.15) -- (-0.15,0.15) -- (0.15,0.15) -- (0.15,-0.15) -- cycle;
			\end{scope}
			\begin{scope}[shift={(3,0)}]
				\draw[dashed, fill=white, rounded corners=3] (-0.15,-0.15) -- (-0.15,0.15) -- (0.15,0.15) -- (0.15,-0.15) -- cycle;
			\end{scope}
		\end{tikzpicture}
		\end{array}$}\right]+\CA_{>}\left[\scalebox{0.7}{$\begin{array}{c}
		
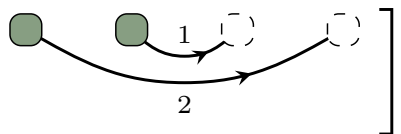
\begin{tikzpicture}
			\draw[thick, postaction=decorate, decoration={markings, mark= at position 0.7 with {\arrow{stealth}}}] (0,0) to[out=330,in=180] (1.5,-0.5) to[out=0,in=210] (3,0);
			\draw[thick, postaction=decorate, decoration={markings, mark= at position 0.7 with {\arrow{stealth}}}] (1,0) to[out=315,in=180] (1.5,-0.25) to[out=0,in=225] (2,0);
			\node[above] at (1.5,-0.25) {$\scriptstyle 1$};
			\node[below] at (1.5,-0.5) {$\scriptstyle 2$};
			\begin{scope}
				\draw[fill=color8, rounded corners=3] (-0.15,-0.15) -- (-0.15,0.15) -- (0.15,0.15) -- (0.15,-0.15) -- cycle;
			\end{scope}
			\begin{scope}[shift={(1,0)}]
				\draw[fill=color8, rounded corners=3] (-0.15,-0.15) -- (-0.15,0.15) -- (0.15,0.15) -- (0.15,-0.15) -- cycle;
			\end{scope}
			\begin{scope}[shift={(2,0)}]
				\draw[dashed, fill=white, rounded corners=3] (-0.15,-0.15) -- (-0.15,0.15) -- (0.15,0.15) -- (0.15,-0.15) -- cycle;
			\end{scope}
			\begin{scope}[shift={(3,0)}]
				\draw[dashed, fill=white, rounded corners=3] (-0.15,-0.15) -- (-0.15,0.15) -- (0.15,0.15) -- (0.15,-0.15) -- cycle;
			\end{scope}
		\end{tikzpicture}
		\end{array}$}\right]\,.
\end{equation} 

We believe that the cascade approach allows the form factors for generic $k$-cell amplitudes to be constructed naturally from simple 1-cell amplitudes.
This approach bypasses the issue of extracting the factors $z$ and $Z$ discussed in Sec.~\ref{sec:subtle}.


\subsection{\texorpdfstring{$\myY(\fs\fl_{2})$}{}, evaluation higher spin representations}\label{sec:sl(2)_hs}

Here we consider evaluation representations of $\myY(\fs\fl_2)$ with spin $\sfrac{\ell}{2}$.
The quiver variety for $\ell>1$ (see Fig.~\ref{fig:quiver}) is \emph{non-Nakajima}, even though the quiver is chiral.
For Nakajima quiver varieties with $\ell=1$, the corresponding supersymmetry is enhanced to $\mathcal{N}=8$, and the quiver variety admits a hyperK\"ahler structure.
The complex combination $\omega_{\mathbb{C}}=\omega_J+i\omega_K$ can be obtained from the complex moment map $d\mu_{\mathbb{C}}=\iota_\xi\omega_{\mathbb{C}}$, where $\mu_{\mathbb{C}}=\partial_{C}W$ is the derivative of the superpotential.
For $\ell>1$, this relation breaks down, and the resulting quiver variety is generically singular.
The standard stable-envelope technique fails in this case, and one must instead use critical stable envelopes \cite{Cao:2026vwf}.
This technique involves cohomology defined in terms of vanishing cycles \cite{yang2020two,Rapcak:2018nsl,Rapcak:2020ueh}.
This technique was applied to the construction of the spin-$\sfrac{\ell}{2}$ representations of $\myY(\fs\fl_2)$ in \cite{Bykov:2019cst}.
Here, however, we rely on the approach of \cite{Galakhov:2020vyb}, in which the ground-state wave functions $\Psi_{\Lambda}$ are renormalized on singular loci.

This example is similar in several respects to the case of the fundamental representation of the Yangian $\myY(\fs\fl_{n+1})$.
The states are now represented by vertical strips of cells, as shown in Fig.~\ref{fig:parti}(c); for convenience, we redraw them here in a more schematic way:
\begin{equation}
	|m\rangle=\left[\begin{array}{c}
		\begin{tikzpicture}
			\draw[thick, postaction=decorate, decoration={markings, mark= at position 0.7 with {\arrow{stealth}}}] (0,0) -- (1,0) node[pos=0.5, below] {$\scriptstyle C$};
			\draw[thick, postaction=decorate, decoration={markings, mark= at position 0.7 with {\arrow{stealth}}}] (1,0) -- (2,0) node[pos=0.5, below] {$\scriptstyle C$};
			\draw[thick, postaction=decorate, decoration={markings, mark= at position 0.7 with {\arrow{stealth}}}] (3,0) -- (4,0) node[pos=0.5, below] {$\scriptstyle C$};
			\draw[thick, dashed] (2,0) -- (3,0);
			\foreach \i in {0, 1, 2, 3, 4}
			{
				\begin{scope}[shift={(\i,0)}]
					\draw[fill=color8, rounded corners=3] (-0.15,-0.15) -- (-0.15,0.15) -- (0.15,0.15) -- (0.15,-0.15) -- cycle;
				\end{scope}
			}
			\foreach \x/\y/\h in {-0.15/4.15/0.2} {
				\draw[ultra thin, fill] (\x,\h) to[out=90,in=180] (\x+0.3,\h+0.17) -- (0.5*\x+0.5*\y-0.3,\h+0.17) to[out=0,in=270] (0.5*\x+0.5*\y,\h+0.3) to[out=270,in=0] (0.5*\x+0.5*\y-0.3,\h+0.13) -- (\x+0.3,\h+0.13) to[out=180,in=90] (\x,\h);
				\draw[ultra thin, fill] (\y,\h) to[out=90,in=0] (\y-0.3,\h+0.17) -- (0.5*\x+0.5*\y+0.3,\h+0.17) to[out=180,in=270] (0.5*\x+0.5*\y,\h+0.3) to[out=270,in=180] (0.5*\x+0.5*\y+0.3,\h+0.13) -- (\y-0.3,\h+0.13) to[out=0,in=90] (\y,\h);
				\node[above] at (0.5*\x+0.5*\y,\h+0.3) {\scriptsize $m$ cells};
			}
		\end{tikzpicture}
	\end{array}\right],\quad I=(1-\delta_{m,0}),\quad m=0,1,\ldots,\ell\,.
\end{equation}

Another point of similarity is that there are no multiple possibilities and no sum over tunneling trajectories: the trajectory is unique.
This is reflected in the following observation.
The catalytic amplitude always involves a single cell jump and is given by \eqref{1-cell}.
Consequently, the recursion relations \eqref{main_amp}, or, equivalently \eqref{recursion}, can be solved explicitly, and yield the following expression for the Berry transport, i.e., the factorization twist:
\begin{equation}
	\raisebox{-5mm}{$\begin{array}{c}
		\begin{tikzpicture}
			\node{$\displaystyle U|a\rangle_{u_1}\otimes|b\rangle_{u_2}=\sum\lm_{k=0}^{\infty}\frac{(2\hbar)^k\theta_{r+1,b+k}\prod\lm_{i=1}^{k}(a-i+1)(\ell-a+i)}{k!\prod\lm_{j=k}^{2k-1}(u_{12}+2\hbar\left(a-b\right)+2\hbar j)}|a-k\rangle_{u_1}\otimes|b+k\rangle_{u_2}\,.$};
			\foreach \x/\y/\h in {-2.9/2.8/-0.9} {
				\draw[ultra thin, fill] (\x,\h) to[out=270,in=180] (\x+0.3,\h-0.17) -- (0.5*\x+0.5*\y-0.3,\h-0.17) to[out=0,in=90] (0.5*\x+0.5*\y,\h-0.3) to[out=90,in=0] (0.5*\x+0.5*\y-0.3,\h-0.13) -- (\x+0.3,\h-0.13) to[out=180,in=270] (\x,\h);
				\draw[ultra thin, fill] (\y,\h) to[out=270,in=0] (\y-0.3,\h-0.17) -- (0.5*\x+0.5*\y+0.3,\h-0.17) to[out=180,in=90] (0.5*\x+0.5*\y,\h-0.3) to[out=90,in=180] (0.5*\x+0.5*\y+0.3,\h-0.13) -- (\y-0.3,\h-0.13) to[out=0,in=270] (\y,\h);
				\node[below] at (0.5*\x+0.5*\y,\h-0.3) {$\scriptstyle \CA^{\rm (fusion)}_{|a\rangle\otimes|b\rangle \to |a-k\rangle\otimes|b+k\rangle}$};
			}
		\end{tikzpicture}
	\end{array}$}
\end{equation}
Compare this expression with \cite[Prop.~3.15]{Pfeiffer2000Yangian} and \cite[Prop.~5]{Bykov:2019cst}.
We can use it to construct the actual $R$-matrices for higher spins.
For example, choosing the following basis in the tensor product of two spin-one representations,
\begin{equation}
	\begin{aligned}
		&|0\rangle_{u_1}\otimes|0\rangle_{u_2},\;|0\rangle_{u_1}\otimes|1\rangle_{u_2},\;|1\rangle_{u_1}\otimes|0\rangle_{u_2},\;|0\rangle_{u_1}\otimes|2\rangle_{u_2},\;|1\rangle_{u_1}\otimes|1\rangle_{u_2}\,,\\
		&|2\rangle_{u_1}\otimes|0\rangle_{u_2},\;|1\rangle_{u_1}\otimes|2\rangle_{u_2},\;|2\rangle_{u_1}\otimes|1\rangle_{u_2},\;|2\rangle_{u_1}\otimes|2\rangle_{u_2}\,.
	\end{aligned}
\end{equation}
the $R$-matrix reads
\begin{equation}
	R^{(j=1)}(u_{12})=
	\begin{pmatrix}
		\frac{u_{12}+4 \hbar }{4 \hbar } & 0 & 0 & 0 & 0 & 0 & 0 & 0 & 0 \\
		0 & \frac{u_{12}}{4 \hbar } & 1 & 0 & 0 & 0 & 0 & 0 & 0 \\
		0 & 1 & \frac{u_{12}}{4 \hbar } & 0 & 0 & 0 & 0 & 0 & 0 \\
		0 & 0 & 0 & -\frac{u_{12} \left(2 \hbar -u_{12}\right)}{4 \hbar  \left(u_{12}+2 \hbar \right)} & \frac{u_{12}}{u_{12}+2 \hbar } & \frac{2 \hbar }{u_{12}+2 \hbar } & 0 & 0 & 0 \\
		0 & 0 & 0 & \frac{u_{12}}{u_{12}+2 \hbar } & \frac{2 u_{12} \hbar +u_{12}^2+8 \hbar ^2}{4 \hbar  \left(u_{12}+2 \hbar \right)} & \frac{u_{12}}{u_{12}+2 \hbar } & 0 & 0 & 0 \\
		0 & 0 & 0 & \frac{2 \hbar }{u_{12}+2 \hbar } & \frac{u_{12}}{u_{12}+2 \hbar } & -\frac{u_{12} \left(2 \hbar -u_{12}\right)}{4 \hbar  \left(u_{12}+2 \hbar \right)} & 0 & 0 & 0 \\
		0 & 0 & 0 & 0 & 0 & 0 & \frac{u_{12}}{4 \hbar } & 1 & 0 \\
		0 & 0 & 0 & 0 & 0 & 0 & 1 & \frac{u_{12}}{4 \hbar } & 0 \\
		0 & 0 & 0 & 0 & 0 & 0 & 0 & 0 & \frac{u_{12}+4 \hbar }{4 \hbar } \\
	\end{pmatrix}\,.
\end{equation}
and satisfies the YBE \eqref{YBE}.

This uniqueness allows us to describe the trajectory explicitly and to compute the amplitude directly as a one-loop determinant.
As discussed in Sec.~\ref{sec:stab}, the union of all instanton trajectories is an attracting set, and stable envelopes are supported on such sets.
Attracting sets in this setting were constructed, in different terms, in \cite{Bykov:2019cst}, as an extension of Pl\"ucker coordinates to the singular quiver varieties considered here.
Instead, we follow the methods of Sec.~\ref{sec:Berry}.

Both crystal-chain states $|a\rangle\otimes|b\rangle$ and $|a-k\rangle\otimes|b+k\rangle$ can be embedded into the quiver representation of dimension $a+b$.
We use the following block embeddings, in which the crystals $|a\rangle$ and $|a-k\rangle$ in the first tensor factor correspond to blocks with \emph{dashed} boundary, and those in the second factor to blocks with \emph{solid} boundary:
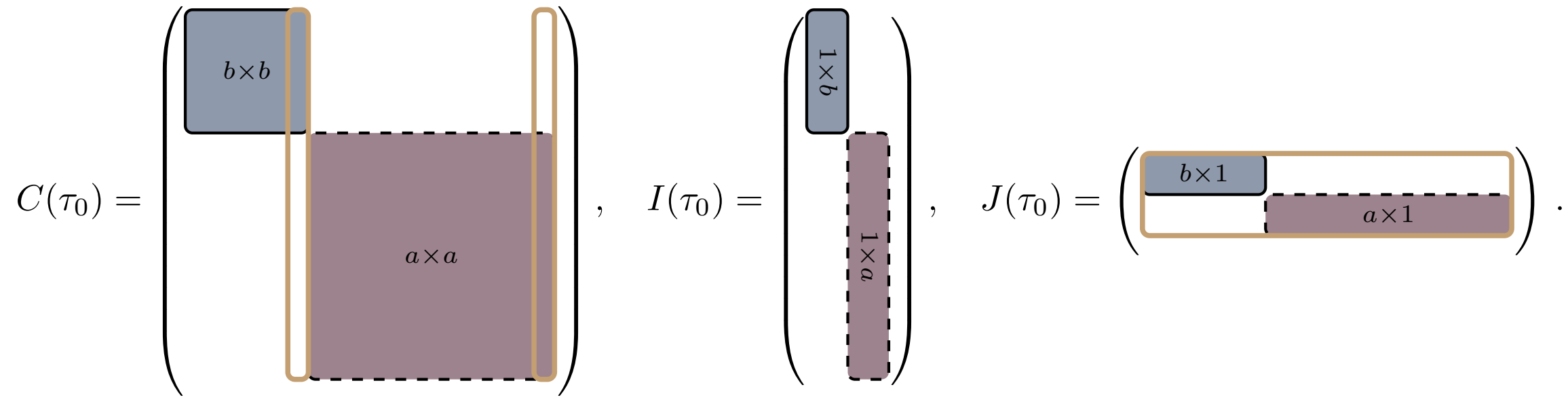
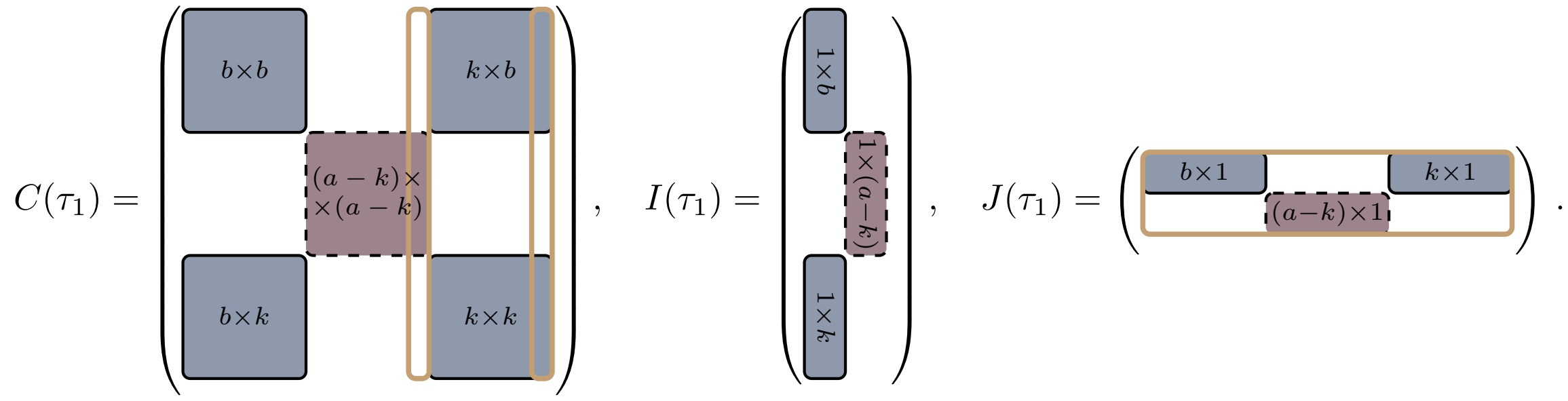
\begin{subequations}
\begin{equation}\label{matrix_a}
	C(\tau_0)=\left(\!\!\!\begin{array}{c}
		\begin{tikzpicture}[scale=0.4]
			\draw[thick, fill=white!20!color4, rounded corners=2] (0,0) -- (3,0) -- (3,-3) -- (0,-3) -- cycle;
			\draw[thick, dashed, fill=white!20!color2, rounded corners=2] (3,-3) -- (9,-3) -- (9,-9) -- (3,-9) -- cycle;
			\draw[color5, line width=0.5mm, rounded corners=2] (2.5,0) -- (3,0) -- (3,-9) -- (2.5,-9) -- cycle;
			\draw[color5, line width=0.5mm, rounded corners=2] (8.5,0) -- (9,0) -- (9,-9) -- (8.5,-9) -- cycle;  
			\node at (1.5,-1.5) {$\scriptstyle b\times b$};
			\node at (6,-6) {$\scriptstyle a\times a$};
		\end{tikzpicture}
	\end{array}\!\!\!\right),\quad I(\tau_0)=\left(\!\!\!\begin{array}{c}
		\begin{tikzpicture}[scale=0.4]
			\draw[thick, fill=white!20!color4, rounded corners=2] (0,0) -- (1,0) -- (1,-3) -- (0,-3) -- cycle;
			\draw[thick, dashed, fill=white!20!color2, rounded corners=2] (1,-3) -- (2,-3) -- (2,-9) -- (1,-9) -- cycle; 
			\node[rotate=-90] at (0.5,-1.5) {$\scriptstyle 1\times b$};
			\node[rotate=-90] at (1.5,-6) {$\scriptstyle 1\times a$};
		\end{tikzpicture}
	\end{array}\!\!\!\right),\quad J(\tau_0)=\left(\!\!\!\begin{array}{c}
		\begin{tikzpicture}[scale=0.4]
			\draw[thick, fill=white!20!color4, rounded corners=2] (0,0) -- (3,0) -- (3,-1) -- (0,-1) -- cycle;
			\draw[thick, dashed, fill=white!20!color2, rounded corners=2] (3,-1) -- (9,-1) -- (9,-2) -- (3,-2) -- cycle;
			\draw[color5, line width=0.5mm, rounded corners=2] (0,0) -- (9,0) -- (9,-2) -- (0,-2) -- cycle;
			\node at (1.5,-0.5) {$\scriptstyle b\times 1$};
			\node at (6,-1.5) {$\scriptstyle a\times 1$};
		\end{tikzpicture}
	\end{array}\!\!\!\right)\,.
\end{equation}
\begin{equation}\label{matrix_b}
	C(\tau_1)=\left(\!\!\!\begin{array}{c}
		\begin{tikzpicture}[scale=0.4]
			\draw[thick, fill=white!20!color4, rounded corners=2] (0,0) -- (3,0) -- (3,-3) -- (0,-3) -- cycle (6,0) -- (9,0) -- (9,-3) -- (6,-3) -- cycle (0,-6) -- (3,-6) -- (3,-9) -- (0,-9) -- cycle (6,-6) -- (9,-6) -- (9,-9) -- (6,-9) -- cycle;
			\draw[thick, dashed, fill=white!20!color2, rounded corners=2] (3,-3) -- (6,-3) -- (6,-6) -- (3,-6) -- cycle;
			\draw[color5, line width=0.5mm, rounded corners=2] (5.5,0) -- (6,0) -- (6,-9) -- (5.5,-9) -- cycle;
			\draw[color5, line width=0.5mm, rounded corners=2] (8.5,0) -- (9,0) -- (9,-9) -- (8.5,-9) -- cycle;  
			\node at (1.5,-1.5) {$\scriptstyle b\times b$};
			\node at (7.5,-1.5) {$\scriptstyle k\times b$};
			\node at (1.5,-7.5) {$\scriptstyle b\times k$};
			\node at (7.5,-7.5) {$\scriptstyle k\times k$};
			\node at (4.5,-4.5) {\scriptsize  $\begin{array}{c}
					(a-k)\times\\
					\times (a-k)
				\end{array}$};
		\end{tikzpicture}
	\end{array}\!\!\!\right),\quad I(\tau_1)=\left(\!\!\!\begin{array}{c}
		\begin{tikzpicture}[scale=0.4]
			\draw[thick, fill=white!20!color4, rounded corners=2] (0,0) -- (1,0) -- (1,-3) -- (0,-3) -- cycle (0,-6) -- (1,-6) -- (1,-9) -- (0,-9) -- cycle;
			\draw[thick, dashed, fill=white!20!color2, rounded corners=2] (1,-3) -- (2,-3) -- (2,-6) -- (1,-6) -- cycle; 
			\node[rotate=-90] at (0.5,-1.5) {$\scriptstyle 1\times b$};
			\node[rotate=-90] at (1.5,-4.5) {$\scriptstyle 1\times (a-k)$};
			\node[rotate=-90] at (0.5,-7.5) {$\scriptstyle 1\times k$};
		\end{tikzpicture}
	\end{array}\!\!\!\right),\quad J(\tau_1)=\left(\!\!\!\begin{array}{c}
		\begin{tikzpicture}[scale=0.4]
			\draw[thick, fill=white!20!color4, rounded corners=2] (0,0) -- (3,0) -- (3,-1) -- (0,-1) -- cycle (6,0) -- (9,0) -- (9,-1) -- (6,-1) -- cycle;
			\draw[thick, dashed, fill=white!20!color2, rounded corners=2] (3,-1) -- (6,-1) -- (6,-2) -- (3,-2) -- cycle;
			\draw[color5, line width=0.5mm, rounded corners=2] (0,0) -- (9,0) -- (9,-2) -- (0,-2) -- cycle;
			\node at (1.5,-0.5) {$\scriptstyle b\times 1$};
			\node at (4.5,-1.5) {$\scriptstyle (a-k)\times 1$};
			\node at (7.5,-0.5) {$\scriptstyle k\times 1$};
		\end{tikzpicture}
	\end{array}\!\!\!\right)\,.
\end{equation}
\end{subequations}
We denote the initial time (the infinite past) by $\tau_0$ and the final time (the infinite future) by $\tau_1$.
Only the fields $I$ and $C$ acquire nonzero vacuum expectation values.
For a single crystal, there is the following explicit expression,
\begin{equation}
	I=E_{1,1},\quad C=\sum\lm_{a}E_{a+1,a}\,,
\end{equation}
where $E_{a,b}$ are the matrix units, and we embed these expressions accordingly.

We now describe the tangent degrees of freedom, with the linearized gauge-group action
\begin{equation}\label{gauge}
	\fg:\quad H_{a\to b}\;\mapsto\; H_{a\to b}+g_b\times\langle H_{a\to b}\rangle-\langle H_{a\to b}\rangle\times g_a
\end{equation}
already modded out \cite{Galakhov:2024bzs}.
They split into light and heavy: light degrees of freedom are those with vanishing charge under the gauge scalar $X$, and heavy ones are those with nonzero charge.
As an explicit example, consider $a=6$, $b=4$, $k=3$.
\begin{subequations}
\begin{equation}\label{mat_1}
	C(\tau_0)=\hspace{-4mm}\begin{array}{c}
		\begin{tikzpicture}
			\node{$\begin{pmatrix}
					0 & 0 & 0 & c_{4}^{(2)} & 0 & 0 & 0 & 0 & 0 & \upsilon _{1,10} \\
					1 & 0 & 0 & c_{2,4}^{(2)} & 0 & 0 & 0 & 0 & 0 & \upsilon _{2,10} \\
					0 & 1 & 0 & c_{3,4}^{(2)} & 0 & 0 & 0 & 0 & 0 & \upsilon _{3,10} \\
					0 & 0 & 1 & c_{4,4}^{(2)} & 0 & 0 & 0 & 0 & 0 & \upsilon _{4,10} \\
					0 & 0 & 0 & \upsilon _{5,4} & 0 & 0 & 0 & 0 & 0 & c_{1,6}^{(1)} \\
					0 & 0 & 0 & \upsilon _{6,4} & 1 & 0 & 0 & 0 & 0 & c_{2,6}^{(1)} \\
					0 & 0 & 0 & \upsilon _{7,4} & 0 & 1 & 0 & 0 & 0 & c_{3,6}^{(1)} \\
					0 & 0 & 0 & \upsilon _{8,4} & 0 & 0 & 1 & 0 & 0 & c_{4,6}^{(1)} \\
					0 & 0 & 0 & \upsilon _{9,4} & 0 & 0 & 0 & 1 & 0 & c_{5,6}^{(1)} \\
					0 & 0 & 0 & \upsilon _{10,4} & 0 & 0 & 0 & 0 & 1 & c_{6,6}^{(1)} \\
				\end{pmatrix}$};
			\draw[rounded corners=3] (-3.1, 2.5) -- (-0.6,2.5) -- (-0.6,0.5) -- (-3.1,0.5) -- cycle;
			\draw[rounded corners=3, dashed] (-0.6,0.5) -- (3,0.5) -- (3,-2.5) -- (-0.6,-2.5) -- cycle;
			\draw[ultra thick, color8] (-1.05,-1.3) circle (0.3) (0.8,-1.3) circle (0.3);
		\end{tikzpicture}
	\end{array}\hspace{-2mm},\quad I(\tau_0)=\hspace{-4mm}\begin{array}{c}
	\begin{tikzpicture}
		\node{$\begin{pmatrix}
				1 & 0\\ 
				0 & 0\\
				0 & 0\\
				0 & 0\\
				0 & 1\\
				0 & 0\\
				0 & 0\\
				0 & 0\\
				0 & 0\\
				0 & 0\\
			\end{pmatrix}$};
		\draw[rounded corners=3]  (-0.4,2.05) -- (0,2.05) -- (0,0.45) -- (-0.4,0.45) -- cycle;
		\draw[rounded corners=3, dashed] (0,0.45) -- (0.4,0.45) -- (0.4,-2.05) -- (0,-2.05) -- cycle;
	\end{tikzpicture}
	\end{array}\hspace{-2mm},\quad J(\tau_0)=\hspace{-4mm}\begin{array}{c}
	\begin{tikzpicture}
		\node{$\begin{pmatrix}
				s_1^{(2)} & \sigma_{2,1}\\ 
				s_2^{(2)} & \sigma_{2,2}\\
				s_3^{(2)} & \sigma_{2,3}\\
				s_4^{(2)} & \sigma_{2,4}\\
				\sigma_{1,5} & s_1^{(1)}\\
				\sigma_{1,6} & s_2^{(1)}\\
				\sigma_{1,7} & s_3^{(1)}\\
				\sigma_{1,8} & s_4^{(1)}\\
				\sigma_{1,9} & s_5^{(1)}\\
				\sigma_{1,10} & s_6^{(1)}\\
			\end{pmatrix}^T$};
		\draw[rounded corners=3] (-0.95,2.35) -- (-0.1,2.35) -- (-0.1,0.4) -- (-0.95,0.4) -- cycle;
		\draw[rounded corners=3, dashed] (-0.1,0.4) -- (0.75,0.4) -- (0.75,-2.4) -- (-0.1,-2.4) -- cycle;
	\end{tikzpicture}
	\end{array}\hspace{-2mm}\,.
\end{equation}
\begin{equation}\label{mat_2}
	C(\tau_1)=\hspace{-4mm}\begin{array}{c}
		\begin{tikzpicture}
			\node{$\begin{pmatrix}
					0 & 0 & 0 & 0 & 0 & 0 & \tilde\upsilon _{1,7} & 0 & 0 & \tilde c_{1,7}^{(2)} \\
					1 & 0 & 0 & 0 & 0 & 0 & \tilde\upsilon _{2,7} & 0 & 0 & \tilde c_{2,7}^{(2)} \\
					0 & 1 & 0 & 0 & 0 & 0 & \tilde\upsilon _{3,7} & 0 & 0 & \tilde c_{3,7}^{(2)} \\
					0 & 0 & 1 & 0 & 0 & 0 & \tilde\upsilon _{4,7} & 0 & 0 & \tilde c_{4,7}^{(2)} \\
					0 & 0 & 0 & 0 & 0 & 0 & \tilde c_{1,3}^{(1)} & 0 & 0 & \tilde\upsilon _{5,10} \\
					0 & 0 & 0 & 0 & 1 & 0 & \tilde c_{2,3}^{(1)} & 0 & 0 & \tilde\upsilon _{6,10} \\
					0 & 0 & 0 & 0 & 0 & 1 & \tilde c_{3,3}^{(1)} & 0 & 0 & \tilde\upsilon _{7,10} \\
					0 & 0 & 0 & 1 & 0 & 0 & \tilde\upsilon _{8,7} & 0 & 0 & \tilde c_{5,7}^{(2)} \\
					0 & 0 & 0 & 0 & 0 & 0 & \tilde\upsilon _{9,7} & 1 & 0 & \tilde c_{6,7}^{(2)} \\
					0 & 0 & 0 & 0 & 0 & 0 & \tilde\upsilon _{10,7} & 0 & 1 & \tilde c_{7,7}^{(2)} \\
				\end{pmatrix}$};
			\draw[rounded corners=3] (-3.05,2.45) -- (-1.05,2.45) -- (-1.05,0.5) -- (-3.05,0.5) -- cycle (-3.05,-1.05) -- (-1.05,-1.05) -- (-1.05,-2.45) -- (-3.05,-2.45) -- cycle (1,2.45) -- (3,2.45) -- (3,0.5) -- (1,0.5) -- cycle (1,-1.05) -- (3,-1.05) -- (3,-2.45) -- (1,-2.45) -- cycle;
			\draw[dashed, rounded corners=3] (-1.05,0.5) -- (1,0.5) -- (1,-1.05) -- (-1.05,-1.05) -- cycle;
			\draw[ultra thick, color8] (-1.3,-1.3) circle (0.3) (0.5,-1.3) circle (0.3);
		\end{tikzpicture}
	\end{array}\hspace{-2mm},\quad I(\tau_1)=\hspace{-4mm}\begin{array}{c}
	\begin{tikzpicture}
		\node{$\begin{pmatrix}
				1 & 0\\ 
				0 & 0\\
				0 & 0\\
				0 & 0\\
				0 & 1\\
				0 & 0\\
				0 & 0\\
				0 & 0\\
				0 & 0\\
				0 & 0\\
			\end{pmatrix}$};
		\draw[rounded corners=3]  (-0.4,2.05) -- (0,2.05) -- (0,0.45) -- (-0.4,0.45) -- cycle;
		\draw[rounded corners=3, dashed] (0,0.45) -- (0.4,0.45) -- (0.4,-0.8) -- (0,-0.8) -- cycle;
		\draw[rounded corners=3]  (-0.4,-0.8) -- (0,-0.8) -- (0,-2.05) -- (-0.4,-2.05) -- cycle;
	\end{tikzpicture}
	\end{array}\hspace{-2mm},\quad J(\tau_1)=\hspace{-4mm}\begin{array}{c}
	\begin{tikzpicture}
		\node{$\begin{pmatrix}
				s_1^{(2)} & \sigma_{2,1}\\ 
				s_2^{(2)} & \sigma_{2,2}\\
				s_3^{(2)} & \sigma_{2,3}\\
				s_4^{(2)} & \sigma_{2,4}\\
				\sigma_{1,5} & s_1^{(1)}\\
				\sigma_{1,6} & s_2^{(1)}\\
				\sigma_{1,7} & s_3^{(1)}\\
				s_5^{(2)} & \sigma_{2,8}\\
				s_6^{(2)} & \sigma_{2,9}\\
				s_7^{(2)} & \sigma_{2,10}\\
			\end{pmatrix}^T$};
		\draw[rounded corners=3] (-0.95,2.35) -- (-0.1,2.35) -- (-0.1,0.4) -- (-0.95,0.4) -- cycle;
		\draw[rounded corners=3, dashed] (-0.1,0.4) -- (0.75,0.4) -- (0.75,-1) -- (-0.1,-1) -- cycle;
		\draw[rounded corners=3] (-0.95,-1) -- (-0.1,-1) -- (-0.1,-2.4) -- (-0.95,-2.4) -- cycle;
	\end{tikzpicture}
	\end{array}\hspace{-2mm}\,.
\end{equation}
\end{subequations}
We denote \emph{light} degrees of freedom by Latin letters such as $c_{i,j}^{(q)}$ or $s_i^{(q)}$.
Here $q$ denotes the position of the crystal in the tensor product to which this embedded degree of freedom belongs.
The indices $i$ and $j$ are the row and column indices of this element in the field matrix before embedding.
\emph{Heavy} degrees of freedom are denoted by Greek letters such as $\upsilon_{i,j}$ or $\sigma_{i,j}$.
Here $i$ and $j$ are simply the row and column indices of the matrix.
To distinguish tangent degrees of freedom at $\tau_0$ from those at $\tau_1$, we place tildes on the latter.
The tangent degrees of freedom appear in the columns with an orange border in \eqref{matrix_a} and \eqref{matrix_b}.

In this case, the instanton trajectory is described by the following $X(\tau)$:
\begin{equation}
	\raisebox{-4.5mm}{$\begin{array}{c}
		\begin{tikzpicture}
			\node{$\displaystyle X(\tau)=(x_2,\ldots,x_2,x_1,\ldots,x_1,\xi(\tau),\ldots, \xi(\tau)),\quad \lim\lm_{\tau_0\to -\infty}\xi(\tau_0)=x_1,\quad \lim\lm_{\tau_1\to +\infty}\xi(\tau_1)=x_2\,.$};
			\foreach \x/\y/\h in {-5.35/-3.85/-0.15} {
				\draw[ultra thin, fill] (\x,\h) to[out=270,in=180] (\x+0.3,\h-0.17) -- (0.5*\x+0.5*\y-0.3,\h-0.17) to[out=0,in=90] (0.5*\x+0.5*\y,\h-0.3) to[out=90,in=0] (0.5*\x+0.5*\y-0.3,\h-0.13) -- (\x+0.3,\h-0.13) to[out=180,in=270] (\x,\h);
				\draw[ultra thin, fill] (\y,\h) to[out=270,in=0] (\y-0.3,\h-0.17) -- (0.5*\x+0.5*\y+0.3,\h-0.17) to[out=180,in=90] (0.5*\x+0.5*\y,\h-0.3) to[out=90,in=180] (0.5*\x+0.5*\y+0.3,\h-0.13) -- (\y-0.3,\h-0.13) to[out=0,in=270] (\y,\h);
				\node[below] at (0.5*\x+0.5*\y,\h-0.3) {$\scriptstyle b\;{\rm times}$};
			}
			\foreach \x/\y/\h in {-3.7/-2.2/-0.15} {
				\draw[ultra thin, fill] (\x,\h) to[out=270,in=180] (\x+0.3,\h-0.17) -- (0.5*\x+0.5*\y-0.3,\h-0.17) to[out=0,in=90] (0.5*\x+0.5*\y,\h-0.3) to[out=90,in=0] (0.5*\x+0.5*\y-0.3,\h-0.13) -- (\x+0.3,\h-0.13) to[out=180,in=270] (\x,\h);
				\draw[ultra thin, fill] (\y,\h) to[out=270,in=0] (\y-0.3,\h-0.17) -- (0.5*\x+0.5*\y+0.3,\h-0.17) to[out=180,in=90] (0.5*\x+0.5*\y,\h-0.3) to[out=90,in=180] (0.5*\x+0.5*\y+0.3,\h-0.13) -- (\y-0.3,\h-0.13) to[out=0,in=270] (\y,\h);
				\node[below] at (0.5*\x+0.5*\y,\h-0.3) {$\scriptstyle a-k\;{\rm times}$};
			}
			\foreach \x/\y/\h in {-2.05/0.05/-0.15} {
				\draw[ultra thin, fill] (\x,\h) to[out=270,in=180] (\x+0.3,\h-0.17) -- (0.5*\x+0.5*\y-0.3,\h-0.17) to[out=0,in=90] (0.5*\x+0.5*\y,\h-0.3) to[out=90,in=0] (0.5*\x+0.5*\y-0.3,\h-0.13) -- (\x+0.3,\h-0.13) to[out=180,in=270] (\x,\h);
				\draw[ultra thin, fill] (\y,\h) to[out=270,in=0] (\y-0.3,\h-0.17) -- (0.5*\x+0.5*\y+0.3,\h-0.17) to[out=180,in=90] (0.5*\x+0.5*\y,\h-0.3) to[out=90,in=180] (0.5*\x+0.5*\y+0.3,\h-0.13) -- (\y-0.3,\h-0.13) to[out=0,in=270] (\y,\h);
				\node[below] at (0.5*\x+0.5*\y,\h-0.3) {$\scriptstyle k\;{\rm times}$};
			}
		\end{tikzpicture}
	\end{array}$}
\end{equation}
The remaining fields transform according to \eqref{homo_eq}.
Essentially, along this trajectory the matrix elements marked by circular contours in \eqref{mat_1} and \eqref{mat_2} transform as follows: 
\begin{equation}
	1\to \int d\tau\, e^{x_1-\xi(\tau)}\cdot 1=\Lambda^{-1}\to \tilde\upsilon_{8,7},\quad \upsilon_{8,4}\to \int d\tau\, e^{x_2-\xi(\tau)}\cdot \upsilon_{8,4}=\Lambda \upsilon_{8,4}\to 1\,.
\end{equation}
A vacuum expectation value $1$ thus turns into a small tangent degree of freedom $\tilde\upsilon_{8,7}$, while a tangent degree of freedom $\upsilon_{8,4}$ is blown up into a vacuum expectation value $1$.
The remaining vacuum expectation values are unchanged, and the remaining tangent degrees of freedom are transported trivially.
To describe the Hecke correspondence in this case, we mod out the tangent space by a linearized gauge transformation \eqref{gauge} that depends on the new vacuum expectation values at $\tau_1$, and compare the result with the tangent space obtained at $\tau_1$ (here we describe explicitly just the field $C$):
\begin{equation}
	\begin{pmatrix}
		0 & 0 & 0 & 0 & 0 & 0 & 0 & 0 & 0 & \upsilon _{1,10} \\
		1 & 0 & 0 & 0 & 0 & 0 & 0 & 0 & 0 & \upsilon _{2,10} \\
		0 & 1 & 0 & 0 & 0 & 0 & 0 & 0 & 0 & \upsilon _{3,10} \\
		0 & 0 & 1 & 0 & 0 & 0 & 0 & 0 & 0 & c_{1,4}^{(2)}+\upsilon _{4,10} \\
		0 & 0 & 0 & 0 & 0 & 0 & 0 & 0 & 0 & c_{1,6}^{(1)} \\
		0 & 0 & 0 & 0 & 1 & 0 & 0 & 0 & 0 & c_{2,6}^{(1)} \\
		0 & 0 & 0 & 0 & 0 & 1 & 0 & 0 & 0 & c_{3,6}^{(1)} \\
		0 & 0 & 0 & 1 & 0 & 0 & \tilde\upsilon _{8,7} & 0 & 0 & c_{4,6}^{(1)}+c_{2,4}^{(2)} \\
		0 & 0 & 0 & 0 & 0 & 0 & 0 & 1 & 0 & c_{5,6}^{(1)}+c_{3,4}^{(2)} \\
		0 & 0 & 0 & 0 & 0 & 0 & 0 & 0 & 1 & c_{6,6}^{(1)}+c_{4,4}^{(2)} \\
	\end{pmatrix}\to\begin{pmatrix}
		0 & 0 & 0 & 0 & 0 & 0 & \tilde\upsilon _{1,7} & 0 & 0 & \tilde c_{1,7}^{(2)} \\
		1 & 0 & 0 & 0 & 0 & 0 & \tilde\upsilon _{2,7} & 0 & 0 & \tilde c_{2,7}^{(2)} \\
		0 & 1 & 0 & 0 & 0 & 0 & \tilde\upsilon _{3,7} & 0 & 0 & \tilde c_{3,7}^{(2)} \\
		0 & 0 & 1 & 0 & 0 & 0 & \tilde\upsilon _{4,7} & 0 & 0 & \tilde c_{4,7}^{(2)} \\
		0 & 0 & 0 & 0 & 0 & 0 & \tilde c_{1,3}^{(1)} & 0 & 0 & \tilde\upsilon _{5,10} \\
		0 & 0 & 0 & 0 & 1 & 0 & \tilde c_{2,3}^{(1)} & 0 & 0 & \tilde\upsilon _{6,10} \\
		0 & 0 & 0 & 0 & 0 & 1 & \tilde c_{3,3}^{(1)} & 0 & 0 & \tilde\upsilon _{7,10} \\
		0 & 0 & 0 & 1 & 0 & 0 & \tilde\upsilon _{8,7} & 0 & 0 & \tilde c_{5,7}^{(2)} \\
		0 & 0 & 0 & 0 & 0 & 0 & \tilde\upsilon _{9,7} & 1 & 0 & \tilde c_{6,7}^{(2)} \\
		0 & 0 & 0 & 0 & 0 & 0 & \tilde\upsilon _{10,7} & 0 & 1 & \tilde c_{7,7}^{(2)} \\
	\end{pmatrix}\,.
\end{equation}

Here we list the types of light fields that contribute to the amplitude:
\begin{enumerate}
	\item\label{item_one} Type $\upsilon_{1,10}\to\tilde c_{1,7}^{(2)}$.
	This process describes the appearance of a light field from a heavy one.
	It imposes the Hecke correspondence constraint $\tilde c_{1,7}^{(2)}=0$.
	These fields constitute $\sV^{(2)}_{\mathrm{zero}}$ in \eqref{decomp}.
	\item Type $c_{1,4}^{(2)}\to\tilde c_{4,7}^{(2)}$ and $c_{4,6}^{(1)}+c_{2,4}^{(2)}\to\tilde c_{5,7}^{(2)}$.
	This imposes the Hecke correspondence constraint $c_{1,4}^{(2)}=\tilde c_{4,7}^{(2)}$.
	These fields constitute $\sV^{(1)}_{\mathrm{eq}}$ and $\sV^{(2)}_{\mathrm{eq}}$ in \eqref{decomp}.
	\item Type $c_{1,6}^{(1)}\to\tilde\upsilon_{5,10}$.
	This is the inverse of the process in item \ref{item_one}.
	These fields constitute $\sV^{(1)}_{\mathrm{zero}}$ in \eqref{decomp}.
	\item\label{item_four} Type $c_{4,6}^{(1)}-c_{2,4}^{(2)}\to 0$.
	These degrees of freedom appear to be transported to zero.
	However, we consider tangent spaces modulo gauge transformations.
	It is therefore more accurate to say that these degrees of freedom become pure gauge on the other side.
	Hence the Hecke correspondence imposes no constraints on them.
	These fields constitute $\sV^{(1)}_{\mathrm{arb}}$ in \eqref{decomp}.
	\item Type $0\to\tilde c_{1,3}^{(1)}$.
	This is the inverse of the process in item \ref{item_four}.
	These fields constitute $\sV^{(2)}_{\mathrm{arb}}$ in \eqref{decomp}.
\end{enumerate}

The tunneling amplitude can then be constructed using \eqref{amplitude}.
There are two additional contributions: one from the fields $s_{a-i+1}^{(1)}$, $i=1,\ldots,k$, which lie in $\sV^{(1)}_{\mathrm{zero}}$, and one from the fields $\upsilon_{a-i+1,a+b-k}$, $i=1,\ldots,k$.
The latter fields are heavy and do not contribute to the amplitude directly.
However, all the heavy fields become light at $x_2=x_3$ in Fig.~\ref{fig:R_phys}, and those fields contribute to the Berry transport according to the prescriptions proposed in \cite[Sec.~5.2]{Galakhov:2025fhd}.
The resulting expression for the amplitude, computed from the Hecke correspondence produced by the instanton trajectory (the attracting leaf), is
\begin{equation}\label{A_leaf}
	\CA^{\rm (leaf)}_{|a\rangle\otimes|b\rangle \to |a-k\rangle\otimes|b+k\rangle}=\prod\lm_{i=1}^{{\rm min}(k,a-k)}\frac{[c_{i,a}^{(1)}]}{[\tilde c_{a-k-i+1,a-k}^{(1)}]}\cdot\prod\lm_{i=1}^k\frac{[s_{a-i+1}^{(1)}]}{[\tilde \upsilon_{a-i+1,a+b-k}]}\,.
\end{equation}
Here $[\ldots]$ denote the corresponding equivariant weights, i.e., the flavor charges.
In particular, we have
\begin{equation}
	[c_{i,j}^{(p)}]=[\tilde c_{i,j}^{(p)}]=2\hbar(i-j-1),\quad [s_{j}^{(p)}]=2\hbar(\ell-j+1),\quad [\tilde \upsilon_{a-i+1,a+b-k}]=u_{12}+2\hbar(a-b+2k-i)\,.
\end{equation}
Expression \eqref{A_leaf} coincides with the expression obtained from the fusion relations:
\begin{tcolorbox}
	\begin{equation}
		\CA^{\rm (fusion)}_{|a\rangle\otimes|b\rangle \to |a-k\rangle\otimes|b+k\rangle}=\CA^{\rm (leaf)}_{|a\rangle\otimes|b\rangle \to |a-k\rangle\otimes|b+k\rangle}\,.
	\end{equation}
\end{tcolorbox}


\section{Conclusion}

The subject of this paper is the factorization of Yangian $R$-matrices, which we find convenient to describe from several complementary perspectives.
A similar logic, however, applies to algebras with additional loop deformations, such as quantum loop algebras, quantum toroidal algebras, and elliptic algebras.
The canonical Gauss decomposition $R=LDU$ of the $R$-matrix admits three interpretations: algebraic, physical, and geometric.

Physically, we interpret the $R$-matrix as an interface across which the real parameters $x_1$ and $x_2$ associated with framing nodes of a quiver supersymmetric QFT are interchanged.
The factorization twist $U$, a ``half'' of the $R$-matrix, is then interpreted as a ``half'' of such an interface, at which the parameters $x_1$ and $x_2$ flow toward each other and become close.
Furthermore, we interpret the interface contribution as Berry transport in the effective infrared quiver SQM.
The \emph{upper-triangular} matrix $U$ is then decomposed into tunneling amplitudes, in which cells representing effective particles migrate in one direction along the flow generated by the height function $h$: from vacua with smaller values of $h$ to vacua with larger values of $h$.
This unidirectional flow explains the absence of instantons that would contribute to the entries of $U$ below the diagonal.

In this paper, we have been most interested in the structure of multi-cell tunneling amplitudes.
We have constructed recursion relations that allow higher multi-cell amplitudes to be obtained from more elementary ones.
We have also discussed the subtleties that arise in these relations.
We have considered several cases in which multi-cell amplitudes are glued from lower-cell ones.
In these cases we have observed finer cancellations among amplitudes, as well as wall-crossing phenomena.
Both are necessary for the recursion relations to hold.

In conclusion, we list several open problems that we believe are worth pursuing further:
\begin{itemize}
	\item As we have seen, tunneling instantons exhibit wall-crossing phenomena.
	Our $\mathcal{N}=4$ SQM is a dimensional reduction of a 2d $\mathcal{N}=(2,2)$ GLSM whose gauge-matter content is described by a quiver.
	One would therefore expect our wall-crossing phenomena to fit into the wider family of soliton wall-crossing in 2d $\mathcal{N}=(2,2)$ theories, or even into higher-dimensional classes of theories once loops are added to the algebra.
	From this point of view, relations between amplitudes such as \eqref{wc} should play the role of wall-crossing relations.
	It is natural to ask whether the spectral parameter introduces a new type of refinement into such relations, namely another charge commuting with the supercharge.
	On the one hand, the $\phi$-parameter decomposes into linear combinations of $\hbar$, whose role in other theories is played by the $\Omega$-background parameters.
	This type of refinement is captured, for example, by the isospin $J_3$ in class-$\CS$ theories \cite{Moore2012FelixKlein}.
	On the other hand, as we have seen, our case contains shifts dictated by the spectral parameters $u_i$ that is not naturally quantized in terms of $\hbar$ in a general situation.
	It would be interesting to study these shifts and this type of wall-crossing formulae further and to compare them with other refinements.
	\item One of the hopes prompted by the interplay of physics and algebra is a generalization of canonical algebraic structures.
	In our context, we ultimately hope to obtain integrability beyond the structures related to (affine) Lie (super)algebras.
	We therefore identify a few natural subdirections:
	\begin{enumerate}
		\item Chiral quivers \cite{Galakhov:2022uyu}, or non-symmetric quivers in the terminology of \cite{Cao:2026vwf}.
		For these quivers, there is no solution of the YBE of the form \eqref{Gauss}, in which the factorization twist has the usual decomposition $U=\bbone\otimes \bbone+\frac{f_0\otimes e_0}{u}+\ldots$.
		Nevertheless, the amplitude construction must still exist.
		Most likely, the YBE is broken because some classical vacua can run to or from infinity during transport.
		In general, this phenomenon breaks the wall-crossing relations, since contributions that run to or from infinity spoil the counting on either the left- or the right-hand side.
		We hope, however, that further study of these phenomena will help to correct these relations.
		\item As we mentioned, the Yangians considered here can be embedded in a wider class of quiver Yangians.
		The corresponding quivers describe toric Calabi-Yau threefolds \cite{Li:2020rij}.
		There has been noticeable recent progress in the study of equivariant cohomological properties of Calabi-Yau fourfolds \cite{Cao:2017swr,Nekrasov:2017cih,Nekrasov:2018xsb,Bonelli:2020gku,Szabo:2023ixw}.
		A natural question therefore arises: can structures analogous to $R$-matrices and Yangians be assigned to fourfolds as well?
		However, despite various proposals in the literature \cite{Bao:2025hfu,Kimura:2024osv} for a BPS algebra to replace the quiver Yangians, the ``right'' algebraic structure and its microscopic physical origin remain open questions.
		\item Another generalization of quiver varieties is that of bow varieties.
		These describe a gluing of the quiver representation of the ADHM equations with the Nahm equations, which together describe combined configurations of instantons and looped monopoles on ALE spaces \cite{Cherkis:2011bh,Takayama:2016bva,Nakajima:2017cbv}.
		Constructions of stable envelopes for bow varieties are available \cite{Rimanyi:2020bgc,Stroppel:2023eos,Botta:2023bvs}.
		It is therefore natural to ask whether this framework can be extended to the algebra and to integrability.
	\end{enumerate}
	\item Geometrically, factorization twists $U$ correspond to stable envelopes and therefore describe a transition from the stable basis in cohomology to the fixed-point basis.
	For the algebras $\mbox{$\myY$}(\widehat{\fg\fl}_n)$, $\mbox{$\myY$}(\widehat{\fg\fl}_{1|1})$, $\mbox{$\myT$}(\widehat{\fg\fl}_{1})$, and $\mbox{$\myT$}(\widehat{\fg\fl}_{1|1})$ corresponding to the affine Dynkin diagrams, the corresponding Fock representations admit bosonization in terms of ``time'' variables $p_k$.
	These $p_k$ correspond to certain cohomology classes, and both bases have a polynomial representation in terms of $p_k$.
	For $\mbox{$\myY$}(\widehat{\fg\fl}_1)$, $\mbox{$\myY$}(\widehat{\fg\fl}_{n+1})$, $\mbox{$\myY$}(\widehat{\fg\fl}_{1|1})$, $\mbox{$\myT$}(\widehat{\fg\fl}_1)$, and $\mbox{$\myT$}(\widehat{\fg\fl}_{1|1})$, the fixed-point basis corresponds to Jack, Uglov, super-Jack, Macdonald, and super-Macdonald polynomials (see, e.g., \cite{Galakhov:2024zqn,Tselousov:2025kdg} and references therein).
	The stable basis for $\mbox{$\myY$}(\widehat{\fg\fl}_1)$ is known \cite{MO} to correspond to Schur polynomials.
	In other words, for $\mbox{$\myY$}(\widehat{\fg\fl}_1)$, the matrix $U$ describes a transition from Jack to Schur polynomials.
	This transition is well known to be triangular, and so is $U$.
	Othertriangular matrices map all the mentioned polynomials to so-called symmetrized monomials; the corresponding matrix coefficients have the combinatorial meaning of Kostka numbers, or of Kostka functions in general.
	It would be interesting to learn whether the basis of Schur polynomials has a distinguished meaning in intrinsically physical terms, and whether triangular matrices of Kostka numbers admit a geometric interpretation in terms of stable envelopes or a physical interpretation in terms of tunneling amplitudes.
	\item It is natural to extend the previous question to other algebraically well-defined natural quantities associated with DIM algebras $\myT(\widehat{\fg\fl}_1)$, such as Baker-Akhiezer functions \cite{Mironov:2024tlc}, non-symmetric Macdonald polynomials \cite{Mironov:2025yak}, and commuting rays of operator modes in the DIM algebra \cite{Mironov:2020pcd,Mironov:2023wga}.
	Whether they have a physical interpretation in terms of quiver QFTs.
	It is expected that a transition between the commuting rays is induced by a form of wall-crossing \cite{Smirnov:2019abc} similar to Weyl reflections \cite{Galakhov:2026cwv}.
\end{itemize}


\section*{Acknowledgments}

The author would like to thank Dmitri Bykov, Alexei Morozov, and Yehao Zhou for illuminating discussions.
The research was carried out within the state
assignment of Ministry of Science and Higher Education of the Russian Federation for IITP RAS.

The author acknowledges the use of a free online AI assistant
solely for language editing and polishing of the manuscript. 
All scientific content, results, and conclusions are the author's own.


\bibliographystyle{utphys}
\bibliography{biblio}

\end{document}